\documentclass{article}
\usepackage[T1]{fontenc}
\usepackage[mathscr]{euscript}
\usepackage{amsmath}
\usepackage{stackrel, amssymb}
\usepackage{physics}
\usepackage{tensor}
\usepackage{graphicx}
\usepackage{mathtools}
\usepackage{slashed}
\usepackage{hyperref}
\hypersetup{ 
	colorlinks=true,
	linkcolor=blue,
	urlcolor=blue,
	citecolor=blue
}
\usepackage{geometry}
\usepackage{sectsty}
\subsectionfont{\normalfont\itshape}
\subsubsectionfont{\normalfont\itshape}
\usepackage[toc, page]{appendix}
\usepackage{xcolor}
\usepackage{biblatex}
\usepackage{authblk}
\title{Coulomb Branches in 3d $\mathscr{N} = 4$ Revisited}
\author{Spencer Tamagni}
\affil{\textit{Center for Theoretical Physics, University of California, Berkeley}}
\date{\today}

\begin{document}
\maketitle 
\begin{abstract}
Using ideas from the gauge theory approach to the geometric Langlands program, we revisit supersymmetric localization with monopole operators in 3d $\mathscr{N} = 4$ supersymmetric gauge theories subject to $\Omega$-deformation. The key novel feature of our setup is a pair of dual boundary conditions, which drastically simplify the dynamics of the theory and the nature of the localization loci. From a careful calculation with these boundary conditions, the mathematical definition of Coulomb branches proposed by Braverman, Finkelberg and Nakajima emerges naturally. It is straightforward to incorporate codimension two defects in the setup, and in this way we gain insight into Webster's construction of tilting bundles on Coulomb branches. 
\end{abstract}
\setcounter{tocdepth}{2}
\tableofcontents 

\section{Introduction}
In this paper, we will make use of ideas from the gauge theory approach to the geometric Langlands program \cite{kw} to revisit supersymmetric localization in $\Omega$-deformed three-dimensional $\mathscr{N} = 4$ gauge theories. The computations necessary for modern applications involve monopole operators, boundary conditions, and codimension two defects, so are quite subtle and require a certain degree of mathematical care. In the end, we will arrive at straightforward quantum field theory derivations of mathematical results of Braverman-Finkelberg-Nakajima \cite{bfn} and Webster \cite{webster2019} on the geometry of the Coulomb branches of these theories. 

\subsection{A pair of dual boundary conditions}
The key idea in our approach is to study the theory on $\mathbb{R}^2 \times [0, 1]$ with a judicious choice of boundary conditions, inspired by the discussion in sections 9 and 10 of \cite{kw}. We show explicitly that the moduli spaces of solutions to the Bogomolny equations considered in \cite{kw} arise as the localization loci of the path integral of a twisted version of 3d $\mathscr{N} = 4$ pure gauge theory in the presence of monopole operators. This (in addition to a straightforward generalization to include matter hypermultiplets) enables us to give a direct determination of the operator product algebra of such operators, a calculation from which the mathematical definition of \cite{bfn} emerges naturally. 

\subsubsection{}
Because some of the mathematical results obtained in this paper are previously known (predominantly by representation theorists), it is worth pointing out explicitly what is new. The most essential point is that we give a fully differential-geometric model for the computations in topologically twisted three-dimensional gauge theory that we perform here. The strategy we use is essentially the same as that pursued in the most classical papers on the subject \cite{wittencohft}, \cite{witten88}, \cite{atiyahjeffrey}, deformed by equivariant parameters as in \cite{nekrasov2002}. 

The formalism developed in \cite{bfn} and various follow-ups to study Coulomb branches, while beautiful, has the drawback that it has been unclear exactly \textit{why} it gives the answers to questions that physicists have previously considered in the context of 3d $\mathscr{N} = 4$ gauge theories. In this paper we will answer this question, in a way which is hopefully satisfactory to physicists and interested mathematicians. 

\subsubsection{}
An amusing result we find in pursuing this goal is that the convolution product defined on the equivariant (K-)homology of the affine Grassmannian \cite{bfm} and its generalization introduced in \cite{bfn} may be computed by simple index theory methods. This can be viewed as an uplift of the results of section 10 of \cite{kw} (see also \cite{witten09}) to the equivariant situation. We also gain an interpretation of homology classes in the ``variety of triples'' of \cite{bfn} in terms of virtual fundamental cycles associated to the moduli problem defined by theories with matter, realizing the paradigm of \cite{lns} in this context. Most considerations in this paper are made at the physical level of rigor, in that we ignore various analytic subtleties that arise in differential-geometric considerations.  

\subsubsection{}
Likewise, the mere fact that we study moduli spaces of solutions to the Bogomolny equations with boundary conditions means that we are able to shed some differential-geometric light on more abstract or conjectural approaches to boundary conditions in three-dimensional topological field theories. 

For example, boundary conditions in $\Omega$-deformed 3d $\mathscr{N} = 4$ theories are often described using deformation quantization modules over Higgs/Coulomb branches (depending on the choice of twist). By considering carefully the behavior of universal characteristic classes constructed by Atiyah and Bott \cite{atiyahbott} in our present context, we are able to (in a meaningful sense) explicitly \textit{derive} the modules over the quantized Coulomb branch algebra which are determined by concrete Dirichlet and Neumann boundary conditions on partial differential equations, confirming some conjectures made in \cite{Bullimore_2016}. 

We will explain some applications of this to enumerative geometry in a companion paper \cite{shiftop}.

\subsubsection{}
One benefit of the formalism we develop is that it is straightforward to extend it to incorporate a codimension two defect at the origin of $\mathbb{R}^2$. The relevant moduli spaces were considered in \cite{gw}, and may also be engineered by a certain $\mathbb{Z}_p$-orbifolding procedure \cite{Nekrasov_2018} from the case with no defect at all. Repeating our analysis in this context leads to a gauge theory realization of the cylindrical KLRW algebras, fully realizing the ``physics motivations'' of \cite{webster2019}. 

The fact that the defects may be realized via $\mathbb{Z}_p$-orbifolds is the basic point of contact with the characteristic $p$ methods of \cite{bk2007}. The projection to the quotient by $\mathbb{Z}_p$ realizing the equivalence between orbifold bundles and parabolic bundles can be viewed as a characteristic zero approximation to the Frobenius pushforward. We will not pursue making this analogy more precise in this work. 

\subsection{Relation to prior work}
In the physics literature on exact computations in supersymmetric gauge theories, there has been of course a lot of prior work on localization with monopole operators, for a very incomplete list of references see \cite{Gomis_2012}, \cite{Ito_2012}, \cite{Brennan_2018}, \cite{Dedushenko_2018}, the review \cite{okuda2014line} and references therein. 

It is more or less immediately clear that applying equivariant localization with respect to a maximal torus to the definition of the quantized Coulomb branch in \cite{bfn}, one will recover formulas for monopole operators obtained previously in the physics literature (this is essentially the relation of the definition of \cite{bfn} to the ``abelianization map'' introduced in \cite{bdg}). In this paper we will go beyond this observation, and explain the geometry behind the construction in \cite{bfn} from physics. 

In any approach to the problem of localizing 3d gauge theories, one has to deal with the following issue: the maximal torus in $SO(3)$ acting on $\mathbb{R}^3$ fixes a line $\mathbb{R} \subset \mathbb{R}^3$, so making an $\Omega$-deformation using the isometries of $\mathbb{R}^3$ only suffices to reduce the problem to an effective quantum mechanics. Our choices of boundary conditions allow this quantum mechanics to in turn be reduced to a problem in zero-dimensional field theory, in other words the analysis of finite-dimensional integrals. In \cite{vorticesandvermas}, \cite{Dimofte_2020}, this quantum mechanics was analyzed directly in terms of moduli spaces of vortices, a technique which tends to work best when the theory in question has enough matter fields to support an interesting Higgs branch. 

Our approach is somewhat complementary to the vortex quantum mechanics, and closer in spirit to \cite{bfn}. To our knowledge, \cite{Dimofte_2020} is the only other place in the physics literature where localization in ($A$-twisted) three dimensional $\mathscr{N} = 4$ theories with codimension two defects has been addressed systematically. 

\subsection{Outline of paper}
This paper is organized as follows. In Section \ref{bckgrd} we set up notations and recall basic notions of 3d $\mathscr{N} = 4$ gauge theories in the ``$A$-type'' topological twist. In Section \ref{loc} we discuss the localization locus of the path integral with our chosen boundary conditions in detail, extending the discussion in \cite{kw} to include equivariance with respect to the maximal torus of symmetries of the problem. In Section \ref{algebras} we use this to compute quantized Coulomb branch algebras, compare to \cite{bfn} and generalize to theories with matter. Section \ref{klrw} extends all this to include codimension two defects and give a physical computation of the KLRW algebras, generalizing in a similar way ideas from \cite{gw} and comparing to \cite{webster2019}. 

While this is a physics paper and the author is a physicist, we have attempted to make it accessible to both physicists and mathematicians. We also do not make any claims to complete rigor on the various analytic details related to infinite-dimensionality issues that must be addressed to properly justify any differential-geometric construction. Our considerations are mostly formal in this sense.

\subsection{Acknowledgements}
My understanding of Coulomb branches and localization with monopole operators was significantly influenced by various useful discussions with Mina Aganagic, Justin Hilburn, Yixuan Li, Nikita Nekrasov, Andrei Okounkov, and Peng Zhou. I am especially grateful to Mina, Justin, and Andrei for encouraging me to think about this problem.

\subsubsection{}
In particular, this paper grew out of efforts to explicitly sum up contributions from a gas of monopoles in $\mathbb{R}^3$ to deduce exact statements about effective actions of 3d $\mathscr{N} = 4$ theories, in parallel to what was done with instantons in 4d $\mathscr{N} = 2$ theories in \cite{no2003}. Unfortunately the moduli spaces of smooth monopoles in $\mathbb{R}^3$ seem to be slightly too non-compact to do this in a naive way, and there is no truly natural compactification we are aware of that remedies the issues. The workaround we use here from \cite{kw} is to place the theory in a box, which has the effect of severely constraining the ability of the dynamical monopoles to fluctuate. This makes the calculation of correlation functions almost trivial, but nontrivial enough to see the algebraic structures that are often associated to 3d $\mathscr{N} = 4$ theories.

\section{Background and Setup} \label{bckgrd}
In this section we recall some basic facts about the 3d $\mathscr{N} = 4$ theories under study, the twist we will use, and the nature of the localization locus. We concentrate on the case of pure gauge theory for simplicity, deferring the discussion of adding matter to section \ref{addhypers}.

\subsection{Basics on twisting and pure gauge theory}
We begin by recalling some basics on the 3d $\mathscr{N} = 4$ vector multiplet, $R$-symmetries and twisting, mainly to establish notations and conventions. We will be performing supersymmetric localization, so the choice of localizing supercharge is essential; we explain this choice in detail here. 

\subsubsection{}
Any three-dimensional quantum field theory with $\mathscr{N} = 4$ supersymmetry has $R$-symmetry group $SU(2)_C \times SU(2)_H$, with subscripts ``C'' and ``H'' signifying that these act respectively by rotation of the twistor spheres of the Coulomb and Higgs branches of the moduli space of vacua. If the theory is obtained by dimensional reduction from a six dimensional $(1, 0)$ theory, $SU(2)_H$ is the R-symmetry already visible in six dimensions, while $SU(2)_C$ arises as the double cover of the $SO(3)_R$ obtained in the course of reduction. 

We will consider a topological twist of the theory, obtained by replacing the group $SU(2)_E$ of rotations of Euclidean spacetime with the diagonal subgroup of $SU(2)_E \times SU(2)_H$. This twist is the same as the dimensional reduction of the Donaldson twist in a four-dimensional $\mathscr{N} = 2$ theory introduced in \cite{witten88}.

The supercharges in the physical theory carry indices as $Q_\alpha^{I\dot{I}}$, where $(\alpha, I, \dot{I})$ are indices for $SU(2)_E \times SU(2)_C \times SU(2)_H$. The twist amounts to identifying the index type of $\alpha$ and $\dot{I}$, leading to $SU(2)_C$ doublets of scalars $Q^I$ and one-forms $G_\mu^I$, $\mu = 0, 1, 2$. We pick out a particular scalar supercharge, $Q$, and denote its action on the fields by $\delta$. 

\subsubsection{}
The basic building block of 3d $\mathscr{N} = 4$ gauge theories is the vector multiplet. The bosonic field content of the vector multiplet is a gauge field $A$ and three real scalars $\vec{\sigma}$, the latter transforming in the three-dimensional representation of $SU(2)_C$. The fermions carry the same indices as the supercharges, so after twisting become two scalars which we will denote by $\psi_3$ and $\eta$, as well as two one-forms that we denote by $\psi$ and $\chi$. It is convenient to view the triplet of scalars as a single real scalar $\sigma$ and a complex scalar $\varphi$ with conjugate $\overline{\varphi}$. This splitting fixes the complex structure on the Coulomb branch. All fields are in the adjoint representation of the gauge group $G$. We consider only $G = U(k)$ in this paper. 

\subsubsection{}
The action of the supercharge $\delta$ on the fields of the vector multiplet is (it is convenient to introduce a bosonic auxiliary field $H$, an adjoint-valued one-form):
\begin{alignat}{3}
& \delta A = \psi \quad   &&   \delta \varphi = 0 \quad && \delta \chi = H \nonumber \\
& \delta \psi = D_A \varphi \quad && \delta \overline{\varphi} = \eta \quad && \delta H = \comm{\chi}{\varphi} \label{topsusy} \\
& \delta \sigma = \psi_3 \quad && \delta \eta = \comm{\overline{\varphi}}{\varphi} \nonumber \\
& \delta \psi_3 = \comm{\sigma}{\varphi} \quad && \quad && \quad \nonumber 
\end{alignat}
This satisfies $\delta^2 = (\text{gauge transformation generated by $\varphi$})$, so is nilpotent on gauge-invariant quantities. We will consider correlation functions of gauge-invariant supersymmetric observables in the cohomology of $\delta$. 

Since the supercharge $\delta$ is a scalar, it is conserved on an arbitrary three-manifold $W$ and the action of the theory on such a three-manifold, with boundary $\partial W$ is given by 
\begin{equation} \label{topaction}
S = -\frac{1}{e^2} \int_{\partial W } \tr( \sigma F_A ) + \delta \Bigg( \frac{1}{e^2} \int_W \tr\Big( \chi \wedge( i(F_A + \star D_A \sigma) + \frac{\lambda}{2} \star H) + \psi \wedge \star D_A \overline{\varphi} + \psi_3 \wedge \star \comm{\sigma}{\overline{\varphi}} + \eta \star \comm{\overline{\varphi}}{\varphi} \Big) \Bigg). 
\end{equation}
In this formula, $e^2$ is the three-dimensional gauge coupling. By invariance under $\delta$-exact variations, we may choose any value for the parameter $\lambda$ that we like. $\lambda = 0$ will eventually be the convenient choice for localization, while when $\lambda = 1$, integrating out $H$ cancels the boundary term and gives for the bosonic part of the action 
\begin{equation}
L_B = \frac{1}{e^2} \Bigg( \frac{1}{2} \norm{F_A}^2 + \frac{1}{2} \norm{D_A \sigma}^2 + \norm{D_A \varphi}^2 + \norm{\comm{\sigma}{\varphi}}^2 + \norm{\comm{\varphi}{\overline{\varphi}}}^2 \Bigg)
\end{equation}
which is the standard one for 3d $\mathscr{N} = 4$ super-Yang-Mills. 

\subsection{KW boundary condition and $\Omega$-deformation} \label{bc}
To fully set the stage for our analysis, we require two additional details. The first is that we want to work on the specific three-manifold $W = \mathbb{R}^2 \times [0, 1]$, with boundary conditions considered in \cite{kw} that we recall presently. The second is that we wish to consider the $\Omega$-deformed version of the theory \cite{nekrasov2002}, \cite{no2003}, \cite{nw2010}, corresponding to working equivariantly with respect to the $U(1)$ action rotating $\mathbb{R}^2$. 

\subsubsection{}
First, the choice of boundary condition. We denote the coordinates on $\mathbb{R}^2 \times [0, 1]$ by $(x, y, t)$, and use a complex coordinate $z = x + iy$ on $\mathbb{R}^2 \simeq \mathbb{C}$ in much of what follows. We record here the boundary conditions on the bosonic fields, see appendix \ref{bcdetails} for the detailed boundary conditions on fermions. 

At the $t = 0$ boundary, we impose the conditions 
\begin{equation} \label{DBC}
\begin{split}
A(t = 0, z, \bar{z}) & = 0 \\
D_t \sigma(t = 0, z, \bar{z}) & = 0 \\
\varphi(t = 0, z, \bar{z}) & = \text{diag}(\varphi_1, \dots, \varphi_k). 
\end{split}
\end{equation}
The last equation is written specifically for the choice of gauge group $G = U(k)$, more generally one takes the boundary value of $\varphi$ to be a fixed generic element in a Cartan subalgebra $\mathfrak{h}_{\mathbb{C}} \subset \mathfrak{g}_{\mathbb{C}}$. 

At the $t = 1$ boundary, we impose 
\begin{equation} \label{NBC}
\begin{split}
i_{\partial_t} F_A(t = 1, z, \bar{z}) & = 0 \\
\sigma(t = 1, z, \bar{z}) & = 0 \\
D_t \varphi(t = 1, z, \bar{z}) & = 0. 
\end{split}
\end{equation}
In English, $A$ and $\varphi$ obey Dirichlet boundary conditions at $t = 0$ and covariant Neumann boundary conditions at $t = 1$, and vice versa for $\sigma$. Slightly more generally, we may fix the restriction of $A$ to $t = 0$ to be an arbitrary flat connection, though for the present geometry in the absence of defects there are not many interesting choices. Note with this choice of boundary conditions, the boundary term in the action \eqref{topaction} vanishes. 

\subsubsection{}
In addition to these boundary conditions, we divide only by gauge transformations which are trivial at $t = 0$ and infinity of $\mathbb{R}^2 \simeq \mathbb{C}$. We view the gauge transformations acting nontrivially on the boundary as part of the global symmetries. These boundary conditions are the simplest ones for which our localization equations may be interpreted as the vanishing of a real moment map, together with some holomorphic equation. 

\subsubsection{}
A related point is that we fix $\varphi$ to the same generic value at $|z| \to \infty$ (and assume all other fields die off sufficiently rapidly there), and do not integrate over the zero mode of $\varphi$ in the path integral. Instead, we view the parameters $\varphi_i$ as equivariant parameters/twisted masses associated to the $G \simeq \mathcal{G}/\mathcal{G}_0$ global symmetry, where $\mathcal{G}$ denotes the group of gauge transformations and $\mathcal{G}_0$ the normal subgroup of those which are trivial on the boundary. That this interpretation is valid is obvious given the explicit form of the supercharge $\delta $ \eqref{topsusy}, as it may be interpreted as a differential in a model for $\mathcal{G}$-equivariant cohomology of the space of fields following \cite{atiyahjeffrey}. This interpretation is essential for the localization analysis. 

\subsubsection{}
With all this in mind, the $\Omega$-deformation is straightforward to describe. One simply deforms the supercharge $\delta$ to a new one $\delta_\varepsilon$, by realizing $\delta_\varepsilon$ as a differential in a model for $U(1)_\varepsilon \ltimes \mathcal{G}$-equivariant cohomology of the space of fields. $\varepsilon$ denotes the equivariant parameter for the $U(1)$ action rotating $\mathbb{R}^2 \simeq \mathbb{C}$. The $\Omega$-deformed theory is roughly obtained by the formal substitution $\varphi \to \varphi - V^\mu_\varepsilon D_\mu$, where $D_\mu$ denotes the covariant derivative with respect to the gauge connection $A$ and $V^\mu_\varepsilon$ is the vector field generating the $U(1)_\varepsilon$ action. 

More precisely, the $\Omega$-deformed supersymmetry $\delta_\varepsilon$ is 

\begin{alignat}{3}
& \delta_\varepsilon A = \psi \quad   &&   \delta_\varepsilon \varphi = i_{V_\varepsilon} \psi \quad && \delta_\varepsilon \chi = H \nonumber \\
& \delta_\varepsilon \psi = D_A \varphi + i_{V_{\varepsilon}} F_A \quad && \delta_\varepsilon \overline{\varphi} = \eta + i_{V_\varepsilon} \psi \quad && \delta_\varepsilon H = \comm{\chi}{\varphi} + D_A i_{V_\varepsilon} \chi + i_{V_\varepsilon} D_A \chi \label{omegasusy} \\
& \delta_\varepsilon \sigma = \psi_3 \quad && \delta_\varepsilon \eta = \comm{\overline{\varphi}}{\varphi} + i_{V_\varepsilon} D_A \overline{\varphi} - i_{V_\varepsilon} D_A \varphi \nonumber \\
& \delta_\varepsilon \psi_3 = \comm{\sigma}{\varphi} + i_{V_\varepsilon} D_A \sigma \quad && \quad && \quad \nonumber 
\end{alignat}
This now squares to a combination of a gauge transformation and a rotation. The $\Omega$-deformed action is 
\begin{equation} \label{omegaaction}
\begin{split}
S_\Omega = -\frac{1}{e^2} \int_{\partial W} \tr(\sigma F_A) + \delta_\varepsilon \Bigg( \frac{1}{e^2}\int_W \tr\Big( \chi \wedge(i(F_A + \star D_A \sigma) + \frac{\lambda}{2} \star H) \\
+ \psi \wedge \star(D_A \overline{\varphi} + i_{V_\varepsilon} F_A) + \psi_3 \wedge \star (\comm{\sigma}{\overline{\varphi}} + i_{V_\varepsilon} D_A \sigma) + \eta \star (\comm{\overline{\varphi}}{\varphi} + i_{V_\varepsilon} D_A \varphi - i_{V_\varepsilon}D_A \overline{\varphi}) \Big) \Bigg). 
\end{split}
\end{equation}
The effect of the $\Omega$-deformation will be the standard one \cite{nekrasov2002}, \cite{no2003}: as we recall briefly, the path integral of this theory can be viewed in a certain cohomological fashion that makes localization onto a moduli space of supersymmetric configurations obvious. The $\Omega$-background (and zero mode of $\varphi$) will introduce a potential on this moduli space, which further localizes the problem to the fixed points of the maximal torus inside of the $U(1)_\varepsilon \times G$ global symmetries, realizing field-theoretically the phenomenon of localization in equivariant cohomology. 

\subsection{Supersymmetric observables}
Our real interest is in the operator product expansion of supersymmetric local operators in this theory, in particular the monopole operators. Specifically, we wish to consider local operators realizing nontrivial cohomology classes of the supercharge $\delta$ (and $\delta_\varepsilon$). 

\subsubsection{}
The most basic supersymmetric local operators are gauge-invariant polynomials in $\varphi(x, y, t)$, as immediately follows from \eqref{topsusy}. In the presence of $\Omega$-deformation, these must be inserted at the origin of $\mathbb{R}^2$ to be supersymmetric. For $G = U(k)$, the generating function for single trace operators built out of $\varphi$ is 
\begin{equation} \label{yobs}
\mathscr{Y}(x, t) = \det(x - \varphi(0, t))
\end{equation}
which is $\delta_\varepsilon$-invariant. In a general situation, the observable obtained may depend (albeit not in a continuous fashion) on the value of the coordinate $t$ along $[0, 1]$; we will explain this more fully in section \ref{univ}.

\subsubsection{}
More interesting are the monopole operators. These are disorder operators, specified by modifying the domain of integration in the path integral to allow for certain singularities in the fields. A supersymmetric monopole operator is labeled by a choice of homomorphism $\rho: U(1) \to G$, and may be inserted at a point $p$. Identify $W$ with $\mathbb{R}^3$ in a local neighborhood of $p$, so that $(r, \theta, \phi)$ are spherical coordinates and $p$ is at $r = 0$. Then insertion of the monopole operator $\mathscr{O}_\rho(p)$ is defined by the instruction to perform the path integral over fields which, relative to some trivialization of the gauge bundle $E$ near $p$, have the singular behavior as $r \to 0$: 
\begin{equation} \label{singmono}
\begin{split}
F_A & \sim  \frac{\mu}{2} \sin \theta d\theta \wedge d\phi \\
\sigma & \sim \frac{\mu}{2r}. 
\end{split}
\end{equation}
We have written the singularity of $A$ in terms of the field strength. Only the fields $(A, \sigma)$ are singular, all others are taken to be regular. Here, $\mu$ is the image of $i \in \text{Lie} \, U(1)$ under the Lie algebra homomorphism induced by $\rho$; for $G = U(k)$, $\mu = \text{diag}(i\mu_1, \dots, i\mu_k)$, where $\mu_1 \geq \dots \geq \mu_k$ are integers. We will also denote such a monopole operator by $\mathscr{O}_\mu(p)$. It depends on $\rho$ only up to conjugation in $G$. The gauge transformations are restricted to those which commute with $\rho(U(1))$ at $p$. This latter fact enables one to additionally insert an arbitrary polynomial in $\varphi$ at $p$, invariant under the subgroup of $G$ commuting with $\rho(U(1))$; such an operator is referred to as a ``dressed'' monopole operator.

\subsubsection{}
Since \eqref{singmono} is a solution to the abelian Bogomolny equations, and $F_A + \star D_A \sigma$ is $\delta$-exact on shell (i.e. after integrating out $H$ and substituting its value in \eqref{topsusy}), this observable is compatible with $\delta$. It is compatible with $\delta_\varepsilon$ provided it is rotation invariant, that is, $p$ has coordinates $(0, 0, t)$ for some $t \in [0, 1]$. 

We will be interested in the operator product algebra of monopole operators in the $\Omega$-deformed theory. We will determine this algebra simply by considering a sufficiently large class of correlation functions involving these operators, and studying the limit where two operators coincide. In this setup, all operators must be inserted at $z = 0$ in $\mathbb{R}^2 \simeq \mathbb{C}$ to preserve supersymmetry, but may be at distinct points on the $t$-interval. Then we consider correlation functions of the type 
\begin{equation} \label{corr}
\langle \mathscr{O}_{\mu_n}(t_n) \dots \mathscr{O}_{\mu_1}(t_1) \mathscr{P} \rangle 
\end{equation}
where $\mathscr{P}$ contains all dressing factors and possible additional insertions of $\varphi$ at other values of $t$. We will show in the following sections that such correlation functions may be computed exactly by explicitly reducing the path integral to a finite-dimensional integral.

\subsubsection{}
Because the three-dimensional theory is placed on $\mathbb{R}^2 \times [0, 1]$ with $\Omega$-deformation, one may expect (as the notation above suggests) that these computations should really be viewed as computing matrix elements in a certain effective quantum mechanics along $[0, 1]$. This is indeed the case, as we explain in section \ref{branes} by making contact with \cite{nw2010} and quantization by branes \cite{Gukov_2009}. This quantum-mechanical formulation provides the most useful way to interpret our calculations. 

\subsection{Coulomb branch}
One motivation to study this problem is that it has become clear in recent years that it is a highly effective way to characterize the Coulomb branch $\mathscr{M}_C$ of the moduli space of vacua of this theory \cite{bdg}. This has become a vast subject that we cannot review completely here, but we mention the important points to motivate the rest of this paper.   

\subsubsection{}
On general grounds \cite{sw1996}, the Coulomb branch is a hyperkahler manifold parameterized by the vacuum expectation values of the triplet of real vector multiplet scalars $\vec{\sigma}$ and a scalar which is dual to the 3d gauge field $A$ called the dual photon. All of these fields are valued in the Cartan $\mathfrak{h} \subset \mathfrak{g}$, considered up to the action of the Weyl group. The duality transform sends the vector multiplet to a hypermultiplet, so the low energy effective action of the theory in the Coulomb phase is that of a nonlinear sigma model with target $\mathscr{M}_C$. 

The distinguished $\mathbb{P}^1$ of complex structures (twistor sphere) of $\mathscr{M}_C$ is rotated by the $SU(2)_C$ group of $R$-symmetries; since our choice of $\delta$ explicitly breaks this symmetry, we have singled out a particular complex structure on the Coulomb branch that we call $I$, corresponding to the splitting of three real scalars into a complex scalar $\varphi$ and a real scalar $\sigma$. After dualizing the gauge field, $\sigma$ is complexified by the dual photon, which fully characterizes the complex structure $I$. 

\subsubsection{}
By similar formal reasoning, the operator product algebra of local operators in the twisted theory (the twist being defined using the supercharge $\delta$) is expected to coincide with the algebra of $I$-holomorphic functions on $\mathscr{M}_C$. A determination of this algebra based on physically reasonable constraints and comparisons with string theory was proposed in \cite{bdg}, and a mathematical definition of the algebra was given in \cite{bfn}. 

In the presence of $\Omega$-deformation, the operator product algebra becomes noncommutative because operators must be inserted along a line at the origin on $\mathbb{R}^2$. This is expected to produce the deformation quantization of the algebra of $I$-holomorphic functions on $\mathscr{M}_C$ \cite{nw2010}, an expectation that is reproduced in a rather beautiful way by the mathematican definition \cite{bfn}. 

\subsubsection{}
While it has become evident that the formalism initiated in \cite{bfn} gives the physically correct answers to various questions concerning Coulomb branches, it has not been clear precisely why this is the case. In this paper we will fill this gap, and argue that a careful computation in the underlying three-dimensional gauge theory using the boundary conditions discussed above will recover the mathematical formalism that has been developed to describe Coulomb branches. 

\section{Localization and monopole moduli spaces} \label{loc}
In this section, we describe the setup and technical ingredients of the localization calculation in the pure gauge theory. We will discuss the actual formulas obtained using these methods in section \ref{algebras}.

\subsection{Path integral formalities}
The correlation functions of monopole operators and functions of the field $\varphi$ that we have decsribed in the previous section are formally computed by path integrals in twisted 3d $\mathscr{N} = 4$ gauge theory
\begin{equation} \label{pathint}
\langle \mathscr{O}_{\mu_n}(t_n) \dots \mathscr{O}_{\mu_1}(t_1) \mathscr{P} \rangle = \int_{\text{BC, singularities}} DA D\sigma D\psi D\psi_3 D\varphi D \overline{\varphi} D \eta D \chi DH e^{-S} \mathscr{P}(\varphi).
\end{equation}
The subscript in the integration means that we integrate over fields with the boundary conditions \eqref{bc} at $t = 0, 1$, and with fixed singular behavior near the insertion points of monopole operators as described in \eqref{singmono}. The action $S$ is taken to be \eqref{topaction} in absence of $\Omega$-deformation, and \eqref{omegaaction} with $\Omega$-deformation. 

To actually make sense out of this prescription, we will argue that the integral on the right hand side formally reduces to a well-defined finite dimensional integral over some moduli space, and then spend the rest of the present section developing the tools to evaluate the finite-dimensional integral. 

\subsubsection{}
The argument for the reduction is a very old and standard one, arising from the following observations (see \cite{atiyahjeffrey} for details, and \cite{moorecohft} for a comprehensive review aimed at physicists). The fields $(A, \sigma, \psi, \psi_3, \varphi)$ together with the action of the supercharge $\delta_\varepsilon$ realize the Cartan model for the $U(1)_\varepsilon \ltimes \mathcal{G}$-equivariant cohomology of the (infinite-dimensional) space of fields $(A, \sigma)$. The part of the path integral involving $(\chi, H)$ is nothing but a Mathai-Quillen integral representation of the Euler class of a certain equivariant vector bundle over the space of $(A, \sigma)$. It is cohomologous under $\delta_\varepsilon$ to a delta form supported on the locus $F_A + \star D_A \sigma = 0$. The part of the path integral involving $(\varphi, \overline{\varphi}, \eta)$ is the projection form in the language of \cite{moorecohft}, which implements the projection onto the quotient by $\mathcal{G}_0$. 

\subsubsection{}
In more physical terms, when $\lambda = 0$ in \eqref{topaction}, $\chi$ and $H$ may be integrated out exactly to produce a supersymmetric delta function on the space of solutions to the equation 
\begin{equation} \label{monopole}
F_A + \star D_A \sigma = 0
\end{equation}
which is nothing but the Bogomolny equation. The insertion of monopole operators instructs one to study solutions to this equation with prescribed Dirac singularities near insertion points. Likewise, one may integrate out $\varphi, \overline{\varphi}, \eta$ in a simple fashion; some details about this procedure can be found in appendix \ref{universalappendix}. The full path integral then reduces to just an integral over the finite-dimensional space of zero modes of $(A, \sigma, \psi, \psi_3)$. 

\subsubsection{}
Specifically, we are led to study the space of solutions to the Bogomolny equations with Dirac singularities, considered up to $\mathcal{G}_0$ gauge equivalence and the boundary conditions discussed in section \ref{bc}. The supercharge $\delta_\varepsilon$ descends to the $U(1)_\varepsilon \times G$-equivariant differential in the Cartan model of the moduli space. Path integrals computing correlation functions of supersymmetric observables collapse to integrals of $U(1)_\varepsilon \times G$-equivariant cohomology classes (the argument we summarized here is almost exactly the same as that justifying the instanton counting \cite{nekrasov2002}, the only new ingredient being boundary conditions). Explicitly, these considerations allow us to reduce the path integral computing the correlation functions \eqref{pathint} to a finite-dimensional integral:
\begin{equation} \label{findimint}
\langle \mathscr{O}_{\mu_n}(t_n) \dots \mathscr{O}_{\mu_1}(t_1) \mathscr{P} \rangle = \int_{\mathscr{M}} \Omega_{\mathscr{P}}
\end{equation}
where $\Omega_{\mathscr{P}}$ is a certain equivariant cohomology class on $\mathscr{M}$ determined by $\mathscr{P}$, and $\mathscr{M}$ is the appropriate moduli space of solutions to Bogomolny equations with singularities determined by the insertions of the $\mathscr{O}_{\mu_i}$. To be more specific, we denote the moduli space of solutions to Bogomolny equations modifications with Dirac singularities at $p_1, \dots, p_n$ with charges $\mu_1, \dots, \mu_n$ as $\mathscr{M}(\{ \mu_i; p_i \})$. Because of the $U(1)_\varepsilon$ equivariance, we always consider $p_i$ with coordinates $(0, t_i)$.  

Thus the first and most essential step is to understand the geometry of $\mathscr{M}$, and after this the problem is reduced to straightforward evaluation of integrals using the localization formula, provided we may determine the classes $\Omega_{\mathscr{P}}$. We explain how to deal with these issues in the present section. 

\subsubsection{}
The reader should view the integrals above as directly analogous to the integrals over instanton moduli spaces that arise in the context of Donaldson theory \cite{witten88} and instanton counting \cite{nekrasov2002}. One difference is that the interesting aspect of our setup is not so much the actual value of the integrals (as we will see soon, they are relatively trivial to compute in the interesting cases), but the ability to reduce a sufficiently large class of correlation functions to such integrals so as to use them as an effective tool in determining the operator product algebra.  

\subsection{Localization locus and Hecke modifications} \label{locushecke}
As discussed above, we wish to study the moduli space of solutions to the monopole equations \eqref{monopole} with the boundary conditions of section \ref{bc} and Dirac singularities \eqref{singmono} of some specified charges $\mu_i$, considered up to gauge equivalence under the group $\mathcal{G}_0$ of gauge transformations trivial at $t = 0$ and at $|z| \to \infty$. 

Fortunately, precisely this situation was studied by Kapustin and Witten in sections 9 and 10 of \cite{kw}. We recall the main points of their analysis in this section.  

\subsubsection{}
Explicitly in coordinates $(z, t)$, \eqref{monopole} becomes 
\begin{equation} \label{holceq}
\begin{split}
\comm{D_t + i \sigma}{\overline{D_z}} & = 0 \\
F_{z\bar{z}} + \frac{i}{2} D_t \sigma & = 0. 
\end{split}
\end{equation}
The second equation may be interpreted as a moment map for the action of $\mathcal{G}_0$ on the space of fields, which we endow with the symplectic structure (recall $z = x + iy$): 
\begin{equation}
\omega_{\mathbb{R}} = \int_W \tr( \delta A_x \wedge \delta A_y + \delta A_t \wedge \delta \sigma) dtdxdy. 
\end{equation}
Verifying that the second equation in \eqref{holceq} is actually the moment map depends in an essential way on the boundary conditions of section \ref{bc}; the calculation requires some integration by parts, and the boundary conditions ensure that the surface terms encountered vanish. The assertion remains valid in the presence of Dirac singularities, when gauge transformations are also restricted to preserve the form of the singularity as in our setup. 

\subsubsection{}
The symplectic form $\omega_{\mathbb{R}}$ is clearly compatible with the complex structure on the space of fields in which $A_{\bar{z}}$ and $A_t + i \sigma$ are holomorphic. By the usual equivalence between symplectic and GIT quotients, we may equivalently obtain the moduli space of solutions by studying the space of $(A_{\bar{z}}, A_t + i \sigma)$ solving the first equation in \eqref{holceq} as a complex manifold, modulo the action of the complexified (that is, $G_{\mathbb{C}}$-valued) gauge transformations. Typically, when dealing with GIT quotients one also needs to take into account a stability condition, but in this case all orbits of the group of complexified gauge transformations turn out to be stable \cite{kw} so there is no need to make such a specification. 

\subsubsection{}
It is straightforward to describe the moduli space in holomorphic language. Denote the gauge bundle over $W$ by $E$. Its restriction $E_t$ to a fixed value of $t$ (away from a Dirac singularity) determines a holomorphic bundle on the spatial slice $\mathbb{C}$, with fixed trivialization near infinity. Equivalently, it extends to a framed holomorphic bundle on $\mathbb{P}^1$. The holomorphic structure is determined by the $(0, 1)$ part of the connection $\overline{D_z}$ (there is no integrability condition for $\overline{\partial}$-operators in one complex dimension, of course). The first equation in \eqref{holceq} says that the $\overline{D_z}$ operator is independent of $t$ up to a $G_{\mathbb{C}}$-valued gauge transformation generated by $A_t + i \sigma$. What this means is that, in a region with no singularities, the holomorphic bundles $E_t$ and $E_{t'}$ for any pair of time slices $t$ and $t'$ are naturally isomorphic. 

\subsubsection{}
Thus, in the presence of Dirac singularities at $z = 0$ and $t_1, \dots, t_n$, after taking the quotient by complexified gauge transformations the moduli space parameterizes tuples of such holomorphic bundles $(E_0, E_1, \dots, E_n)$, where $E_i$ is the bundle in the region $t_i < t < t_{i + 1}$. The only moduli of the solution are those which encode the ``jumping'' behavior of $E$ across the singularities, relating $E_{i - 1}$ and $E_i$.

It is explained in section 9 of \cite{kw} that upon crossing a singularity, $E_t$ undergoes a Hecke modification at the point $z = 0$. In more explicit terms, this means the following. We consider holomorphic bundles with fixed trivialization near $|z| \to \infty$. Such a bundle is encoded by a $G_{\mathbb{C}}$-valued holomorphic transition function $g(z)$ on $\mathbb{C} \setminus \{ 0 \}$ relating the bundle over a patch covering a neighborhood of $z = 0$, and the bundle on $\mathbb{C} \setminus \{ 0 \}$. The bundle is holomorphically trivial if and only if $g(z)$ is nonsingular at $z = 0$. It is obtained from the trivial bundle by a Hecke modification of type $\mu$ if $g(z)$ behaves as $z^\mu$ as $z \to 0$. 

The Dirichlet boundary condition at $t = 0$ says that $E_0$ is fixed to be the trivial bundle. Then $E_1$ is obtained from $E_0$ by a Hecke modification of type $\mu_1$, $E_2$ from $E_1$ by a Hecke modification of type $\mu_2$, and so on. The Neumann boundary condition at $t = 1$ says that there is no constraint on the bundle $E_n$; it is encoded simply by the data of Hecke modifications which came from earlier values of $t$. 

The conclusion of the discussion in sections 9 and 10 of \cite{kw}, summarized in the briefest possible terms here, is that the moduli space of solutions of the Bogomolny equations $\mathscr{M}(\{ \mu_i; p_i \})$ with our by now familiar boundary conditions and Dirac singularities at points $p_i$ with coordinates $(z, t) = (0, t_i)$, is precisely the space of successive Hecke modifications of an otherwise trivial holomorphic bundle, of types determined by the charges $(\mu_1, \dots, \mu_n)$. 

The description of Hecke modifications in terms of transition functions leads to an explicit description of the relevant moduli spaces via the affine Grassmannian, which is the basic point of connection with \cite{bfn} and the source of our gauge-theoretic understanding of the results there.  

\subsection{Holomorphic data and affine Grassmannian} \label{grg}
Now we will formalize what we mentioned above about the transition functions $g(z)$. 

\subsubsection{}
Let $\mathscr{K} = \mathbb{C}((z))$ denote the field of formal Laurent series in the variable $z$ and $\mathscr{O} = \mathbb{C}[[z]]$ denote the ring of formal power series. Then the (set of $\mathbb{C}$-points of the) affine Grassmannian of $G$ is defined as 
\begin{equation}
\text{Gr}_G := G_{\mathbb{C}}(\mathscr{K})/G_{\mathbb{C}}(\mathscr{O}).
\end{equation}
This infinite-dimensional space can be viewed as parameterizing all possible Hecke modifications at $z = 0$ of a fixed trivial $G_{\mathbb{C}}$-bundle on $\mathbb{C}$, of unspecified Dirac charge (strictly speaking, since we use formal series we should replace $\mathbb{C}$ by the formal disk). We also understand all of our bundles as trivialized at infinity, or equivalently bundles on $\mathbb{P}^1$ with framing. 

The correspondence is as explained in the previous section: any such bundle is characterized by a transition function $g(z) \in G_{\mathbb{C}}(\mathscr{K})$, and two functions $g(z)$ and $g'(z)$ describe equivalent bundles if and only if $g^{-1}(z) g'(z) \in G_{\mathbb{C}}(\mathscr{O})$. This shows that such bundles are in natural correspondence with points in $\text{Gr}_G$. 

The space of Hecke modifications of type $\mu$ is identified with a left $G_{\mathbb{C}}(\mathscr{O})$-orbit in $\text{Gr}_G$. In our notation, these can be identified with our moduli spaces for $n = 1$ Dirac singularity of charge $\mu$ at some point $p$:
\begin{equation}
\mathscr{M}(\mu; p) \simeq \text{Gr}_G^\mu := G_{\mathbb{C}}(\mathscr{O}) z^\mu \subset \text{Gr}_G
\end{equation}
which just formalizes what was said in the previous section. See appendix \ref{affineflags} for more details on the decomposition of $\text{Gr}_G$ into $G_{\mathbb{C}}(\mathscr{O})$-orbits. 

\subsubsection{}
Viewing the $G_{\mathbb{C}}(\mathscr{O})$-orbits as attracting manifolds as in appendix \ref{affineflags}, it is clear that the moduli space is generically non-compact and a natural compactification is provided by taking the orbit closure $\overline{\mathscr{M}}(\mu; p) := \overline{\text{Gr}}_G^\mu$. These orbit closures are referred to as Schubert varieties, or sometimes affine Schubert varieties. These compactifications are generically singular, but are smooth when $\mu$ is minuscule. In that situation, the orbit is already closed. 

\subsubsection{}
The spaces $\mathscr{M}(\{ \mu_i; p_i \})$ with multiple Dirac singularities parameterize successive Hecke modifications. In terms of transition functions this means the following. Such successive modifications correspond to $n$-tuples $(g_1(z), \dots, g_n(z)) \in G_{\mathbb{C}}(\mathscr{K})^n$, modulo the action of $G_{\mathbb{C}}(\mathscr{O})^{n - 1}$ given by $g_i(z) \sim h_{i - 1}^{-1}(z) g_i(z) h_i(z)$, with $h_i(z) \in G_{\mathbb{C}}(\mathscr{O})$ and $h_0(z) := 1$. This quotient is sometimes called a convolution Grassmannian and written $\text{Gr}_G \widetilde{\times} \dots \widetilde{\times} \text{Gr}_G$. In the notation of the previous section, the transition function of the bundle $E_i$ is $g_1(z) \dots g_i(z)$. Restricting to Hecke modifications of specified types gives convolutions of $G_{\mathbb{C}}(\mathscr{O})$-orbits: 
\begin{equation}
\mathscr{M}(\mu_1, \dots, \mu_n; p_1, \dots, p_n) \simeq \text{Gr}_G^{\mu_1} \widetilde{\times} \dots \widetilde{\times} \text{Gr}_G^{\mu_n}
\end{equation}
and taking closures gives a compactification
\begin{equation}
\overline{\mathscr{M}}(\mu_1, \dots, \mu_n; p_1, \dots, p_n) := \overline{\text{Gr}}_G^{\mu_1} \widetilde{\times} \dots \widetilde{\times} \overline{\text{Gr}}_G^{\mu_n}. 
\end{equation}
In this language, there is an obvious map (induced from a map $\text{Gr}_G \widetilde{\times} \dots \widetilde{\times} \text{Gr}_G \to \text{Gr}_G$):
\begin{equation}
\begin{split}
m: \, \, \, & \overline{\mathscr{M}}(\mu_1, \dots, \mu_n; p_1, \dots, p_n) \to \overline{\mathscr{M}}\Big( \sum_{i = 1}^n \mu_i; p \Big) \\
& (g_1(z), \dots, g_n(z)) \mapsto g_1(z) g_2(z) \dots g_n(z). 
\end{split}
\end{equation}
In terms of the gauge-theoretic description of the previous section, this map just sends $(E_0, E_1, \dots, E_n) \mapsto (E_0, E_n)$. This map is a partial resolution of singularities, and acting by $m$ corresponds physically to colliding the singular Dirac monopoles. Thus, $m$ plays an important role in the operator product expansion for monopole operators. If all $\mu_i$ are minuscule, then the convolution variety is smooth and this map is a full resolution of singularities; this is the case we will focus on. 

\subsubsection{}
The ideas in this section have been reviewed for completeness to make contact with \cite{bfn}, and also to give a more concise and precise description of the spaces of Hecke modifications encountered differential-geometrically in the previous section. Readers familiar with the affine Grassmannian are encouraged to verify the various statements we make using differential-geometric methods in the following sections. 

\subsection{Fixed point locus} \label{fixedpts}
To complete the program outlined in the beginning of the present section, it is of course necessary to understand the structure of the torus-fixed points in the moduli spaces $\overline{\mathscr{M}}( \{ \mu_i; p_i \} )$, and be able to understand the tangent spaces at these points as representations of the torus. As discussed in section \ref{grg}, the compactified moduli spaces are generically singular, so we will restrict ourselves to the fixed points in the smooth locus. In the examples we will actually compute with minuscule monopole operators, the compactified moduli spaces will be smooth, so this accounts for all the fixed points. 

\subsubsection{}
The torus in question is $U(1)_\varepsilon \times T$, the maximal torus inside the $U(1)_\varepsilon \times G$ global symmetries. The fixed points in $\mathscr{M}(\mu; p)$ are straightforward to describe. For simplicity we assume $G = U(k)$, so that we may view the restriction of $E$ to a fixed time slice as a rank $k$ holomorphic vector bundle over $\mathbb{C}$, with fixed trivialization at infinity, and regard $\mu = \text{diag}(\mu_1, \dots, \mu_k)$. 

The fixed points are isolated. Requiring $U(1)_\varepsilon \times T$-invariance means that $E$ has a fixed decomposition into a sum of line bundles 
\begin{equation}
E = \bigoplus_{i = 1}^k \mathscr{L}_i
\end{equation}
and the only Hecke modification compatible with $U(1)_\varepsilon \times T$ is that which sends $\mathscr{L}_i \to \mathscr{L}_i \otimes \mathscr{O}(\lambda_i)$, where $\lambda$ is in the Weyl orbit of $\mu$. Here, we have denoted by $\mathscr{O}(\lambda_i)$ the line bundle which extends to a framed bundle on $\mathbb{P}^1$ that is isomorphic to the usual bundle $\mathscr{O}_{\mathbb{P}^1}(\lambda_i)$. It carries a natural equivariant structure under $U(1)_\varepsilon$ which rotates the fiber over $0$ with weight $\lambda_i$; this fact will be essential in all that follows. 

\subsubsection{}
These Hecke modifications simply describe the abelian solutions to the Bogomolny equations in the presence of a single Dirac singularity with the boundary conditions of section \ref{bc}. These are labeled by cocharacters $\lambda$ in the Weyl orbit of the dominant cocharacter $\mu$. 

In terms of holomorphic data/transition functions, the fixed points correspond to the monomials $g(z) = z^\lambda$. In this language, fixed points in the moduli spaces $\mathscr{M}(\mu_1, \dots, \mu_n; p_1, \dots, p_n)$ are described by tuples of transition functions which are all monomials; explicitly, fixed points are $(g_1(z), \dots, g_n(z)) = (z^{\lambda_1}, \dots, z^{\lambda_n})$ in the notation of the previous section, where $\lambda_i$ is in the Weyl orbit of $\mu_i$. The fact that the special, toric Hecke modifications are of this form is the fastest way to understand the nontrivial shift in equivariant weight $\lambda$ under $U(1)_\varepsilon$ mentioned above. 

To reiterate, this classification holds only for the fixed points in the \textit{smooth} part of the compactified moduli spaces; there are in general additional fixed points in the singular locus which arise due to monopole bubbling effects. We will be primarily focused on operator products of minuscule monopole operators, where this phenomenon does not arise, so can in large part avoid this. 

\subsection{Instantons/monopoles correspondence} \label{instmon}
An essential technical tool for the calculations we perform is a correspondence between monopole solutions with Dirac singularities, and $U(1)$-invariant instantons on $\mathbb{R}^4$ (or a Taub-NUT space; since we will be interested in the moduli space only as a complex manifold, the distinction is not relevant). This correspondence was discovered by Kronheimer \cite{kronheimer} and explained in the context of the present problem in \cite{kw}. It provides a very useful differential-geometric local model to understand various aspects of the moduli spaces in our problem. We will use it in the next section to compute the character of the tangent space at the $U(1)_\varepsilon \times T$-fixed points.  

\subsubsection{}
The simplest way to understand this statement is the following. It is well-known that the monopole equations \eqref{monopole} are the dimensional reduction of the instanton equations $F_{\mathscr{A}}^+ = 0$ for a four-dimensional gauge field $\mathscr{A}$. An equivalent statement is that smooth monopoles in $\mathbb{R}^3$ are $S^1$-invariant instantons on $\mathbb{R}^3 \times S^1$. 

To engineer monopoles with singularities from smooth instantons, one replaces $\mathbb{R}^3 \times S^1$ by an ALE (or ALF) space, regarded as an $S^1$ fibration over $\mathbb{R}^3$. Then smooth instantons on such a four-manifold, invariant under the rotation of the $S^1$ fiber, will turn out to be the same as monopoles on $\mathbb{R}^3$ with Dirac singularities at the points where the $S^1$ fiber shrinks to zero size. For just a single Dirac singularity we can consider smooth instantons on $\mathbb{R}^4 \simeq \mathbb{C}^2$. One proves the correspondence between instantons and monopoles by a direct calculation in coordinates that we presently outline. 

\subsubsection{}
If $(z_1, z_2)$ denote coordinates on $\mathbb{C}^2$, viewed as a hyperkahler manifold, the circle action $(z_1, z_2) \to (e^{i\theta} z_1, e^{-i\theta}z_2)$ preserves the standard triplet of symplectic forms and has hyperkahler moment maps 
\begin{equation}
\begin{split}
\mu_{\mathbb{C}} & = z_1 z_2 := x^1 + ix^2 = r \sin \theta e^{i \phi} \\
\mu_{\mathbb{R}} & = \frac{1}{2}(|z_1|^2 - |z_2|^2) := x^3 = r\cos \theta.  
\end{split}
\end{equation}
We get a map $\mathbb{C}^2 \to \mathbb{R}^3$ given by $(z_1, z_2) \mapsto (\Re \mu_{\mathbb{C}}, \Im \mu_{\mathbb{C}}, \mu_{\mathbb{R}})$ with generic fiber $S^1$ which shrinks to a point over the origin of $\mathbb{R}^3$. Letting $\psi$ be the fiber coordinate, a short calculation shows that the metric takes the standard Gibbons-Hawking form 
\begin{equation}
ds^2_{\mathbb{C}^2} = |dz_1|^2 + |dz_2|^2  = V(r) (dr^2 + r^2 d\theta^2 + r^2 \sin^2 \theta d\phi^2) + V(r)^{-1}(d\psi + \alpha)^2 
\end{equation}
where 
\begin{equation}
\begin{split}
    V(r) & = \frac{1}{2r} \\
    \alpha & = \frac{1}{2} \cos \theta d\phi. 
\end{split}
\end{equation}
Note that these solve $d\alpha = \star dV$. Then we consider a four-dimensional gauge field $\mathscr{A}$, parameterized as 
\begin{equation}
\mathscr{A} = g\Big( A + \sigma \Big( \frac{d\psi + \alpha}{V(r)} \Big) \Big) g^{-1} + g dg^{-1}
\end{equation}
for a three dimensional gauge field $A$ and scalar $\sigma$, independent of the $\psi$ coordinate. Using the explicit form of the metric it follows immediately that $F_{\mathscr{A}} + \star_4 F_{\mathscr{A}} = 0$ if and only if $F_A + \star D_A \sigma = 0$. Moreover $\mathscr{A}$ is smooth if and only if $(A, \sigma)$ have asymptotics 
\begin{equation}
\begin{split}
A & \sim - \frac{\lambda}{2} \cos \theta d\phi \\ 
\sigma & \sim \frac{\lambda}{2r} 
\end{split}
\end{equation}
as $r \to 0$ for some cocharacter $\lambda \neq 0$, and $g = e^{\lambda \psi}$. 

\subsubsection{}
From the form of $g$ we see that the gauge bundle on $\mathbb{C}^2$ acquires a nontrivial equivariant structure with respect to the circle action, so that the fiber over the origin becomes a $U(1)$-module determined by $\lambda$. 

This shows that monopoles on $\mathbb{R}^3$ with a Dirac singularity at the origin specified by $\lambda$, are the same as instantons on $\mathbb{C}^2$ invariant with respect to the circle action $(z_1, z_2) \to (e^{i\theta}z_1, e^{-i\theta}z_2)$, lifted to an action on the total space of the gauge bundle in such a way that the fiber over the origin transforms according to $\lambda$. 

While this correspondence was established here only for monopoles in $\mathbb{R}^3$, and not our global geometry, it is clear from the explicit discussion above that it is local. Thus, we may use it to model local properties of our moduli spaces, which is enough for our purposes. 

\subsection{Tangent space to $\mathscr{M}(\mu; p)$} \label{tangent}
Now we are able to explain, following \cite{kw} and \cite{pauly}, how to use equivariant index theory to determine the character of the tangent space to our moduli spaces at the $U(1)_\varepsilon \times T$-fixed points. We focus here on the simplest case of a single Dirac singularity, and explain in the next section how to generalize to multiple singularities. 

We give the algebra-geometric computation of the character in appendix \ref{affineflags}. The reader may enjoy comparing the two discussions. 

\subsubsection{}
The tangent space to $\mathscr{M}(\mu; p)$ is naturally identified with $H^1$ of the following elliptic complex:
\begin{equation}
0 \to \Omega^0(\text{ad} E) \to \Omega^1(\text{ad} E) \oplus \Omega^0(\text{ad} E) \to \Omega^2(\text{ad} E) \to 0
\end{equation}
obtained by linearizing the Bogomolny equations \eqref{monopole} and gauge symmetries. $\Omega^i(\text{ad} E)$ denotes adjoint valued $i$-forms on $W$: the middle term in the complex arises from the linearized fields $(\delta A, \delta \sigma)$, the first arrow is linearized gauge transformations and the second arrow is the operator entering the linearized Bogomolny equations. 

At a fixed point $\lambda$ of $U(1)_\varepsilon \times T$, this complex is $U(1)_\varepsilon \times T$-equivariant and one may compute its equivariant index. This equivariant index will coincide up to a sign with the character of the tangent space to $\mathscr{M}(\mu; p)$, provided that $H^0$ and $H^2$ vanish, which they indeed do. 

In the next few sections, we will outline how to use the instanton/monopole correspondence to compute the index, our main result being \eqref{char1}.  

\subsubsection{}
It is straightforward to show by a vanishing theorem \cite{kw} that the index vanishes in the absence of Dirac singularities. Moreover, the contribution of a Dirac singularity to the index is purely local, and can be obtained by replacing $W$ with $\mathbb{R}^3$. This follows generally by the excision properties of index theory.  

For the computation on $\mathbb{R}^3$, we may use the instanton/monopole correspondence. The correspondence directly implies that the $U(1)$-invariant part of the contribution of the origin of $\mathbb{R}^4$ to the equivariant index of the complex linearizing the instanton equations will be the contribution of a Dirac singularity to the index of the complex linearizing the Bogomolny equations, see also Section 4 in \cite{pauly}. 

In fact, we are interested in the moduli space as a complex manifold, and the character of its holomorphic tangent space. The discussion of section \ref{instmon} makes it clear that the complex structure on $\mathscr{M}(\mu; p)$ is compatible with the complex structure on the instanton moduli space obtained by identifying $\mathbb{R}^4 \simeq \mathbb{C}^2$. The instanton equations for a connection $\mathscr{A}$, in complex notation, become 
\begin{equation}
\begin{split}
F_{\mathscr{A}}^{0, 2} & = 0 \\
F_{\mathscr{A}}^{1, 1} \wedge \omega_{\mathbb{R}} & = 0
\end{split}
\end{equation}
where $\omega_{\mathbb{R}}$ denotes the standard Kahler form on $\mathbb{C}^2$. The second equation may be interpreted as a real moment map, so one may replace it with the instruction to divide by the complexified gauge transformations and impose some stability conditon. The first equation then says that the connection $\mathscr{A}$ gives the gauge bundle $\mathscr{E}$ a holomorphic structure. This identification of instantons with stable holomorphic bundles is of course one of the simplest incarnations of Donaldson-Uhlenbeck-Yau theorem.

\subsubsection{}
The complex arising from linearizing the holomorphic equations and gauge symmetries is 
\begin{equation}
0 \to \Omega^{0, 0}(\text{ad} \mathscr{E}) \xrightarrow{\overline{\partial}_{\mathscr{A}}} \Omega^{0, 1}(\text{ad} \mathscr{E}) \xrightarrow{\overline{\partial}_{\mathscr{A}}} \Omega^{0, 2}(\text{ad} \mathscr{E}) \to 0.
\end{equation}
The problem has an obvious $U(1) \times U(1) \times T$ symmetry, where $U(1) \times U(1)$ acts on $\mathbb{C}^2$ by rotations and $T$ acts on the gauge bundle in the obvious way. Then, if $q_1$ and $q_2$ denote equivariant variables for the $U(1) \times U(1)$ action, and $a_i$, $i = 1, \dots, n$ denote equivariant variables for the $T$ action (we choose the notation to mimic that which is traditional in instanton counting), the contribution of the origin of $\mathbb{C}^2$ to the equivariant index is given by formal application of Hirzebruch-Riemann-Roch:
\begin{equation}
-\int_{\mathbb{C}^2} \text{td}_{U(1)^2 \times T}(\mathbb{C}^2) \text{ch}_{U(1)^2 \times T}(\text{ad} \mathscr{E}) = - \sum_{i, j = 1}^k \frac{a_i/a_j}{(1 - q_1)(1 - q_2)} 
\end{equation}
and the integral over $\mathbb{C}^2$ is understood by equivariant residue. We have included the minus sign so that this formally coincides with $H^1 - H^0 - H^2$ of the deformation complex. Some readers may recognize this computation as essentially the contribution of the vector multiplet to the perturbative part of Nekrasov's instanton partition function \cite{nekrasov2002}. 

\subsubsection{}
To complete the calculation, we project to invariants. Write $q_1 = q^{1/2} x$, $q_2 = q^{1/2}x^{-1}$. It is clear from the discussion in section \ref{instmon} that the equivariant parameter $q = q_1q_2$ is identified with $e^{i\varepsilon}$, the parameter for $U(1)_\varepsilon$, upon reduction to $\mathbb{R}^3$. The parameter $x$ is thus the equivariant variable for the $U(1)$ we wish to quotient by. To reflect the fact that the fiber of $\mathscr{E}$ over the origin becomes a module for this $U(1)$ with weight determined by $\lambda$, we also write  $a_i = e^{i (\varphi_i + \frac{\lambda_i \varepsilon}{2})} x^{\lambda_i}$. The $\varepsilon$-dependent shift is included for later convenience, mainly just to avoid square roots of $q$. We will gain a greater apprecition for the significance of such shifts in the following sections. 

The projection to $U(1)$-invariants is implemented by contour integration over $x$ (this statement requires a certain care due to non-compactness of $\mathbb{C}^2$, but the result of these considerations is just reflected in the choice of integration contour below). On the other hand, the $U(1)$-invariant part is simply the index of the Bogomolny complex, which in turn is the character of the tangent space $T_\lambda \mathscr{M}(\mu; p)$ at the fixed point corresponding to $\lambda$, under $U(1)_\varepsilon \times T$. Then we find 
\begin{equation}
T_\lambda \mathscr{M}(\mu; p) = - \sum_{i, j = 1}^k e^{i(\varphi_i - \varphi_j)} q^{\frac{\lambda_i - \lambda_j}{2}} \int_{\gamma} \frac{dx}{2\pi i x} \frac{x^{\lambda_i - \lambda_j}}{(1 - q^{1/2}x)(1 - q^{1/2}x^{-1})}. 
\end{equation}
The contour $\gamma$ extracts the residue at infinity. The residue computation is straightforward, leading to 
\begin{equation} \label{char1}
T_\lambda \mathscr{M}(\mu; p) = \sum_{\substack{i, j \, \, \, s.t. \\ \lambda_i > \lambda_j}} e^{i(\varphi_i - \varphi_j)}(1 + e^{i\varepsilon} + e^{2i\varepsilon} + \dots + e^{i(\lambda_i - \lambda_j - 1)\varepsilon}) := \chi_\lambda(\varepsilon, \varphi). 
\end{equation}
In writing this and all further equations for characters, we follow a standard convention in instanton counting/K-theoretic enumerative geometry and identify $U(1)_\varepsilon \times T$-modules with their characters, reading all equations in $K_{U(1)_\varepsilon \times T}(\text{pt})$. We have also introduced the function $\chi_\lambda(\varepsilon, \varphi)$ that will be useful in writing formulas later on. 

\subsubsection{}
While this equation has been written for $G = U(k)$, it has an obvious generalization to general $G$ since the only thing playing a role is the adjoint representation (we use $\alpha$ to denote the roots of $G$):
\begin{equation}
T_\lambda \mathscr{M}(\mu; p) = \sum_{\substack{\alpha \, \, \, s.t. \\ \alpha \cdot \lambda > 0}} e^{i \alpha \cdot \varphi} (1 + e^{i\varepsilon} + \dots + e^{i(\alpha \cdot \lambda - 1)\varepsilon}).  
\end{equation}
These formulas may also be obtained purely algebra-geometrically using the geometry of the affine Grassmannian; we review this calculation in appendix \ref{affineflags}. The discussion in the past few sections amounts to a differential-geometric derivation of the same formula. In particular, turning off the equivariant parameters one finds the well-known dimension formula for Schubert varieties; this calculation was summarized in \cite{kw} and one may view the present section as its equivariant upgrade. 

\subsubsection{}
As the computation explained above is purely local, one may reasonably expect that for the analogous computation with multiple Dirac singularities, one may simply add the contributions due to the individual monopoles. This is essentially true, although the precise statement requires a certain care. We address this issue presently. 

\subsection{Universal bundle} \label{univ}
Suppose we are in a situation with $n$ monopole operators of various charges $\mu_i$, and consider the total space $\overline{\mathscr{M}}(\{ \mu_i; p_i \}) \times W$. Because $\overline{\mathscr{M}}$ is a moduli space of bundles on $W$, we may construct the corresponding universal bundle $\mathscr{E}$ over $\overline{\mathscr{M}} \times W$. Its fiber over a given point $(m, w)$ is the fiber of the gauge bundle over $w \in W$ determined by the point $m \in \overline{\mathscr{M}}$.  

Understanding some key properties of the universal bundle will complete our preparations for the calculations in Section \ref{algebras}. 

\subsubsection{}
We begin by understanding the supersymmetric observables built out of $\varphi$, that is to say, fully specifying the integrand in \eqref{findimint}. The restrictions of $\mathscr{E}$ to points in the $U(1)_\varepsilon$-fixed set in $W$, $\mathscr{E} \eval_{(t, 0)}$ give $U(1)_\varepsilon \times G$-equivariant holomorphic vector bundles over $\overline{\mathscr{M}}$; precisely because the holomorphic type of the gauge bundle is locally constant in $t$, so too are these. 

That is to say, we get only finitely many distinct bundles in this way depending on the interval in which $t$ lies. For $0 < t < t_1$, the corresponding bundle is topologically trivial, but has nontrivial $G$-equivariant structure. For $t_1 < t < t_2$, we get a distinct bundle since it has undergone a Hecke modification on $\mathbb{C}$ of type determined by the point in $\overline{\mathscr{M}}$, and so on for the successive Hecke modifications. 

\subsubsection{}
This has the following very explicit consequence for localization formulas. Suppose all $\mu_i$ are minusucle, so that $\overline{\mathscr{M}}$ is smooth with fixed points $(\lambda_1, \dots, \lambda_n)$, where $\lambda_i$ is in the Weyl orbit of $\mu_i$. Then the restriction of $\mathscr{E} \eval_{(t, 0)}$, for $0 < t < t_1$ to $(\lambda_1, \dots, \lambda_n)$ has weights $\varphi_1, \dots, \varphi_k$ under $U(1)_\varepsilon \times T$, for any $\lambda_i$. After crossing $t = t_1$, while staying at the point $(\lambda_1, \dots, \lambda_n)$ in $\overline{\mathscr{M}}$, the gauge bundle undergoes a Hecke modification implemented by the singular holomorphic gauge transformation $z^{\lambda_1}$; reflecting this, the restriction of $\mathscr{E} \eval_{(t, 0)}$ to $(\lambda_1, \dots, \lambda_n)$ for $t_1 < t < t_2$ has weights $\varphi_i + (\lambda_1)_i \varepsilon$ under $U(1)_\varepsilon \times T$ for $i = 1, \dots, k$, and so on for successive modifications. 

\subsubsection{}
It follows from the discussion in appendix \ref{universalappendix} that the observables \eqref{yobs} are identified on the localization locus with the equivariant Chern polynomials: 
\begin{equation} \label{chernpoly}
\mathscr{Y}(x, t) = \text{det}(x - \varphi(t, 0)) = c_x^{U(1)_\varepsilon \times G}\Big( \mathscr{E}^*\eval_{(t, 0)} \Big) \in H^*_{U(1)_\varepsilon \times G}(\overline{\mathscr{M}}).
\end{equation}
Note that if $\varepsilon$ is set to $0$, these observables become independent of $t$ since one can move $\varphi$ away from $z = 0$ to vary $t$ without hitting a singularity, without altering the cohomology class of $\mathscr{Y}$. However, the $\Omega$-background leads to locally constant $t$-dependence on $[0, t_1) \cup (t_1, t_2) \cup \dots \cup (t_n, 1]$. These characteristic classes are the incarnations of the universal characteristic classes defined by Atiyah and Bott \cite{atiyahbott} in the present context, see also the discussion in \cite{witten09}. 

\subsubsection{}
Thus far we have avoided the actual positions of the singularities. $\varphi$ is nonsingular at those points, so it still makes sense to have insertions there, and in fact these are the dressing factors which may enter the more general dressed monopole operators. Given what we have already said it is easy to describe such factors. At such a point, the gauge group is reduced to the subgroup $H_i$ preserving the form of the singularity $\mu_i$. If $\mu_i$ is generic, this subgroup will just be $T$. At the other extreme, if $G = U(k)$ and $\mu_i$ is a minuscule coweight, $H_i = U(\ell) \times U(k - \ell)$ for some $\ell$. 

At $t_i$, one may insert any $H_i$-invariant polynomial in $\varphi(t_i, 0)$. While the gauge bundle does not make sense as a $G$-bundle at the position of a singularity, it does have a canonical extension to an $H_i$-bundle there, so $\mathscr{E} \eval_{(t_i, 0)}$ naturally determines an $H_i$-bundle over the moduli space. The dressing factors are simply the equivariant characteristic classes of these bundles. 

\subsubsection{}
To make this perhaps more explicit, consider the case $G = U(k)$ of a single monopole operator with charge $\mu = (1, 1, \dots, 1, 0, \dots, 0)$ where there are $\ell$ $1$'s appearing. Then the moduli space is an ordinary Grassmannian $\overline{\mathscr{M}}(\mu; p) \simeq \text{Gr}(\ell, k)$, and $H = U(\ell) \times U(k - \ell)$. The $H$-bundle determined from the universal bundle is simply the direct sum of the usual tautological and quotient bundles over the Grassmannian $\text{Gr}(\ell, k)$. The dressing factors are just the characteristic classes of the tautological and quotient bundles, which are in fact known to generate the equivariant cohomology ring of $\text{Gr}(\ell, k)$. 

\subsubsection{}
We now explain how to use what we learned about $\mathscr{E}$ to compute the character of the tangent space to $\overline{\mathscr{M}}$ at a fixed point in the presence of multiple monopole insertions. The tangent space to the moduli space at any smooth point $m$ is identified with $H^1$ of the deformation complex
\begin{equation}
0 \to \Omega^0 \Big( \text{ad} \mathscr{E} \eval_{m \times W} \Big) \to \Omega^1 \Big( \text{ad} \mathscr{E} \eval_{m \times W} \Big) \to \Omega^2 \Big( \text{ad} \mathscr{E} \eval_{m \times W} \Big) \to 0.
\end{equation}
Then the parameters $\varphi_i$ appearing in the character formula \eqref{char1} of the previous section acquire a more obvious geometric significance: they are the $U(1)_\varepsilon \times T$-weights of the fiber of the universal bundle over a point $(\lambda, (t, 0)) \in \overline{\mathscr{M}}(\mu; p) \times W$ where $0 \leq t < t_1$. 

Now suppose $m$ is a fixed point of the $U(1)_\varepsilon \times T$ action on $\overline{\mathscr{M}}(\{ \mu_i; p_i \})$ in the smooth locus. Such points are labeled by $n$-tuples $(\lambda_1, \dots, \lambda_n)$ where $\lambda_i$ is in the Weyl orbit of $\mu_i$. We suppose the positions of the monopole operators on the $t$ axis are chosen in the chamber $t_1 < \dots < t_n$. Then we have, for the character of the tangent space,
\begin{equation} \label{charn}
T_{(\lambda_1, \dots, \lambda_n)}\overline{\mathscr{M}}(\{ \mu_i; p_i \}) = \sum_{i = 1}^n \chi_{\lambda_i}\Big( \varepsilon, \varphi + \varepsilon \sum_{j < i} \lambda_j \Big)
\end{equation}
using the function $\chi_\lambda$ defined in \eqref{char1}. 

\subsubsection{}
This formula is deduced as follows. Each Dirac singularity contributes to the index in a purely local way, as in the previous section, and the total index is simply the sum of these contributions. However, the weights in the fiber of the universal bundle over the origin of $\mathbb{C}$ \text{jump} along the $t$-interval. This is because, for $t_i < t < t_{i + 1}$, the gauge bundle has undergone successive Hecke modifications at the origin of types $\lambda_1, \lambda_2, \dots, \lambda_i$. As explained in section \ref{fixedpts}, such Hecke modifications are implemented by singular holomorphic gauge transformations $z^{\lambda_i}$. Then it is clear that the $U(1)_\varepsilon \times T$ weights in the fiber must shift accordingly, since $z$ transforms nontrivially under $U(1)_\varepsilon$. Adding together the local contributions, using the correct local weights for $\mathscr{E}$, gives precisely \eqref{charn}. 

\subsubsection{}
There is also the following physical justification for this formula: it arises by applying field theory factorization. The integral over $\overline{\mathscr{M}}$ is what remains of the field theory path integral, which in turn is independent of the metric length of $[0, 1]$.\footnote{While this statement follows from the formal decoupling of $\delta_\varepsilon$-exact terms, we can also see explicitly that there is no anomaly. The distances between monopole operators on $[0, 1]$ enter as Kahler moduli for $\overline{\mathscr{M}}$, and integrals of equivariant cohomology classes are of course insensitive to those parameters.} Therefore, we may scale the length up, making the monopole operators arbitrarily separated, and apply factorization. The path integral reduces to a product of integrals over fields defined in $\mathbb{R}^3$ neighborhoods of the individual monopoles, where the argument of the previous section may be applied. The jumps in $\varphi$ along the interval may be understood by inspecting \eqref{omegasusy}: on the localization locus, we must have $D_A \varphi + i_{V_\varepsilon} F_A = 0$. This means that turning on a magnetic flux shifts the effective value of $\varphi$ away from its VEV by an $\varepsilon$-dependent amount. The fluxes from the monopole operators lead exactly to the shifts in \eqref{charn}.  

\subsubsection{}
While the physical argument is more intuitive, the discussion in terms of universal bundles is more mathematically precise and is better preparation for the generalization in Section \ref{klrw}. 

\section{Coulomb Branch Algebras} \label{algebras}
We can use what we have set up in Section \ref{loc} to give a gauge-theoretic derivation of the Coulomb branch algebras. Upon recalling the algebra-geometric description of our localization loci, the computations recover the definition given in \cite{bfn}. We also discuss the generalization to include matter, $K$-theoretic analogs for four-dimensional gauge theories, and explicitly characterize the modules over the quantized Coulomb branch determined by the Dirichlet and Neumann boundary conditions in gauge theory. In addressing the final point we confirm some conjectures of \cite{Bullimore_2016}. 

\subsection{Pure gauge theory Coulomb branch}
Here we determine the quantized Coulomb branch algebra by direct analysis of the equivariant integrals set up in the previous section. We consider the case of a unitary gauge group $G = U(k)$, in which case it is known a priori on mathemathical or physical grounds that dressed monopole operators of minuscule charge generate the Coulomb branch algebra. On mathematical grounds, this follows from a dimension count of the monopole moduli spaces. Physically this is equivalent to certain considerations about $R$-symmetry charges.

\subsubsection{Difference operator realization} \label{diffop}
Let us begin by introducing a notation for some of the equivariant integrals we explained how to compute in section \ref{loc}. Suppose we consider a correlation function of $n$ monopole operators of charges $\mu_i$, $i = 1, \dots, n$, and a gauge-invariant polynomial in the adjoint scalar $f(\varphi)$ inserted at $t = 1$. We write 
\begin{equation}
    \bra{f} \mathscr{O}_{\mu_n}(t_n) \dots \mathscr{O}_{\mu_1}(t_1) \ket{\varphi} := \langle f(\varphi(0, t = 1)) \mathscr{O}_{\mu_n}(t_n) \dots \mathscr{O}_{\mu_1}(t_1) \rangle.
\end{equation}
On the left hand side of this equation, $\varphi$ is shorthand for the tuple of parameters $(\varphi_1, \dots, \varphi_k)$ which are part of the data of the boundary condition at $t = 0$ (see \eqref{bc}), while in the right hand side the local observable $f(\varphi(z, \bar{z}, t))$ appears inserted at the point $z = 0, t = 1$ (see Figure \ref{fig:monopole_op_correlator} below). We explained that the path integral computing the right hand side formally reduces to a finite dimensional equivariant integral in \eqref{findimint}, which in this particular instance is 
\begin{equation}
    \langle f(\varphi(0, t = 1)) \mathscr{O}_{\mu_n}(t_n) \dots \mathscr{O}_{\mu_1}(t_1) \rangle = \int_{\overline{\mathscr{M}}(\{ \mu_i; p_i \})} f(\text{Chern roots of $\mathscr{E}|_{t = 1}$}). 
\end{equation}
We used the discussion in section \ref{univ} identifying $\Omega_\mathscr{P}$ in the notation of \eqref{findimint} with the characteristic class of $\mathscr{E} \eval_{(t = 1, 0)}$ determined by $f$.

\begin{figure}
    \centering
    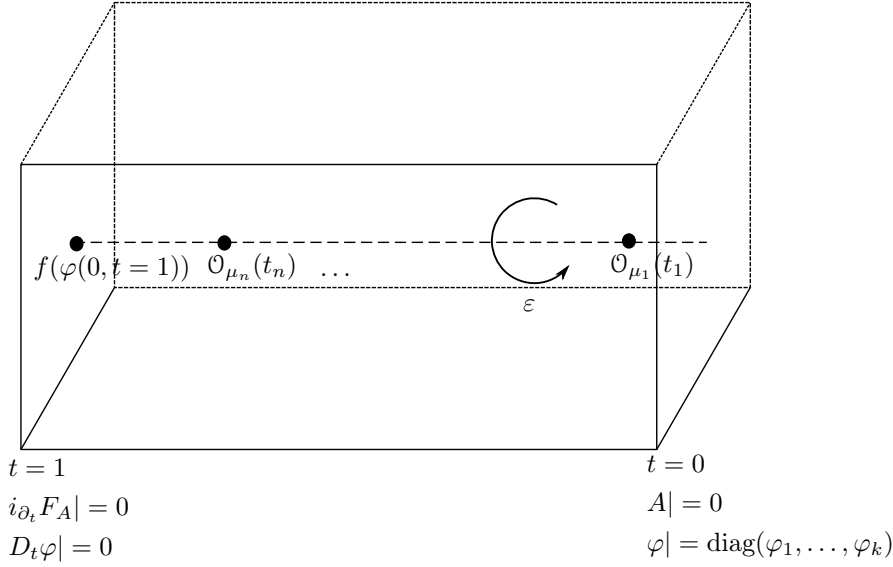
    \caption{Geometry of $\mathbb{R}^2_\varepsilon \times [0, 1]$ for computation of correlation functions $\langle f(\varphi(0, t = 1) \mathscr{O}_{\mu_n}(t_n) \dots \mathscr{O}_{\mu_1}(t_1) \rangle = \bra{f} \mathscr{O}_{\mu_n}(t_n) \dots \mathscr{O}_{\mu_1}(t_1) \ket{\varphi}$. The coordinate $t$ runs from right to left.} 
    \label{fig:monopole_op_correlator}
\end{figure}

\subsubsection{}
This notation is intentionally chosen to mimic that of operators and states in elementary quantum mechanics. Heuristically, the reason we might do this is as follows. The $\Omega$-background on $\mathbb{R}^2 \times [0, 1]$ effectively localizes the problem to a one-dimensional field theory on $[0, 1]$, in other words a problem in quantum mechanics. The boundary conditions at $t = 0, 1$ determine certain ``states''; at $t = 0$, from \eqref{bc} we see that the boundary condition is labeled by the parameters $\varphi_1, \dots, \varphi_k$ so we get an ``in-state'' $\ket{\varphi}$. At $t = 1$, the boundary condition supports the insertion of the nontrivial boundary local operator $f(\varphi(0, t = 1))$, giving the ``out state'' $\bra{f}$. The operator insertions in the middle become operators in the quantum mechanics. We will give a more high-level discussion of the quantum mechanical interpretation in section \ref{branes}. 

\subsubsection{}
While the notations reflect these somewhat heuristic quantum field-theoretic considerations, they can also be approached from the following direct route which readers from a mathematical background may find more useful. $f$ is a gauge-invariant polynomial, so it can be viewed as an element $\bra{f} \in H^*_{U(1)_\varepsilon \times G}(\text{pt})$. Associated to the bundle $\mathscr{E}\eval_{(t = 1, 0)}$ on $\overline{\mathscr{M}}$ we have the characteristic class homomorphism $C(\mathscr{E}): H^*_{U(1)_\varepsilon \times G}(\text{pt}) \to H^*_{U(1)_\varepsilon \times G}(\overline{\mathscr{M}})$ given by pullback under the classifying map to $BG$. As $\overline{\mathscr{M}}$ is a compact space for minuscule coweights $\mu_i$ (this follows from its identification with a convolution Grassmannian in section \ref{grg}), integration is defined in non-localized equivariant cohomology and gives rise to a degree $-2\dim \overline{\mathscr{M}}$ map $\int_{\overline{\mathscr{M}}}: H^*_{U(1)_\varepsilon \times G}(\overline{\mathscr{M}}) \to H^{* - 2\dim \overline{\mathscr{M}}}_{U(1)_\varepsilon \times G}(\text{pt})$. The composition $\int_{\overline{\mathscr{M}}} C(\mathscr{E})$ gives rise to a linear map $H^*_{U(1)_\varepsilon \times G}(\text{pt}) \to H^{* - 2\dim \overline{\mathscr{M}}}_{U(1)_\varepsilon \times G}(\text{pt})$. Thus we can view the products of monopole operators $\prod_i \mathscr{O}_{\mu_i}(t_i)$, which determine $\overline{\mathscr{M}}$ and $\mathscr{E}$, as defining explicit linear maps on $H^*_{U(1)_\varepsilon \times G}(\text{pt}) \simeq \mathbb{C}[\varepsilon, \varphi_1, \dots, \varphi_k]/W$. We will characterize the operator product algebra of $\mathscr{O}_{\mu_i}$ by determining the span of the $\prod_i \mathscr{O}_{\mu_i}(t_i)$ as a subalgebra in $\text{End}(H^*_{U(1)_\varepsilon \times G}(\text{pt}))$.

\subsubsection{}
After all the preparations, we may compute. Consider the simplest case of $n = 1$ monopole operator, for the simplest minuscule dominat coweights for $U(k)$, $\mu^+ = (1, 0, \dots, 0)$ and $\mu^- = (0, \dots, 0, -1)$. We study the monopole moduli spaces $\overline{\mathscr{M}}(\mu^\pm; p)$ with Dirichlet and Neumann conditions, with Dirac singularities, discussed extensively in Section \ref{loc}. The fixed points in these moduli spaces are labeled by integers $i = 1, \dots, k$, corresponding to weights in the Weyl orbit of $\mu^\pm$. In the language of Hecke modifications, we fix a decomposition of the bundle $E = \oplus_{j = 1}^k \mathscr{L}_j$ as a sum of line bundles, then the modification corresponding to the $i$-th fixed point is $\mathscr{L}_j \to \mathscr{L}_j \otimes \mathscr{O}(\pm \delta_{ij})$. Using \eqref{char1} we can read off the weights in the tangent space at each fixed point, and using the discussion in Section \ref{univ} we may determine the weights in $\mathscr{E} \eval_{t = 1}$ at each fixed point to be $\varphi_j \pm \varepsilon \delta_{ij}$, $j = 1, \dots, k$. Then we get explicit formulas for the correlation functions via the localization formula:
\begin{equation}
\begin{split}
\bra{f} \mathscr{O}_{\mu^+}(t_1) \ket{\varphi} = \int_{\overline{\mathscr{M}}(\mu^+; p)} f(\text{Chern roots of $\mathscr{E} |_{t = 1}$}) = \sum_{i = 1}^k \frac{f(\varphi_1, \dots, \varphi_i + \varepsilon, \dots, \varphi_k)}{\prod_{j (\neq i)}(\varphi_i - \varphi_j)} \\
\bra{f} \mathscr{O}_{\mu^-}(t_1) \ket{\varphi} = \int_{\overline{\mathscr{M}}(\mu^-; p)} f(\text{Chern roots of $\mathscr{E} |_{t = 1}$}) = \sum_{i = 1}^k \frac{f(\varphi_1, \dots, \varphi_i - \varepsilon, \dots, \varphi_k)}{\prod_{j (\neq i)}(\varphi_j - \varphi_i)}.
\end{split}
\end{equation}
These formulas are also easy to understand more directly, because the moduli spaces in question can be shown to be isomorphic to projective spaces $\mathbb{P}^{k - 1}$. Intuitively, they are spaces of Hecke modifications which look in some gauge like multiplication by $\text{diag}(z^{\pm 1}, 1, \dots, 1)$. The choice of gauge is a choice of one-dimensional subspace in the fiber $\mathbb{C}^k$ of the gauge bundle over $0$, and there is a $\mathbb{P}^{k - 1}$ of these. The $\varepsilon$-shifts in the numerators are deduced using the jumping behavior of the universal bundle discussed in section \ref{univ}. 

\subsubsection{}
In view of the quantum mechanical notation, these formulas, in addition to the trivial insertion (integral over a point) $\braket{f}{\varphi} = f(\varphi_1, \dots, \varphi_k)$ giving the ``wavefunction'', suggest the identification in the ``$\varphi$-basis'' 
\begin{equation}
\begin{split}
\mathscr{O}_{\mu^+} & \longrightarrow \sum_{i = 1}^k \exp(-\varepsilon \pdv{}{\varphi_i}) \frac{1}{\prod_{j (\neq i)} (\varphi_i - \varphi_j)} \\
\mathscr{O}_{\mu^-} & \longrightarrow \sum_{i = 1}^k \exp(+ \varepsilon \pdv{}{\varphi_i}) \frac{1}{\prod_{j (\neq i)} (\varphi_j - \varphi_i)}. 
\end{split}
\end{equation}

\subsubsection{}
There is one more thing to check, which is that this identification actually respects the operator product algebra of the monopole operators. The action of the operator product $\mathscr{O}_{\mu_2}(t_2) \mathscr{O}_{\mu_1}(t_1)$ on $\bra{f}$ has a natural geometric definition via taking characteristic classes and integration over the moduli space $\overline{\mathscr{M}}(\mu_1, \mu_2; p_1, p_2)$. We need to check that the linear operator on $H^*_{U(1)_\varepsilon \times G}(\text{pt})$ determined geometrically by $\mathscr{O}_{\mu_2}(t_2) \mathscr{O}_{\mu_1}(t_1)$ is the same as the product of the difference operators associated to the individual $\mathscr{O}_{\mu_i}$. 

We can explicitly check this using the results in section \ref{univ}. For concreteness and to lighten notations we consider the operator product of $\mathscr{O}_{\mu^+}$ and $\mathscr{O}_{\mu^-}$, though the argument may be repeated line by line for any pair of minuscule coweights. 

The path integral computing the correlator $\bra{f} \mathscr{O}_{\mu^-}(t_2) \mathscr{O}_{\mu^+}(t_1) \ket{\varphi}$ reduces to an integral over the space $\overline{\mathscr{M}}(\mu^+, \mu^-; p_1, p_2)$. Note that the moduli space itself depends on the ordering of $t_i$. It has a Kahler modulus proportional to $|t_2 - t_1|$ (see section 10.3 in \cite{kw}), it is singular at $t_1 = t_2$ and when $t_1, t_2$ pass through each other it undergoes a flop transition. This reflects the general fact that when matrix elements in quantum mechanics are computed by path integrals, operators come out time ordered. We pick here the ordering $0 < t_1 < t_2 < 1$; the moduli space for the other ordering is related by a flop. The relevant equivariant integral may be computed using \eqref{charn} for $n = 2$:
\begin{equation} \label{m+m-}
\int_{\overline{\mathscr{M}}(\mu^+, \mu^-; p_1, p_2)} f(\text{Chern roots of $\mathscr{E}|_{t = 1}$}) = \sum_{i_1, i_2 = 1}^k \frac{f(\varphi_1, \dots, \varphi_{i_1} + \varepsilon, \dots, \varphi_{i_2} - \varepsilon, \dots, \varphi_k)}{\prod_{j (\neq i_2)}(\varphi_j - \varphi_{i_2} + \varepsilon(\delta_{i_1 j} - \delta_{i_1 i_2})) \prod_{j (\neq i_1)}
(\varphi_{i_1} - \varphi_j)} . 
\end{equation}
The $\varepsilon$-shifts in the numerator reflect the fact that at $t = 1$, the universal bundle has undergone two Hecke modifications, and for the toric modifications these shift one of the $\varphi_i$ by $\pm \varepsilon$. We then sum over the toric modifications with weights determined by \eqref{charn}.

By inspection, the left hand side of \eqref{m+m-} is equal to 
\begin{equation}
    f(\varphi_1, \dots, \varphi_k) \Bigg(\sum_{i_2 = 1}^k \exp(+ \varepsilon \pdv{}{\varphi_{i_2}}) \frac{1}{\prod_{j (\neq i_2)} (\varphi_j - \varphi_{i_2})} \Bigg) \Bigg(\sum_{i_1 = 1}^k \exp(-\varepsilon \pdv{}{\varphi_{i_1}}) \frac{1}{\prod_{j (\neq i_1)} (\varphi_{i_1} - \varphi_j)} \Bigg)
\end{equation}
where we understand that difference operators act from the right. This shows that the geometric definition of composition of monopole operators agrees with the algebraic definition of composing difference operators, so we do indeed get a realization of the operator product algebra of monopole operators via difference operators. In other words, the quantum-mechanical notation is rigorously justified. The embedding of the Coulomb branch algebra into difference operators we have defined is the quantized ``abelianization map'' introduced in \cite{bdg}. 

\subsubsection{}
It is hopefully obvious how to incoporate dressed monopole operators into all this; the dressing factors just give rise to extra numerators in localization formulas. The above analysis then fully determines the operator product algebra of the dressed monopole operators in terms of difference operators: there is a faithful embedding of the quantized Coulomb branch into the ring of difference operators in $\varphi_i$ by formulas similar to those deduced above. The images of the minuscule dressed monopole operators generate the image of the Coulomb branch algebra. 

\subsubsection{Comparison to BFN} \label{BFNcomp}
At this point we have fully recovered the definition of the Coulomb branch given in \cite{bfn}, at least in the pure gauge theory case, in an almost tautological way. Let us recall why this is so. The authors there define the quantized Coulomb branch algebra as the equivariant Borel-Moore homology of the affine Grassmannian $H_*^{\mathbb{C}^\times_\varepsilon \ltimes G_{\mathbb{C}}(\mathscr{O})}(\text{Gr}_G)$, where notations are as in Section \ref{grg}. As a vector space, 
\begin{equation}
H_*^{\mathbb{C}^\times_\varepsilon \ltimes G_{\mathbb{C}}(\mathscr{O})}(\text{Gr}_G) = \bigoplus_{\text{$\mu$ dominant}} [\overline{\text{Gr}}^\mu_G] \mathbb{C}[\mathfrak{h}_{\mathbb{C}}]^{W_\mu}. 
\end{equation}
$W_\mu$ denotes the parabolic Weyl group associated to $\mu$. We now see that the discussion in Section \ref{loc} amounts to producing a natural isomorphism between this and the space of dressed monopole operators: the orbit closures are the moduli spaces of solutions to Bogomolny equations $\overline{\mathscr{M}}(\mu; p)$ which arise as the localization locus of the path integral in the presence of a single monopole operator of charge $\mu$ with our boundary conditions, and the invariant polynomials are the allowed dressing factors. The equivariant integrals computing correlators depend on the integration domain only via its equivariant fundamental class.  

The reason why it is physically natural to identify a monopole operator with the equivariant fundamental class of the localization locus is precisely because this is how these operators are defined in the ultraviolet, by changing the domain of path integration. 

This homology group has a ring structure defined by convolution \cite{bfm, bfn}. The convolutions of Schubert varieties were explained in section \ref{grg} to be identical to the moduli spaces we consider with multiple Dirac singularities, and these determine the product structure on the homology. Thus we see the ``geometric operator product'' we referred to in the previous section is really the convolution product (that the field-theoretically defined operator product agrees with the convolution product was already noticed in \cite{kw} in the context of geometric Satake isomorphism). Then we may say, at least for the gauge group $G = U(k)$, that the subalgebra of $\text{End}(H^*_{U(1)_\varepsilon \times G}(\text{pt}))$ we explained how to construct in the previous section is isomorphic in a tautological way to the convolution algebra $H_*^{\mathbb{C}^\times_\varepsilon \ltimes G_{\mathbb{C}}(\mathscr{O})}(\text{Gr}_G)$---just identify $\mathscr{O}_{\mu_n}(t_n) \dots \mathscr{O}_{\mu_1}(t_1) \mapsto m_*[\overline{\text{Gr}}^{\mu_1}_G \widetilde{\times} \dots \widetilde{\times} \overline{\text{Gr}}^{\mu_n}_G]$ for $\mu_i$ minuscule, where $m_*$ is pushforward under the map $m$ introduced in section \ref{grg} (a similar formula may be written if some of the monopole operators have dressing factors). 

Finally, equivariant localization by $\mathbb{C}^\times_\varepsilon \times T_{\mathbb{C}}$ gives rise to an injective homomorphism of convolution algebras $H_*^{\mathbb{C}^\times_\varepsilon \ltimes G_{\mathbb{C}}(\mathscr{O})}(\text{Gr}_G) \xhookrightarrow{} H_*^{\mathbb{C}^\times_\varepsilon \times T_{\mathbb{C}}}(\text{Gr}_T)^W_{\text{loc}}$ (the subscript denotes localization at certain hyperplanes), and the latter may be shown to be isomorphic to a ring of difference operators in the $\varphi_i$ variables (see appendix A in \cite{bfnslice} for details on this). The calculation we performed leading to \eqref{m+m-} in the previous section can be reinterpreted as verifying that this map is indeed a homomorphism. This embedding is precisely what we discovered via our quantum mechanical formulation. Of course, the difference operators also make contact with abelianized coordinates introduced in \cite{bdg} based on physical motivations; the analysis we gave here explains why this had to be the case a priori. 
\subsection{Coupling to hypermultiplets} \label{addhypers}
Now we will explain variations on the theme we have explored in the context of the pure gauge theory. The first is the incorporation of matter. As we have presented the details rather exhaustively for the pure gauge theory, and the structure is rather similar to the incorporation of matter in four-dimensional $\mathscr{N} = 2$ theories in the context of instanton counting \cite{lns, nekrasov2002}, we will mostly state results and just explain some novel aspects of our current situation. 

\subsubsection{Definition and boundary conditions}
The bosonic field content of a hypermultiplet in a complex representation $R$ of the gauge group is a pair of complex scalars. One scalar, $Q$, is taken to be valued in $R$, and the other scalar $\widetilde{Q}$ is taken to be valued in $R^*$ (we do not consider half-hypermultiplets in this work). The $R$-valued pair $(Q, \widetilde{Q}^\dagger)$ transforms as a doublet under the $R$-symmetry $SU(2)_H$. Upon twisting, this doublet becomes a two-component spinor $Q_\alpha$ of $SU(2)$ spacetime rotations, $\alpha = 1, 2$.  

The whole analysis of section \ref{loc} can be repeated by writing explicitly the fermionic field content of the hypermultiplet, extending the supercharge $\delta, \delta_\varepsilon$ to act on the hypermultiplet, and adding hypermultiplet contributions to the action. We will just state what we expect the consequence of this formal procedure to be for the finite-dimensional equivariant integrals.

Let $(A, \sigma)$ be a pair solving the Bogomolny equation $F_A + \star D_A \sigma = 0$. We have the Dirac-like operator $\slashed{\mathscr{D}} := \gamma^\mu D_\mu + i \sigma$ acting on $Q_\alpha$, and may study the Dirac equation $(\slashed{\mathscr{D}} Q)_\alpha = 0$. The reader should compare with the dimensional reduction of the four-dimensional chiral Dirac operator in the background of an instanton, along the lines of section \ref{instmon}. 

Since we work on $W = [0, 1] \times \mathbb{R}^2$, we need to supplement the Dirac equation with boundary conditions. Good (elliptic) boundary conditions for Dirac equations typically set half of the components of the spinor to zero. If we write $Q_\alpha = (Q, \widetilde{Q}^\dagger)$, then a sensible boundary condition is 
\begin{equation} \label{hypbdry}
\begin{split}
Q(t = 0, z, \bar{z}) & = 0 \\
\widetilde{Q}(t = 1, z, \bar{z}) & = 0.
\end{split}
\end{equation}
The Neumann boundary conditions on the other half of the components at either end are just determined by consistency of the Dirac equation. Clearly this is not the most general boundary condition possible: the detail is in what one means precisely by ``half'' of the components of $Q_\alpha$. Such a choice is encoded in a $G$-invariant Lagrangian splitting of the representation $R \oplus R^*$. The point is that any representation which is a ``double'' of a representation $R$ in this sense comes with a canonical invariant Lagrangian splitting, as part of its definition, so our choice is in this sense natural. 

While we did not explicitly introduce the fermionic content of the hypermultiplet, the boundary conditions above together with supersymmetry uniquely determine the boundary conditions on the fermionic fields as was the case with the vector multiplet in appendix \ref{bcdetails}.

The above data alone are enough to determine the contribution of the hypermultiplets to the finite-dimensional integrals. 

\subsubsection{Integration of Euler class and hypermultiplet contribution to the index}
We find ourselves in the following situation, typical in topological field theory as applied to enumerative geometry. In the presence of matter, the localization locus remains one of the spaces $\overline{\mathscr{M}}(\{ \mu_i; p_i \})$ considered in section \ref{loc}. Over this space we have the virtual vector bundle $\mathscr{F} = - \text{Index}(\slashed{\mathscr{D}}) = \text{coker} \slashed{\mathscr{D}} - \text{ker} \slashed{\mathscr{D}}$. We suppress the explicit dependence on the representation $R$, for notational simplicity. In fact, it is reasonable to expect that $\ker \slashed{\mathscr{D}} = 0$ in our situation because the boundary conditions at $t = 0$ and $t = 1$ have been chosen in a complementary way, to lift the zero modes of $Q_\alpha$. In this situation $\mathscr{F}$ is an actual vector bundle, not just a virtual one. 

When computing the path integral, the hypermultiplet zero modes lead to an insertion of the equivariant Euler class of the bundle $\mathscr{F}$:
\begin{equation}
\bra{f} \mathscr{O}_{\mu_n}(t_n) \dots \mathscr{O}_{\mu_1}(t_1) \ket{\varphi} = \int_{\overline{\mathscr{M}}(\{ \mu_i; p_i \})} f(\text{Chern roots of $\mathscr{E} |_{t = 1}$}) \cup \text{Euler}_{U(1)_\varepsilon \times G \times G_F}(\mathscr{F}). 
\end{equation}
In this formula, we have also included equivariance with respect to a flavor symmetry group $G_F$, corresponding to turning on complex masses for the hypermultiplets. $G_F$ is just a symmetry of the representation $R$ that commutes with $G$. One can think of $\mathscr{F}$ as an obstruction bundle and its Euler class as determining a virtual fundamental cycle for $\overline{\mathscr{M}}$. 

\subsubsection{}
Thus, the problem of theories with matter is readily solved provided we can compute the equivariant Euler class of $\mathscr{F}$. What is needed for this is essentially knowledge of the fibers over the torus fixed points as modules over the maximal torus of $U(1)_\varepsilon \times G \times G_F$. These can be obtained readily by the following observation: the Dirac operator $\slashed{\mathscr{D}}$ is the dimensional reduction of the Dirac operator acting on spinors of positive chirality in four dimensions. Thus, the Dirac equation in a nontrivial monopole background can be obtained by reduction along the $U(1)$ fiber of $\mathbb{R}^4 \to \mathbb{R}^3$ as in section \ref{instmon}. The problem is further simplified by the fact that $\mathscr{F}$ is a holomorphic bundle; this means that we can rephrase the computations on $\mathbb{R}^4 \simeq \mathbb{C}^2$ in complex geometry. 

\subsubsection{}
Let us first consider the case of a single Dirac singularity $\overline{\mathscr{M}}(\mu; p)$. By the same arguments of section \ref{instmon} and \ref{tangent}, the contribution of a fixed point $\lambda$ in the smooth locus is obtained from the $U(1)$-invariant part of the contribution of the origin of $\mathbb{C}^2$ to the equivariant index of the Dirac complex in four dimensions. It is well-known that in holomorphic terms, the Dirac complex becomes the Dolbeault complex twisted by the square root of the canonical bundle:
\begin{equation}
0 \to \Omega^{0, 0}(\mathscr{E}_R \otimes \mathscr{K}^{1/2}) \xrightarrow{\overline{\partial}_{\mathscr{A}}} \Omega^{0, 1}(\mathscr{E}_R \otimes \mathscr{K}^{1/2}) \xrightarrow{\overline{\partial}_{\mathscr{A}}} \Omega^{0, 2}(\mathscr{E}_R \otimes \mathscr{K}^{1/2}) \to 0.
\end{equation}
We denote by $\mathscr{E}_R$ the associated bundle to the four-dimensional gauge bundle $\mathscr{E}$ via the representation $R$. To simplify the formulas a bit, we specialize to the case that $G = U(k)$ and $R$ is given by $n$ copies of the fundamental representation, $R = \text{Hom}(M, \mathbb{C}^k)$ for a multiplicity space $M \simeq \mathbb{C}^n$. The flavor group $G_F$ is $U(n)$ acting on $M$ in the obvious way, with equivariant parameters (complex masses) $m_\alpha$, $\alpha = 1, \dots, n$. Readers may readily generalize the discussion to arbitrary representations $R$ of $G \times G_F$. 

Then the contribution of the origin of $\mathbb{C}^2$ in this case is easily read off from the equivariant integral: 
\begin{equation}
\int_{\mathbb{C}^2} \text{td}(\mathbb{C}^2) \text{ch}(\mathscr{E}_R \otimes \mathscr{K}^{1/2}) = \frac{(q_1q_2)^{1/2}}{(1 - q_1)(1 - q_2)} \sum_{i = 1}^k \sum_{\alpha = 1}^n a_i/m_\alpha
\end{equation}
and as in section \ref{tangent} the character at a fixed point $\lambda$ is given as a contour integral (we redefine $m_\alpha$ as an additive parameter and absorb an $\varepsilon/2$ into it, purely for convenience):
\begin{equation}
\mathscr{F} \eval_\lambda = -\sum_{i = 1}^k \sum_{\alpha = 1}^n e^{i(\varphi_i - m_\alpha)} q^{\frac{\lambda_i}{2}} \int_\gamma \frac{dx}{2\pi ix} \frac{x^{\lambda_i}}{(1 - q^{1/2}x)(1 - q^{1/2}x^{-1})}. 
\end{equation}
The residue is once again easily evaluated, leading to 
\begin{equation} \label{charmatter1}
\mathscr{F} \eval_\lambda = \sum_{\substack{i \, \, \, s.t. \\ \lambda_i > 0}} \sum_{\alpha = 1}^n e^{i(\varphi_i - m_\alpha)} (1 + q + \dots  + q^{\lambda_i - 1}) := \chi_{\mathscr{F}, \lambda}(\varepsilon, m, \varphi). 
\end{equation}
As the only thing really entering this computation are the weights of the representation $R$, it is easy to state the general formula: the sum runs over weights $\omega$ of $R$ under the maximal torus of $G$ such that $\langle \omega, \lambda \rangle > 0$ using the natural pairing between weights and coweights. It is easy, in this language, to explain what corresponds to a more general choice of Lagrangian splitting in \eqref{hypbdry}. If we have picked a splitting $R \oplus R^* = L \oplus L^*$, simply replace the weights of $R$ by the weights of $L$. 

\subsubsection{}
Imitating the arguments in section \ref{univ}, the formula in a situation with $n$ singularities $\overline{\mathscr{M}}(\{\mu_i; p_i \})$ at a fixed point $(\lambda_1, \dots, \lambda_n)$ is 
\begin{equation} \label{charmattern}
\mathscr{F} \eval_{(\lambda_1, \dots, \lambda_n)} = \sum_{i = 1}^n \chi_{\mathscr{F}, \lambda_i}\Big(\varepsilon, m, \varphi + \varepsilon\sum_{j < i} \lambda_j  \Big). 
\end{equation}
Readers familiar with the construction of \cite{bfn} will recognize \eqref{charmatter1}, up to conventions, as the fixed point contributions of the bundle (in our notation, $N = R^*$) $gN(\mathscr{O})/(N(\mathscr{O}) \cap gN(\mathscr{O}))$ restricted to the Schubert cell labeled by dominant coweight $\mu$, $g(z) \in G_{\mathbb{C}}(\mathscr{O})z^\mu$. This object appears via the excess intersection formula whenever one performs computations with the definition in \cite{bfn}. I believe $\mathscr{F}$ should in general coincide with this object as a coherent sheaf on the Schubert variety, but proving this would require a more careful analysis of the cokernel of the Dirac operator than I am willing or able to give here.

The discussion above is good enough to establish an identification in equivariant $K$-theory, which is all that is really needed for computations. Then essentially the same arguments in section \ref{diffop} together with the results in \cite{bfn} complete the physical derivation of the construction of \cite{bfn} for theories with matter of cotangent type. This in particular includes all quiver gauge theories, and for quivers of ADE type it was proven in \cite{bfnslice} that this definition agrees with what was expected based on string theory brane engineering constructions going all the way back to \cite{hananywitten}. 

\subsubsection{Example: SQCD}
While the results above were enough to make contact with the mathematical definition, we will also briefly repeat the explicit calculations of section \ref{diffop} in the presence of matter. We consider our running example, $G = U(k)$ with $n$ hypermultiplets in the fundamental. Incorporating the Euler class contribution leads to a modification in the formulas via \eqref{charmatter1}: 
\begin{equation} \label{m+-sqcd}
\begin{split}
\bra{f} \mathscr{O}_{\mu^+}(t_1) \ket{\varphi} & = \int_{\overline{\mathscr{M}}(\mu^+; p)} f(\text{Chern roots of $\mathscr{E}|_{t = 1}$}) \cup \text{Euler}(\mathscr{F}) = \sum_{i = 1}^k f(\varphi_1, \dots, \varphi_i + \varepsilon, \dots, \varphi_k) \frac{\prod_{\alpha = 1}^n(\varphi_i - m_\alpha)}{\prod_{j (\neq i)}(\varphi_i - \varphi_j)}  \\
\bra{f} \mathscr{O}_{\mu^-}(t_1) \ket{\varphi} & = \int_{\overline{\mathscr{M}}(\mu^-; p)} f(\text{Chern roots of $\mathscr{E}|_{t = 1}$}) \cup \text{Euler}(\mathscr{F}) = \sum_{i = 1}^k  \frac{f(\varphi_1, \dots, \varphi_i - \varepsilon, \dots, \varphi_k)}{\prod_{j (\neq i)}(\varphi_j - \varphi_i)}.
\end{split}
\end{equation}
This leads once again to a difference operator realization as in section \ref{diffop}: 
\begin{equation} \label{diffopmatter}
\begin{split}
\mathscr{O}_{\mu^+} & \longrightarrow \sum_{i = 1}^k \exp(- \varepsilon \pdv{}{\varphi_i}) \frac{\prod_{\alpha = 1}^n(\varphi_i - m_\alpha)}{\prod_{j (\neq i)}(\varphi_i - \varphi_j)}  \\
\mathscr{O}_{\mu^-} & \longrightarrow \sum_{i = 1}^k \exp(\varepsilon \pdv{}{\varphi_i}) \frac{1}{\prod_{j (\neq i)}(\varphi_j - \varphi_i)}. 
\end{split}
\end{equation}
We will give a general explanation for this phenomenon in section \ref{qmechbranes}. It is not difficult to generalize the discussion here to obtain difference operator realizations of Coulomb branch algebras for general quiver gauge theories (as long as all gauge groups are unitary), but we will not write these explicitly to keep the formulas simple. 

\subsection{K-theoretic analogs}
In section we briefly outline how this construction may be repeated for the case of of four-dimensional gauge theories on the geometry $\mathbb{R}^2 \times [0, 1] \times S^1$. Our main purpose is to highlight the key features: namely to exhibit the appropriate complex structure of the four-dimensional Coulomb branch, and to make contact with the original construction \cite{kw} in the maximally supersymmetric case. 

\subsubsection{Main example: pure gauge theory}
To avoid unnecessary complications in the formulas, we only write equations explicitly in the pure gauge theory case. The four-dimensional gauge theory on a circle can be viewed as a loop space version of three-dimensional gauge theory \cite{Nekrasov_1998}, where we deform the supercharge $\delta$ of \eqref{topsusy} to be equivariant with respect to the rotation of the loops. Call the coordinate along $S^1$ by $s$. Then promote all fields to functions of $s$ (so they can be viewed as fields on $\mathbb{R}^2 \times [0, 1] \times S^1$), and replace $\varphi$ by $A_s + i \phi$ for a real scalar $\phi$. The action of the supercharge $\delta$ in four dimensions is (we drop the differential form notation and restore the index $\mu = 0, 1, 2$ for the $\mathbb{R}^2 \times [0, 1]$ directions):
\begin{alignat}{3}
& \delta A_\mu = \psi_\mu \quad   &&   \delta (A_s + i \phi) = 0 \quad && \delta \chi_\mu = H_\mu \nonumber \\
& \delta \psi_\mu = i D_\mu \phi + F_{\mu s} \quad && \delta (A_s - i \phi) = \eta \quad && \delta H_\mu = i \comm{\chi_\mu}{\phi} - D_s \chi_\mu \label{kthsusy} \\
& \delta \sigma = \psi_3 \quad && \delta \eta = 2 i D_s \phi \nonumber \\
& \delta \psi_3 = i\comm{\sigma}{\phi} - D_s \sigma \quad && \quad && \quad \nonumber 
\end{alignat}
The supercharge is no longer nilpotent, but squares to the action of $D_s + i \phi$, in other words a combined gauge transformation and translation on $S^1$. Viewing $\delta$ as a twisted supercharge leads to a theory which is \textit{not} topological in four dimensions; it is topological along $\mathbb{R}^2 \times [0, 1]$, but behaves as a quantum mechanics along $S^1$ and has nontrivial time evolution there. One may view $A_\mu dx^\mu + A_s ds$ as a four-dimensional gauge field and identify $\eta = 2 \psi_s$. 

\subsubsection{}
The action on $W \times S^1$ is a straightforward generalization of \eqref{topaction} (we suppress the explicit volume form $\star 1$ in some places):
\begin{equation} \label{kthaction}
\begin{split}
S = & - \frac{1}{e^2} \int_{\partial(W \times S^1)} ds \wedge \tr(\sigma F_A) \\
& + \delta\Bigg( \frac{1}{e^2} \int_{W \times S^1} \tr\Big( ds \wedge \chi \wedge (i(F_A + \star D_A \sigma) + \frac{\lambda}{2} \star H) + \psi^\mu(F_{\mu s} - i D_\mu \phi) + \psi_3(-D_s \sigma + i \comm{\phi}{\sigma}) + \eta (-\frac{i}{4} D_s \phi) \Big) \Bigg). 
\end{split}
\end{equation}
When $\lambda = 1$, integrating out the auxiliary field leads to the bosonic part of the action (which has a full four-dimensional symmetry: all field strengths and covariant derivatives are taken in the four-dimensional sense): 
\begin{equation}
L_B = \frac{1}{2} \norm{F_A}^2 + \frac{1}{2} ( \norm{D_A \sigma}^2 + \norm{D_A \phi}^2) + \norm{\comm{\sigma}{\phi}}^2
\end{equation}
which is indeed the bosonic part of the action for four dimensional $\mathscr{N} = 2$ super-Yang-Mills, written in terms of two real scalars $\phi, \sigma$ rather that one complex scalar as is more standard. 

\subsubsection{}
The boundary conditions have an obvious uplift to the four-dimensional setting: at the $t = 0$ boundary, one puts a Dirichlet boundary on the gauge field and turns on a generic background Wilson line (flat connection) along the $S^1$, while also putting Dirichlet boundary conditions on the real scalar $\phi$. The holonomy of the complexified flat connection $A + i \phi ds$ along $S^1$ at $t = 0$ is a diagonal matrix whose entries are multiplicative uplifts of the parameters $\varphi_i$ appearing throughout the paper. $\sigma$ has Neumann boundary conditions at $t = 0$ as before. The opposite is true at $t = 1$: $(A, \phi)$ obey Neumann boundary conditions and $\sigma$ obeys a Dirichlet condition. 

We now turn to the supersymmetric observables. Gauge-invariant polynomials in the complex scalar $\varphi$ lift to supersymmetric Wilson line operators: a Wilson line in representation $R$ given by 
\begin{equation} \label{wilson}
\tr_R\Bigg( \mathcal{P} \exp( - \int_{S^1} ds(A_s + i \phi) )\Bigg)
\end{equation}
is automatically supersymmetric as a consequence of \eqref{kthsusy}. Likewise, monopole operators lift automatically to this setting by simply taking the singular behavior to be independent of the coordinate $s$; this leads to the standard definition of a supersymmetric 't Hooft line in the 4d $\mathscr{N} = 2$ theory. Thus our localization computations readily lift to compute the algebra of mixed Wilson-'t Hooft line operators in 4d $\mathscr{N} = 2$ theory on $\mathbb{R}^2 \times [0, 1] \times S^1$, where all line operators wrap the $S^1$ direction. 

\subsubsection{Reduction to quantum mechanics}
Imitating the arguments in the appendix to \cite{Nekrasov_1998}, the theory with action \eqref{kthaction} in the presence of monopole singularities localizes to the minimal supersymmetric quantum mechanics on the compactified moduli space of Hecke modifications $\overline{\mathscr{M}}(\{ \mu_i; p_i \})$. The path integral of the four-dimensional theory reduces to the path integral along $S^1$ of this quantum mechanics, which naively computes the index of the Dirac operator on $\overline{\mathscr{M}}(\{ \mu_i; p_i \})$. However there is an immediate problem: inspection of the simplest examples illustrates that $\overline{\mathscr{M}}$ does not in general admit a spin structure (even ignoring the issue of singularities). In physics language, this quantum mechanics suffers from a global anomaly. 

\subsubsection{}
The workaround proposed in \cite{bfn} is to note that $\overline{\mathscr{M}}$ is always Kahler, so the Dirac operator is the $\overline{\partial}$ operator twisted by the non-existent $\mathscr{K}^{1/2}_{\overline{\mathscr{M}}}$. The authors of \cite{bfn} propose, effectively, that one should take the supersymmetric quantum mechanics to be the one computing the index of the $\overline{\partial}$ operator. The space of supersymmetric ground states in this quantum mechanics is the cohomology of the structure sheaf: 
\begin{equation}
\mathcal{H} = H^*(\overline{\mathscr{M}}; \mathscr{O}_{\overline{\mathscr{M}}}). 
\end{equation}
The effect of the background Wilson lines is easy to see in the quantum mechanics: the group $G$ acts on $\mathcal{H}$, and the background Wilson line amounts to an insertion of the group element generating this action in the trace computed by supersymmetric quantum mechanics. Likewise, in this setting the $\Omega$-deformation is easy to describe: it is simply a twisted boundary condition on $S^1$, where the fields only come back to themselves along $S^1$ up to a rotation of the $\mathbb{R}^2$ plane through an angle $\varepsilon$. If there are no insertions of supersymmetric Wilson lines along $[0, 1]$, the four-dimensional partition function in this situation computes the character of $\mathcal{H}$ under the group of all symmetries of the problem: 
\begin{equation}
Z_{\text{4d}}(q, \varphi) = \Tr_{\mathcal{H}}(-1)^F g_{\text{rot}} g_{\text{gauge}} = \chi(\overline{\mathscr{M}}, \mathscr{O}_{\overline{\mathscr{M}}})
\end{equation}
where $g_{\text{gauge}}$ is associated to the action of $G$ and $g_{\text{rot}}$ is associated to the action of the spacetime rotation; both of these become internal symmetries in the effective quantum mechanics description. When writing K-theoretic formulas, it is implicit that the variables $\varphi_i$ are multiplicative, and we use $q = e^{iR\varepsilon}$ where $R$ is the radius of the circle parameterized by the $s$ coordinate. In the final line, we have rewritten the formula algebra-geometrically as the equivariant Euler character of the structure sheaf of the moduli space. 

\subsubsection{}
Inclusion of additional supersymmetric Wilson lines is easiest to state in the algebra-geometric notation. As explained in depth in section \ref{univ}, there are universal bundles over the moduli space and in the present context the eigenvalues of the holonomies of the supersymmetric Wilson observables \eqref{wilson} become the $K$-theoretic Chern roots, and the Wilson observables themselves can be interpreted as equivariant $K$-theory classes which are polynomials in the universal bundles. The path integral with such insertions computes the equivariant Euler characteristic of those $K$-theory classes: 
\begin{equation}
\chi(\overline{\mathscr{M}}, \mathscr{O}_{\overline{\mathscr{M}}} \otimes (\text{universal})). 
\end{equation}
Then using the standard localization of quantum mechanical path integrals, or the equivalent K-theoretic index theorem, one computes the four-dimensional partition function for the pure gauge theory as, for example in the case of a single insertion of a 't Hooft line with charge $\mu^+ = (1, 0, \dots, 0)$
\begin{equation}
\chi(\overline{\mathscr{M}}(\mu^+; p), \mathscr{O}_{\overline{\mathscr{M}}} \otimes (\text{universal})) = \sum_{i = 1}^k \frac{1}{\prod_{j (\neq i)}(1 - \varphi_j/\varphi_i)} (\text{universal}) \eval_{(0, \dots, 1, \dots, 0)}. 
\end{equation}
This is just the naive multiplicative uplift of the computations from section \ref{diffop}. One may also couple to matter in the four-dimensional case, by replacing the Euler class by its $K$-theoretic analog $\Lambda^*(\mathscr{F})$. 

\subsubsection{}
The $K$-theoretic Coulomb branches coincide with the Coulomb branch of the 4d $\mathscr{N} = 2$ theory compactified on a circle, viewed in a particular complex structure. In the complex structure $I$ used to describe the three-dimensional Coulomb branches, the four-dimensional one is the total space of a Seiberg-Witten integrable system \cite{sw1996} and fibers over the space of expectation values of the complex scalars with fibers parameterized by Wilson lines complexified by dual photons. The $K$-theoretic Coulomb branch described here corresponds to the Coulomb branch in complex structure usually called $J$; it fibers holomorphically over a base parameterized by Wilson lines and real scalars. The fibers are parameterized by real scalars complexified by dual photons; since the complex scalar in the 4d $\mathscr{N} = 2$ vector multiplet is split into two real scalars which are paired in distinct ways with the gauge field and dual photon, this amounts to a hyperkahler rotation relative to complex structure $I$. Thus, one finds in this way the same underlying hyperkahler manifold viewed in a distinct complex structure. If the 4d $\mathscr{N} = 2$ theory is of class $\mathcal{S}$, its Coulomb branch on a circle is a Hitchin system, and the complex structure called $J$ here coincides with the usual complex structure $J$ of the Hitchin system, where it is natural to view the Hitchin moduli space as the moduli space of local systems. 

\subsection{Reinterpretation via branes} \label{branes}
In this section we give a more conceptual explanation for why the quantum mechanical structure discovered in section \ref{diffop} emerged. As a result of this analysis we will give a more careful determination of the deformation quantization modules arising microscopically from Dirichlet and Neumann boundary conditions in gauge theory, confirming expectations of \cite{Bullimore_2016}. 

The main idea will be to follow \cite{nw2010} and reduce the three-dimensional gauge theory to a two-dimensional sigma model with target space $\mathscr{M}_C$. Upon recalling the relationship between the $A$-model and quantization \cite{Gukov_2009}, this will render transparent the quantum-mechanical interpretation of our results. 

\subsubsection{On the effective sigma model}
Here we recall the main points of \cite{nw2010} in a fashion which makes contact with our present problem. The situation we are studying is very closely related to the one in section 3 of that paper, so we can borrow much of the analysis. The main difference is that the role of the Hitchin moduli space there will be played by the Coulomb branch here.  

\subsubsection{}
We begin with the 3d $\mathscr{N} = 4$ gauge theory on $[0, 1] \times \mathbb{R}^2$, in the twist we have been studying throughout the paper. Eventually we will turn on the $\Omega$-deformation along $\mathbb{R}^2$, but we begin by discussing the undeformed case. As the theory is topological, one may deform the metric on $\mathbb{R}^2$ to that of a ``cigar'' with fixed asymptotic radius, and consider compactification of the theory on the circle direction of the cigar. The effective two-dimensional geometry is $\Sigma = \mathbb{R}_{\geq 0} \times [0, 1]$ (call the coordinates $(r, t)$), which is a two-manifold with corners. We will be interested primarily in the effective boundary condition along $r = 0$, and return in section \ref{qmechbranes} to discuss the boundary conditions at $t = 0, 1$. 

\subsubsection{}
The gauge theory on the cigar geometry, after dualizing the three-dimensional gauge fields to scalars, reduces at long distances to a two-dimensional sigma model on $\Sigma$ with target $\mathscr{M}_C$. This sigma model has $\mathscr{N} = (4, 4)$ supersymmetry, as $\mathscr{M}_C$ is hyperkahler. We refer to the three distinguished complex structures as $I, J$ and $K$. The tip of the cigar determines a half-BPS boundary condition in the sigma model (in other words, a brane), simply by constraining the fields in such a way that captures the fact that there is no boundary at $r = 0$ in the microscopic description. In the absence of $\Omega$-deformation, this is simply a space-filling brane on $\mathscr{M}_C$ with trivial Chan-Paton bundle. This is a brane of type $(B, B, B)$, meaning that it is compatible with the $B$-type topological twist of the sigma model to $\mathscr{M}_C$ in all three complex structures $(I, J, K)$. 

Moreover, the supercharge $\delta$ of \eqref{topsusy} descends to that of the $B$-model in complex structure $I$. This is easy to see: $\varphi$ obeys a $B$-model condition in \eqref{topsusy}, while (if we call the angular coordinate along the cigar $\theta$), $\sigma + i A_\theta$ obeys an $A$-model condition (the condition for a supersymmetric field configuration $F_A + \star D_A \sigma = 0$ reduces, for abelian gauge fields, to the Cauchy-Riemann equations along $\Sigma$ for $\sigma, A_\theta$---see \cite{nw2010} for more detail). Upon dualizing $A_\theta$ to the dual photon $\gamma$, this becomes a $B$-model condition for $\sigma + i \gamma$. 

\subsubsection{}
The deformation of the metric on $\mathbb{R}^2$ from the flat one to the cigar is compatible with the $\Omega$-deformation, so one may consider the $\Omega$-deformed theory in this setting, and ask for the brane at $r = 0$ in the effective sigma model. The $\Omega$-deformation is a drastic deformation of the theory, so one expects a rather different answer. In \cite{nw2010} it was shown that, away from the tip of the cigar, the $\Omega$-deformation can be undone by a field redefinition. If one writes $\varphi = A_4 + i A_5$ and considers the limit of large asymptotic radius of the cigar, the field redefinition is just $A_4 \to \varepsilon A_\theta$, $\varepsilon A_\theta \to -A_4$. The effective asymptotic radius of the cigar in the undeformed language is $1/\varepsilon$. 

Then in this language, and given our conventions for the complex structures $(I, J, K)$, the $\Omega$-deformed supercharge $\delta_\varepsilon$ becomes the supercharge of the $A$-model in symplectic structure $\omega_K$ (the Kahler form for complex structure $K$). This is readily verified: originally $A_4 + i A_5$ obeyed a $B$-model condition and $\sigma + i A_\theta$ obeyed an $A$-model condition. Under the field redefinition, $A_\theta + i A_5$ obeys a $B$-model condition which dualizes to an $A$-model condition for $\gamma + i A_5$, and $\sigma - i A_4$ satisfies an $A$-model condition (we define the complex structure $K$ such that these are the holomorphic coordinates; in the complex structure $J$ the roles of $A_4$ and $A_5$ are reversed). The brane from the tip of the cigar is identified in \cite{nw2010} with the so-called canonical coisotropic brane $\mathcal{B}_{cc}$ in this $A$-model, which is an $(A, B, A)$ brane in the earlier terminology. It is a space-filling brane with a rank one Chan-Paton bundle carrying a connection with curvature proportional to $\omega_J$. See section 11 of \cite{kw} for a review of the basic properties of these branes. 

\subsubsection{}
The space of boundary local operators $\text{End}(\mathcal{B}_{cc})$ at the $r = 0$ boundary, as a vector space, is the space of $I$-holomorphic functions on $\mathscr{M}_C$---in the microscopic gauge theory these were just the supersymmetric local operators preserved by $\delta_\varepsilon$. As a ring, $\text{End}(\mathcal{B}_{cc})$ is the deformation quantization of this ring of functions with respect to the holomorphic symplectic form $\Omega_I = \omega_J + i \omega_K$, with $\varepsilon$ playing the role of Planck's constant. 

\subsubsection{Quantum mechanical reformulation} \label{qmechbranes} 
$A$-models with canonical coisotropic branes are closely related to quantum mechanics and representation theory, as elucidated in \cite{Gukov_2009}, \cite{wittenqm}. The basic reason for this is that the $\mathcal{B}_{cc}$ brane gives a natural framework for deformation quantization. Moreover, given any other brane $\mathscr{L}$ in the $A$-model of type $\omega_K$, the space of open string states $\text{Hom}(\mathcal{B}_{cc}, \mathscr{L})$ is naturally a right module for the deformation quantization $\text{End}(\mathcal{B}_{cc})$, simply by joining open strings. On the other hand, we have reviewed how $\Omega$-deformed 3d $\mathscr{N} = 4$ gauge theories reduce to such $A$-models. We will now explain how the structure of our formulas can be understood using the relation to $A$-models and quantization. 

The fact that we can compute microscopically by integrating over spaces of Hecke modifications allows for a much more explicit determination of such $A$-model amplitudes than is usually possible, in other words the 3d gauge theory provides a useful UV completion of the effective $A$-model with target $\mathscr{M}_C$. 

When we reduce to the $A$-model with target $\mathscr{M}_C$, there are three boundaries: $r = 0$ which supports the $\mathcal{B}_{cc}$ brane as explained in the previous section, and $t = 0, 1$ which support some branes $\mathscr{L}_{0, 1}$ descending from the Dirichlet and Neumann boundary conditions. See Figure \ref{fig:branesfig} below.

\begin{figure}
    \centering
\begingroup%
  \makeatletter%
  \providecommand\color[2][]{%
    \errmessage{(Inkscape) Color is used for the text in Inkscape, but the package 'color.sty' is not loaded}%
    \renewcommand\color[2][]{}%
  }%
  \providecommand\transparent[1]{%
    \errmessage{(Inkscape) Transparency is used (non-zero) for the text in Inkscape, but the package 'transparent.sty' is not loaded}%
    \renewcommand\transparent[1]{}%
  }%
  \providecommand\rotatebox[2]{#2}%
  \newcommand*\fsize{\dimexpr\f@size pt\relax}%
  \newcommand*\lineheight[1]{\fontsize{\fsize}{#1\fsize}\selectfont}%
  \ifx\svgwidth\undefined%
    \setlength{\unitlength}{196.06955492bp}%
    \ifx\svgscale\undefined%
      \relax%
    \else%
      \setlength{\unitlength}{\unitlength * \real{\svgscale}}%
    \fi%
  \else%
    \setlength{\unitlength}{\svgwidth}%
  \fi%
  \global\let\svgwidth\undefined%
  \global\let\svgscale\undefined%
  \makeatother%
  \begin{picture}(1,1.16022272)%
    \lineheight{1}%
    \setlength\tabcolsep{0pt}%
    \put(0,0){\includegraphics[width=\unitlength,page=1]{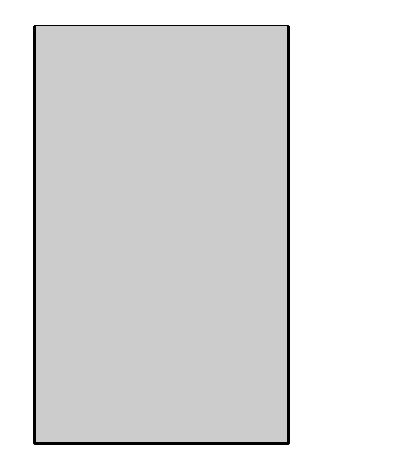}}%
    \put(0.65349377,1.12097495){\color[rgb]{0,0,0}\makebox(0,0)[lt]{\lineheight{1.25}\smash{\begin{tabular}[t]{l}$t = 0$\end{tabular}}}}%
    \put(0.03371139,1.11986296){\color[rgb]{0,0,0}\makebox(0,0)[lt]{\lineheight{1.25}\smash{\begin{tabular}[t]{l}$t = 1$\end{tabular}}}}%
    \put(0.38004295,0.00976214){\color[rgb]{0,0,0}\makebox(0,0)[lt]{\lineheight{1.25}\smash{\begin{tabular}[t]{l}$\mathcal{B}_{cc}$\end{tabular}}}}%
    \put(0.74686188,0.61341351){\color[rgb]{0,0,0}\makebox(0,0)[lt]{\lineheight{1.25}\smash{\begin{tabular}[t]{l}$\mathscr{L}_{0}$\end{tabular}}}}%
    \put(-0.00189832,0.62106388){\color[rgb]{0,0,0}\makebox(0,0)[lt]{\lineheight{1.25}\smash{\begin{tabular}[t]{l}$\mathscr{L}_{1}$\end{tabular}}}}%
  \end{picture}%
\endgroup%

    \caption{The worldsheet of the effective $A$-model with target $\mathscr{M}_C$ upon performing cigar reduction.}
    \label{fig:branesfig}
\end{figure}
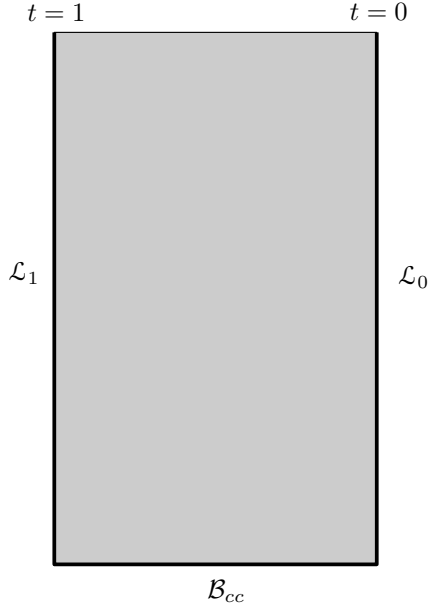

The possible boundary operators which can be inserted at $r = 0, t = 0$ determine elements of $\text{Hom}(\mathscr{L}_0, \mathcal{B}_{cc})$ by state/operator correspondence, and likewise the operators at $r = 0, t = 1$ determine elements of $\text{Hom}(\mathcal{B}_{cc}, \mathscr{L}_1)$. Then the full amplitudes are canonically elements of $\text{Hom}(\mathscr{L}_0, \mathscr{L}_1)$. On the other hand, the amplitudes are of course valued in the complex numbers, allowing us to conclude $\text{Hom}(\mathscr{L}_0, \mathscr{L}_1) \simeq \mathbb{C}$. We will give an alternative geometric explanation of this later. 

\subsubsection{}
It is strictly speaking only true that amplitudes are valued in numbers if we view $\varepsilon, \varphi_i$ as fixed at some generic background values, as the amplitudes are actually functions of these variables. The value of $\varphi_i$ is in fact part of the data determining the brane $\mathscr{L}_0$, due to \eqref{DBC}. Then let us denote by $\ket{\varphi} \in \text{Hom}(\mathscr{L}_0, \mathcal{B}_{cc})$ the element corresponding to the insertion of the trivial operator $1$ at $r = 0, t = 0$. The space $\text{Hom}(\mathcal{B}_{cc}, \mathscr{L}_1)$ is spanned by the allowed boundary operators at $r = 0, t = 1$, which just consist of gauge-invariant polynomials in $\varphi(t = 1, 0)$ (Neumann boundary conditions do not allow for the insertion of monopole operators at the boundary). Denote by $\bra{f} \in \text{Hom}(\mathcal{B}_{cc}, \mathscr{L}_1)$ the state corresponding to $f(\varphi(t = 1, 0))$. Including insertions of monopole operators in the interior of $t \in [0, 1]$, the correlation functions under study can be reinterpreted as $A$-model amplitudes essentially quantum-mechanical in nature:
\begin{equation} \label{qmamps}
\langle \mathscr{O}_{\mu_n}(t_n) \dots \mathscr{O}_{\mu_1}(t_1) f(\varphi(t = 1, 0)) \rangle = \bra{f} \mathscr{O}_{\mu_n}(t_n) \dots \mathscr{O}_{\mu_1}(t_1) \ket{\varphi} \in \text{Hom}(\mathscr{L}_0, \mathscr{L}_1) \simeq \mathbb{C}. 
\end{equation}
Moreover, the normalization is determined by the fact that with no insertions, the gauge theory path integral reduces to the integral of $1$ over a point
\begin{equation}
\braket{f = 1}{\varphi} = 1. 
\end{equation}
On the other hand, by our discussion in section \ref{univ} we know that in the pure gauge theory (with a straightforward generalization to include matter):
\begin{equation}
    \bra{f} \mathscr{O}_{\mu_n}(t_n) \dots \mathscr{O}_{\mu_1}(t_1) \ket{\varphi} = \int_{\overline{\mathscr{M}}(\{ \mu_i; p_i \})} f \Big(\text{Chern roots of $\mathscr{E} \eval_{t = 1}$} \Big).
\end{equation}
The left hand side is written in a notation intentionally reminiscent of elementary quantum mechanics; the goal of the present section was to explain that this analogy is quite literal. 

\subsection{Dirichlet and Neumann conditions as branes in $\mathscr{M}_C$}
Using what we learned thus far, we can also determine the support of the branes $\mathscr{L}_{0, 1}$ inside the Coulomb branch $\mathscr{M}_C$. We will, for concreteness, consider $U(k)$ gauge theory with $n$ flavors with generic complex masses $m_\alpha$, so that the Coulomb branch is a smooth affine variety. It is not difficult to generalize the statements to (at least) arbitrary quiver gauge theories. 

Let us start with $\mathscr{L}_0$, the brane determined by the Dircihlet boundary condition. Inside any correlation function, suppose we insert a gauge-invariant polynomial $P(\varphi(t_*, 0))$, where the precise value of $t_*$ is irrelevant so long as it is smaller than any other value of $t$ supporting an operator insertion. Then, by the discussion of the universal bundle in section \ref{univ}, $\mathscr{E} |_{t = t_*}$ is a topologically trivial but $G$-equivariantly nontrivial vector bundle over the localization locus, allowing one to perform the manipulations 
\begin{equation}
\bra{f} \dots P(\varphi(t_*, 0)) \ket{\varphi} = \int_{\overline{\mathscr{M}}} (\dots) P(\text{Chern roots of $\mathscr{E}|_{t = t_*}$} )= P(\varphi_1, \dots, \varphi_k) \bra{f} \dots \ket{\varphi}. 
\end{equation}
Since this is true inside arbitrary correlators, we conclude 
\begin{equation}
P(\varphi(t_*, 0)) \ket{\varphi} = P(\varphi_1, \dots, \varphi_k) \ket{\varphi}
\end{equation}
so that, as the notation suggests, $\ket{\varphi}$ is a state with definite value of $\varphi$. The support of the corresponding brane $\mathscr{L}_0$ is thus easily found by considering the semiclassical $\varepsilon \to 0$ limit: $\mathscr{L}_0$ wraps the fiber of the integrable system $\mathscr{M}_C \to \mathfrak{h}_{\mathbb{C}}/W$ over the corresponding point $(\varphi_1, \dots, \varphi_k)$. So long as $\varphi_i$ are generic, this fiber should be smooth. As the fiber of an $I$-holomorphic complex integrable system, such a brane is an $I$-holomorphic Lagrangian. Thus it is a $(B, A, A)$ brane, due to the formula $\Omega_I = \omega_J + i \omega_K$. 

\subsubsection{}
This result could have been guessed based on the simple fact that the boundary value of $\varphi$ is fixed at $t = 0$ by \eqref{DBC}, while (after dualizing the gauge field to the dual photon) the boundary values of other fields are unconstrained. By contrast, $\varphi$ is unconstrained at $t = 1$, while $\sigma$ and the dual photons are fixed. Because this boundary condition involves fixing the dual photon, it is difficult to determine the support of $\mathscr{L}_1$ based purely on semiclassical reasoning; fortunately, we have already performed exact calculations in the full gauge theory, so we can repeat the more precise derivation given above without much effort. 

\subsubsection{}
Since we wish to determine the support of $\mathscr{L}_1$, and the role of $\varphi_i$ and the monopole operators have been effectively reversed, what one wishes to consider now is the insertion of a (say, minuscule) monopole operator inside some suitably general class of quantum mechanical amplitudes. The semiclassical $\varepsilon \to 0$ limit as we bring the monopole operator close to $t = 1$ will encode the support of $\mathscr{L}_1$. In fact, a suitably general class is just given by \eqref{m+-sqcd}. To facilitate the computation and compare with earlier literature, let us introduce abelianized coordinates $(\varphi_i, u^\pm_i)$ following \cite{bdg}, \cite{Bullimore_2016} to which we refer for more details. In terms of abelianized coordinates the monopole operators are 
\begin{equation}
    \mathscr{O}_{\mu^\pm} := \sum_{i = 1}^k \widehat{u}_{\pm, i}
\end{equation}
so in particular the abelianized monopole operators are essentially shift operators by \eqref{diffopmatter}. The abelianized coordinates $u_{\pm, i}$ are not quite good functions on the Coulomb branch, as they make sense only after a localization in the $\varphi_i$ variables and break the Weyl group symmetry. However, certain invariant combinations of them like $\mathscr{O}_{\mu^\pm}$ extend to regular functions on the whole $\mathscr{M}_C$.

\subsubsection{}
In the semiclassical $\varepsilon \to 0$ limit, the shift operators in \eqref{diffopmatter} act trivially, and the Lagrangian $\mathscr{L}_1$ is characterized by the following relations in abelianized coordinates (again, this is stated in the SQCD case, with boundary conditions \eqref{hypbdry} on the hypermultiplets---different choices of Lagrangian splitting for the matter representation in \eqref{hypbdry} would lead to different numerators):
\begin{equation} \label{l1supp}
\begin{split}
u_{+, i} & = \frac{\prod_{\alpha = 1}^n(\varphi_i - m_\alpha)}{\prod_{j(\neq i)}(\varphi_i - \varphi_j)} \\
u_{-, i} & = \frac{1}{\prod_{j (\neq i)}(\varphi_j - \varphi_i)}. 
\end{split}
\end{equation}
While the abelianized coordinates are not good coordinates on the Coulomb branch, the functions generating the coordinate ring $\mathbb{C}[\mathscr{M}_C]$ can be constructed from them, so these equations fully characterize the support of $\mathscr{L}_1$ as a brane in $\mathscr{M}_C$. It is a holomorphic section of the integrable system $\mathscr{M}_C \to \mathfrak{h}_{\mathbb{C}}/W$; correspondingly, as an $I$-holomorphic Lagrangian, it is another $(B, A, A)$ brane. The formula \eqref{l1supp} had been previously proposed in \cite{Bullimore_2016}, describing in our language the support of $\mathscr{L}_1$, on the basis of some guesswork and passing it through several consistency checks. We have derived it here from first principles, as \eqref{m+-sqcd} were computed directly in the microscopic gauge theory, and we obtained \eqref{l1supp} just by analyzing them from the point of view of the effective sigma model. 

\subsubsection{}
With this geometric understanding of the branes $\mathscr{L}_{0, 1}$, we can give another derivation of the claim that $\text{Hom}(\mathscr{L}_0, \mathscr{L}_1) \simeq \mathbb{C}$. As both $\mathscr{L}_{0, 1}$ are $(B, A, A)$ branes, they are in particular $B$-branes in complex structure $I$. Then the morphism space $\text{Hom}(\mathscr{L}_0, \mathscr{L}_1)$ is interpreted in the $B$-model as $\text{Ext}^*(\mathscr{O}_{\mathscr{L}_0}, \mathscr{O}_{\mathscr{L}_1}) \simeq \mathbb{C}$ since $\mathscr{L}_0, \mathscr{L}_1$ are smooth subvarieties intersecting transversely at a single point. 

What we have described here is a small piece of the mathematical theory of deformation quantization modules for the Coulomb branches $\mathscr{M}_C$. Any boundary condition for the effective sigma model, so any $A$-brane of type $\omega_K$, determines a module over the deformation quantization $\text{End}(\mathcal{B}_{cc})$ as described in the beginning of this section. We identified the Dirichlet and Neumann boundary conditions in the microscopic gauge theory as branes in the effective sigma model, and by direct computation identified the corresponding deformation quantization modules. 

Readers familiar with the mathematical construction of \cite{bfn} may note that the deformation quantization module $\text{Hom}(\mathcal{B}_{cc}, \mathscr{L}_1)$, spanned as a vector space by symmetric polynomials in $\varphi_i$, is one of the modules that can be constructed essentially tautologically from the presentation of the quantized Coulomb branch algebra as a convolution algebra (modulo certain infinite-dimensionality technicalities). One of the goals of this section was to show that it arises naturally from physics, by simply rethinking the computation of the same gauge theory correlators.  

In a companion paper \cite{shiftop}, we will apply the machinery developed here to enumerative geometry of quasimaps in quiver varieties. This gives rise to an interesting connection with multiplication morphisms for generalized affine Grassmannian slices introduced in \cite{bfnslice} and actions of Hecke operators on geometric Eisenstein series \cite{braverman2000geometric}. In the $\varepsilon \to 0$ limit this recovers some observations of Teleman about the relationship between quantum cohomology and Coulomb branches \cite{teleman2014gauge}, \cite{teleman2019}.

\section{Line operators and the KLRW algebras} \label{klrw}
Using what we learned thus far, we can understand the cylindrical KLRW algebras and their relation to the derived category of coherent sheaves on the Coulomb branch \cite{webster2019} from a physical point of view. This was in fact the main motivation for the analysis done in this paper. We will arrive at what we hope is, for a physicist, a more satisfactory understanding of one half of the mirror symmetry underlying the constructions \cite{minabmodel}, \cite{minaamodel}. In the language of the present paper, this arises from considering the 3d $\mathscr{N} = 4$ theory on $\mathbb{R}^2 \times [0, 1]$, with a codimension two defect included at the origin of $\mathbb{R}^2$. The KLRW algebra will arise as the algebra of supersymmetric local operators along the line defect. Showing this by a direct quantum field theory calculation will be the main goal of this section. 

Just as the constructions thus far had a direct relation to \cite{kw}, the calculations in the present section have a direct relation to \cite{gw}. 

\subsection{Pure gauge theory and the cylindrical nilHecke algebra} \label{puregaugepara}
Paralleling the rest of the paper, we will first explain all the calculations in the pure gauge theory case. As we have explained the philosophy and technique of the calculations at some length in the rest of the paper, we will skip the analogous details here, explaining only the new features. 

\subsubsection{Gukov-Witten line defect}
We once again consider the theory with action \eqref{topaction} and supersymmetry \eqref{topsusy} (eventually subjected to the $\Omega$-deformation), on the geometry $\mathbb{R}^2 \times [0, 1]$. We place a codimension two defect at the origin of $\mathbb{R}^2$ by the following construction \cite{gw}: if $(r, \theta)$ denote polar coordinates in $\mathbb{R}^2$, we integrate only over gauge fields with the prescribed singularity
\begin{equation} \label{lineopsing}
    A \sim \alpha d\theta 
\end{equation}
as $r \to 0$, where $\alpha \in \mathfrak{h}$ is a fixed generic element of the Cartan. For $G = U(k)$, $\alpha = \text{diag}(\alpha_1, \dots, \alpha_k)$ with distinct eigenvalues $\alpha_i$. The more invariant characterization of this behavior is that the holonomy of $A$ around small enough loops $\gamma$ linking $r = 0$ is conjugate to $\exp(-2\pi \alpha)$. In this description, it is clear that the parameter space of the defect is $\alpha \in \mathfrak{h}/W_{\text{aff}}$, where $W_{\text{aff}} = W \ltimes \Lambda_{\text{cochar}}$. This is because the action of the Weyl group preserves the conjugacy class, and translation of $\alpha$ by a cocharacter does not affect $\exp(-2\pi\alpha)$. 

\subsubsection{}
Just as the form of the Dirac singularities reduced the gauge group at the position of insertion to the subgroup preserving the form of the singularity, the line defect (with generic $\alpha$) breaks the gauge group down to its maximal torus $T$ along $r = 0$. This simple fact will underly much of the structure in what follows. 

The form of the boundary conditions \eqref{NBC} is unchanged, while in \eqref{DBC} we now must fix $A(t = 0, z, \bar{z}) = \alpha d\theta$ due to the endpoint of the line operator. We are interested in the correlation functions of supersymmetric local operators inserted along the axis $r = 0$ of the line operator. Since the gauge group has been broken down there, and there is a prescribed singularity for the gauge field in that region, the first step is to describe new supersymmetric local operators which may arise in such a situation. This has been done essentially in \cite{gw}, but we recall the important points for completeness. 

\subsubsection{}
With no line operator, any gauge-invariant polynomial $P(\varphi(t, 0))$ is a good observable. These can be viewed equivalently as $W$-invariant polynomials in the eigenvalues $\varphi_i(t, 0)$. Since the line operator breaks the gauge group down to $T$ along $r = 0$, one may now relax the $W$-invariance condition; any polynomial in $\varphi_i(t, 0)$ will be a good supersymmetric observable. We will give a more geometric explanation for this in section \ref{univpara} discussing the behavior of the universal bundle in this context. 

More subtle are the analogs of monopole operators inserted along the line operator. One way to motivate them is the following: to create a further local singularity at some point $(t = t_*, r = 0)$ along the line operator, one can consider varying the parameters $\alpha(t)$ of the line operator within some small region $t_* - \epsilon < t < t_* + \epsilon$ along a loop in parameter space \cite{gukov2014surface}. Taking $\epsilon \to 0$ produces a junction between two equivalent line defects, with a disorder-type local operator at $t = t_*$. These are the natural generalizations of monopole operators in this setting, and are clearly labeled by $\pi_1(\mathfrak{h}/W_{\text{aff}}) = W_{\text{aff}}$. We denote the operator corresponding to $\widehat{w} \in W_{\text{aff}}$ by $\mathscr{O}_{\widehat{w}}(t)$, paralleling the earlier notation for monopole operators.  

Because this characterization is a bit more abstract than the explicit Dirac singularity \eqref{singmono}, we refer readers to section 5.3 of \cite{gw} for an explicit characterization of the singular behavior of the gauge fields in the presence of such a generalized defects (at least for $G = SU(2)$). The basic idea is that the form of the singularity can be captured by a certain model solution to the Yang-Mills equations on $S^2$ linking the point of operator insertion, with punctures at the north and south pole of $S^2$ due to the line operator. We do not believe that the explicit form of the field singularity is particularly illuminating, so we opt for the description in the previous paragraph, and explain its consequence for the moduli space of solutions in the next section and in appendix \ref{affineflags}.

\subsubsection{}
Then our goal, paralleling the discussion in section \ref{loc}, is to understand the evaluation of correlators of the form 
\begin{equation} \label{puregaugelineop}
\langle \mathscr{O}_{\widehat{w}_n}(t_n) \dots \mathscr{O}_{\widehat{w}_1}(t_1) \Omega \rangle. 
\end{equation}
We follow exactly the same strategy: the path integral computing any such correlator can be collapsed to a finite-dimesional equivariant integral over the moduli space of solutions to the Bogomolny equations in the presence of the line singularity \eqref{lineopsing}. To analyze the resulting finite-dimensional integrals, we first discuss the geometry of the localization locus and behavior of the universal bundle in sections \ref{heckepara} and \ref{univpara}. As we will see, the situation is in fact simpler than the case with no line operator. Just as in sections \ref{loc} and \ref{algebras}, this will be enough to explicitly evaluate the relevant integrals. We will use the quantum-mechanical perspective of \ref{qmechbranes} to gain a clear understanding of the meaning of the results and the algebra of the $\mathscr{O}_{\widehat{w}}(t)$ operators. 

\subsubsection{Localization locus and parabolic Hecke modifications} \label{heckepara}
Much of the discussion of section \ref{locushecke} goes through in the presence of the line operator: the localization locus for the path integral computing \eqref{puregaugelineop} reduces to a moduli space of appropriate Hecke-type modifications. The only adaptation necessary is to interpret the singularity \eqref{lineopsing} in holomorphic terms, and understand the generalizations of Hecke modifications labeled by $\widehat{w} \in W_{\text{aff}}$. In fact, just as we could largely restrict to minuscule coweights $\mu$ when studying the Coulomb branch algebras, here we can restrict to the generators of $W_{\text{aff}}$. 

As is well-known and reviewed in section 3.4 of \cite{gw}, the meaning of the singularity \eqref{lineopsing} in holomorphic terms is the following: the restriction of the gauge bundle $E_t$ to each time slice is a holomorphic $G_{\mathbb{C}}$ bundle with parabolic structure at the origin of $\mathbb{C}$. A parabolic structure at a point is simply a reduction of the structure group to a parabolic subgroup $P \subset G_{\mathbb{C}}$ at that point. Since we assume the parameter $\alpha$ in \eqref{lineopsing} is generic, in our case this means the bundle admits a reduction of the structure group to a Borel $B \subset G_{\mathbb{C}}$ at $0 \in \mathbb{C}$. This is the holomorphic analog of the breaking of the gauge group to the maximal torus along the line singularity. 

\subsubsection{}
As before, the Bogomolny equations simply assert that $E_t$, as a parabolic bundle, is constant away from the singularities, and jumps in a prescribed way when crossing them. The jumping is an analog of Hecke modification for parabolic bundles. The moduli space of solutions is the space of such successive modifications. Following our earlier notation, we denote the compactified moduli spaces by $\overline{\mathscr{M}}(\{ \widehat{w}_i; p_i \})$. 

While the parabolic analogs of Hecke modifications are well-known in geometric representation theory, they are less familiar to physicists. We will describe them in the present section explicitly in examples, at the same time explicitly describing the relevant localization loci for the present problem. We will follow the elementary approach taken in section 9 of \cite{kw}. We consider the gauge group $G = U(k)$, so $G_{\mathbb{C}} = GL_k$ and $B$ is the subgroup of upper triangular matrices. 

Consider the insertion of just a single operator $\mathscr{O}_{\widehat{w}}(t_1)$, corresponding to a simple reflection $\widehat{w} = s_{i, i+1}$ in the affine Weyl group (viewing $W_{\text{aff}} = W \ltimes \Lambda_{\text{cochar}}$, this element is $(s_{i, i + 1}, 0)$). In differential-geometric terms, for $t < t_1$ we have a line singularity with some parameters $(\alpha_1, \dots, \alpha_k)$, and in crossing $t_1$ the values of $\alpha_i$ and $\alpha_{i + 1}$ are exchanged. In complex geometry, we have a fixed parabolic bundle $E_t$ on $\mathbb{C}$ for $t < t_1$, with a reduction of the structure group to $B \subset GL_k$ at $0 \in \mathbb{C}$. We aim to understand what happens when we cross $t_1$ in the language of complex geometry.

\subsubsection{}
The first step is to note that a reduction of the structure group to $B$ at the origin is the same as a choice of a flag $\{0\} \subset V_1 \subset V_2 \subset \dots \subset V_k \simeq \mathbb{C}^k$ in the fiber of $E_t$ over $0$, where $\dim V_j = j$. To see this a bit more explicitly, note that for any rank $k$ holomorphic vector bundle, we are free to choose a trivialization in a neighborhood of the origin, which is just a collection of $k$ sections $f_j(z)$, $j = 1, \dots, k$ such that $f_1(0), \dots, f_k(0)$ are linearly independent. Fixing a parabolic structure by choosing a such a flag of subspaces means that we in addition require that 
\begin{equation} \label{flagchoice}
    V_m = \text{Span}\{ f_1(0), f_2(0), \dots, f_m(0)\}. 
\end{equation}
This reduces the structure group to $B$ at the origin, because the condition is only preserved by taking triangular linear combinations of the $f_j(0)$. In this language, it is fairly obvious what must happen when we cross $t_1$: we obtain a new bundle $E'_t$ for $t > t_1$, with a parabolic structure corresponding to a flag $\{ 0 \} \subset W_1 \subset \dots \subset W_k \simeq \mathbb{C}^k$, with $W_j = V_j$ for $j \neq i$, and 
\begin{equation}
W_i = \text{Span}\{ f_1(0), \dots, f_{i - 1}(0), f_{i + 1}(0) \}. 
\end{equation}
In other words, we exchange $f_i(0)$ and $f_{i + 1}(0)$. This flag is preserved by a subgroup $B' \subset GL_k$ abstractly isomorphic to $B$, but obtained from it by conjugation by the permutation matrix exchanging $i$ and $i + 1$. This is the meaning of a ``parabolic Hecke modification of type $s_{i, i + 1}$'': relative to some trivialization of $E_t$ by $f_j(z)$ satisfying \eqref{flagchoice}, it acts by $f_i(z) \leftrightarrow f_{i + 1}(z)$. 

\subsubsection{}
Then the space of solutions to the Bogomolny equations in this situation is just the space of parabolic bundles $E'$ on $\mathbb{C}$, with fixed trivialization at $|z| \to \infty$, related by such a Hecke modification to a fixed reference parabolic bundle $E$ and considered up to isomorphism. From the discussion above, it must be clear that this is the same as the space of flags $\{ 0 \} \subset W_1 \subset \dots \subset W_k \simeq \mathbb{C}^k$, such that $W_j = V_j$ for $j \neq i$ and $W_i \neq V_i$ for some fixed reference flag $\{ 0 \} \subset V_1 \subset \dots \subset V_k \simeq \mathbb{C}^k$. Compactifying the space of Hecke modifications simply means dropping the constraint $V_i \neq W_i$. 

Using the description based on flags, it is easy to determine the moduli space $\overline{\mathscr{M}}(s_{i, i + 1}; p)$ explicitly: such pairs of flags are uniquely determined by a choice of one-dimensional subspace $L \subset V_{i + 1}/V_{i - 1} \simeq \mathbb{C}^2$. Clearly from any such subspace we can reconstruct the appropriate flag of $W_j$'s, and to go in the other direction simply identify $L = W_i/W_{i - 1}$. So the relevant moduli space is the space of one-dimensional subspaces of $\mathbb{C}^2$, better known as $\mathbb{P}^1$. Thus, the relevant moduli space of Hecke modifications is $\overline{\mathscr{M}}(s_{i, i + 1}; p) \simeq \mathbb{P}^1$. 

In fact, this is the most fundamental example that we will actually need for computations. Just as we could restrict ourselves largely to minuscule monopole operators in the consideration of the Coulomb branch algebras, we can consider here just those Hecke modifications corresponding to generators of the affine Weyl group $W_{\text{aff}}$. For $G = U(k)$, the simple reflections describe almost all the generators. There is just one more given by the affine simple reflection $\widehat{w} = \widehat{s}_{1, k} = (s_{1, k}, (1, 0, \dots, 0, -1)) \in W \ltimes \Lambda_{\text{cochar}}$ (and another generator of length zero, which will not be so important for us, see appendix \ref{affineflags}). 

Given our discussion above, it is easy to give an explicit description of a parabolic Hecke modification corresponding to a general element $\widehat{w} = (w, \mu) \in W_{\text{aff}}$: $E'$ is obtained from $E$ by a Hecke modification of type $\widehat{w}$ if, relative to some trivialization of $E$ by sections $f_j(z)$ satisfying \eqref{flagchoice}, $E'$ is generated near $0$ by sections $z^{\mu_j}f_{w(j)}(z)$. This can be used to identify the other relevant moduli space: for the affine reflection $\widehat{s}_{1, k}$, one finds again $\overline{\mathscr{M}}(\widehat{s}_{1, k}; p) \simeq \mathbb{P}^1$. See appendix \ref{affineflags} for a more complete and formal treatment of the concepts discussed above along the lines of section \ref{grg}.

\subsubsection{}
To perform computations, we then only really need two more things: a discussion of the structure of the $U(1)_\varepsilon \times T$-fixed points in the moduli spaces, and knowledge of the jumping behavior of the universal bundle along the $t$ axis in the presence of these generalized monopole operators. We will discuss the structure of the fixed points here, and take up the discussion of the universal bundle in the next section. 

It is of course rather trivial to see the structure of $U(1)_\varepsilon \times T$ fixed points for the moduli spaces associated to $s_{i, i + 1}, \widehat{s}_{1, k}$: there are two in each. Let us concentrate on the simple reflections. If we identify once and for all $V_j \simeq \mathbb{C}^j = \text{Span}\{ e_1, \dots, e_j \}$ with a coordinate flag, then one fixed point is $W_i = \text{Span}\{e_1, \dots, e_{i - 1}, e_{i + 1} \}$, and the other is $W_i = \text{Span}\{e_1, \dots, e_{i - 1}, e_i \}$ (corresponding to no Hecke modification at all). We can thus label the first fixed point as $[s_{i, i + 1}]$, as corresponds to this reflection acting on the coordinate basis, and the second fixed point by $[1]$ since it corresponds to trivial action on the coordinate basis. In an analogous fashion, one labels one fixed point in $\overline{\mathscr{M}}(\widehat{s}_{1, k}; p) \simeq \mathbb{P}^1$ by $[s_{1, k} z^{(1, 0, \dots, 0, -1)}]$, and the other by $[1]$. 

\subsubsection{}
Because the geometry is so simple, it is likewise straightforward to determine the weights in the tangent spaces: for the fixed point $[s_{i , i + 1}]$, the weight is $\varphi_{i + 1} - \varphi_i$ (from the discussion above, it should be evident that this reduces to a computation with flag varieties). For the fixed point $[s_{1, k}z^{(1, 0, \dots, 0, -1)}]$, the weight is $\varphi_1 - \varphi_k - \varepsilon$. We will give an alternative derivation of these, closer to the index theory techniques used in section \ref{loc}, in section \ref{orbifolds}. For a geometric derivation see appendix \ref{affineflags}. 

We are interested in considering moduli spaces of successive Hecke modifications as well, though exactly as in section \ref{fixedpts}, the fixed points just correspond to tuples of the fixed points described thus far. Likewise, the weights in the tangent spaces can be obtained along the lines of section \ref{univ}, using knowledge of the universal bundle and the locality/excision property enabling us to add up local contributions as in \eqref{charn}. We turn to this issue now. 

\subsubsection{Universal bundle, again} \label{univpara}
Let $W = \mathbb{C} \times [0, 1]$, and consider as in section \ref{univ} the total space $\overline{\mathscr{M}}(\{ \widehat{w}_i; p_i \}) \times W$. We take $\widehat{w}_i$ to each be one of the simple generators of $W_{\text{aff}}$ considered in the previous section; then the moduli space is smooth and we may integrate over it straightforwardly. Denote by $t_1 < t_2 < \dots < t_n$ the positions along the $t$-axis of the generalized monopole operators, as usual. Let $\mathscr{E}$ denote the universal bundle over $\overline{\mathscr{M}} \times W$. The restrictions $\mathscr{E} \eval_{(t, 0)}$ give $U(1)_\varepsilon \times T$-equivariant holomorphic vector bundles over the moduli space, as before. The goal of this section will be to generalize the relevant parts of section \ref{univ} to include the line operator. 

The novel feature relative to section \ref{univ} is that the gauge bundle is in fact a parabolic bundle, so that as in section \ref{heckepara} the fiber over $0 \in \mathbb{C}$ comes with a distinguished flag of subspaces for each value of $t$. This means in terms of the universal bundle that we get a flag of subbundles $\{ 0 \} \subset \mathscr{E}_1 \eval_t \subset \mathscr{E}_2 \eval_t \subset \dots \subset \mathscr{E}_k \eval_t \simeq \mathscr{E} \eval_{(t, 0)}$, such that the quotients $\mathscr{E}_{i} \eval_t / \mathscr{E}_{i - 1} \eval_t := \mathscr{L}_i \eval_t$ are line bundles on $\overline{\mathscr{M}}$. 

The first thing this explains is the meaning of the supersymmetric observables built from $\varphi(t, 0)$ in the presence of the line operator. We explained previously that these are in one-to-one correspondence with polynomials on the Cartan $\mathbb{C}[\mathfrak{h}]$, with no Weyl invariance condition. The reason is that the individual diagonal elements $\varphi_i(t, 0)$ of $\varphi$ now acquire significance on the localization locus: as cohomology classes on $\overline{\mathscr{M}}$, 
\begin{equation} \label{chernrootpara}
    \varphi_i(t, 0) = c_1^{U(1)_\varepsilon \times T}\Big( \mathscr{L}_i \eval_t \Big). 
\end{equation}

\subsubsection{}
As before, there is locally constant behavior on $[0, t_1) \cup (t_1, t_2) \cup \dots \cup (t_n, 1]$ due to the jumping behavior of the $\mathscr{L}_i$ upon undergoing Hecke modifications. One may identify the equivariant variables $\varphi_i$ themselves in this language as simply the $T$-weights of the line bundles $\mathscr{L}_i \eval_t$ for $0 \leq t < t_1$. As $t$ is varied, the Hecke modifications will both act in the previous fashion (shifting the local weights by multiples of $\varepsilon$), but in the parabolic setting the local weights may now be meaningfully permuted by permuting the line bundles $\mathscr{L}_i$. This is just a restatement of the discussion of section \ref{heckepara}. 

The second important thing one learns from this is the generalization of \eqref{charn} to the setting including the line operator. To state the formula, it's best to introduce a bit of notation. Let $\sigma_i \in \{1, s_{i, i + 1} \} \simeq \mathbb{Z}_2$ denote the elements of the $\mathbb{Z}_2$ subgroups of $W$ which we identified canonically with the fixed points inside of the spaces $\overline{\mathscr{M}}(s_{i, i + 1}; p) \simeq \mathbb{P}^1$. Let us consider the moduli space $\overline{\mathscr{M}}(\{ \widehat{w}_j; p_j \})$, with $\widehat{w}_j = s_{i_j, i_j + 1}$, $j = 1, \dots, n$, and some indices $i_j$ labeling which simple reflection we place at each point $t_j$. We omit the affine reflection here just for simplicity in writing the formulas. The fixed points inside $\overline{\mathscr{M}}(\{ \widehat{w}_j; p_j \})$ correspond to $n$-tuples $(\sigma_{i_1}, \dots, \sigma_{i_n}) \in \mathbb{Z}_2^n$. Then, combining the discussion in sections \ref{univ}, \ref{heckepara}, and the present section, the character of the tangent space at such a fixed point is 
\begin{equation} \label{charnpara}
T_{(\sigma_{i_1}, \dots, \sigma_{i_n})} \overline{\mathscr{M}}(\{ \widehat{w}_j; p_j \}) = \sum_{j = 1}^n \exp(i \sigma_{i_j} \sigma_{i_{j - 1}} \dots \sigma_{i_1} \cdot(\varphi_{i_j} - \varphi_{i_j + 1})). 
\end{equation}
The origin of this formula is the same as \eqref{charn}, namely one simply adds the contributions from the individual singularities using the correct local weights for the universal bundle. To include the affine reflection, one just has to add in appropriate $\varepsilon$-shifts. We leave this as an exercise. 

\subsubsection{Quantum mechanical interpretation}
We now have everything we need to compute formulas, but before doing this we briefly remark on how the inclusion of the line operator interacts with the quantum mechanical view on the formulas we developed in section \ref{qmechbranes}, based on compactification to an effective sigma model with target $\mathscr{M}_C$. This will both illuminate the formulas and also illustrate the relation to the derived category of coherent sheaves on $\mathscr{M}_C$. 

In the presence of the $\Omega$-deformation, most of the discussion of section \ref{branes} goes through with only one difference: instead of obtaining the canonical coisotropic brane $\mathcal{B}_{cc}$ along the $r = 0$ boundary, one obtains some other brane $\widehat{\mathcal{T}}$ there. This brane is the image in the effective sigma model of the line singularity in three dimensions. Unlike the canonical coisotropic brane, it is rather difficult to characterize explicitly, though one can show that it is space-filling and that its Chan-Paton bundle must have rank $|W|$ (see the discussion near the end of \cite{nw2010}). 

While the brane $\widehat{\mathcal{T}}$ is somewhat mysterious, its endomorphism algebra $\text{End}(\widehat{\mathcal{T}})$ is more straightforward: it is spanned as a vector space by the supersymmetric local operators along the line defect that we have constructed in the past few sections. We will compute the product structure shortly. 

Apart from this, the situation is the same: we have the state $\ket{\varphi} \in \text{Hom}(\mathscr{L}_0, \widehat{\mathcal{T}})$ corresponding to the trivial operator $1$ from the $t = 0, r = 0$ corner, and states $\bra{f} \in \text{Hom}(\widehat{\mathcal{T}}, \mathscr{L}_1)$ from the $t = 1, r = 0$ corner associated to polynomials $f(\varphi(t = 1, 0))$. Regarded as a polynomial in $\varphi_i(t = 1, 0)$, this is no longer required to be symmetric. Then from the structure of the $A$-model amplitudes, we will be able to deduce the structure of the noncommutative algebra $\text{End}(\widehat{\mathcal{T}})$ and construct its action in a certain module $\text{Hom}(\widehat{\mathcal{T}}, \mathscr{L}_1)$.  

We can also consider the situation where we remove the $\Omega$-deformation. It is a somewhat less drastic operation, because the line operator forces some local operators to remain at $r = 0$ since they only make sense when supported on it. One obtains from the line operator a $(B, B, B)$ brane $\mathcal{T}$ at $r = 0$, which corresponds to a hyperholomorphic vector bundle of rank $|W|$ over $\mathscr{M}_C$. The algebra $\text{End}(\mathcal{T})$ will turn out to remain noncommutative even in the absence of $\Omega$-deformation, as expected for the endomorphism algebra of a vector bundle. In fact, $\mathcal{T}$ will be the first example of a very special class of vector bundles on Coulomb branches, an issue to which we return after discussing the incorporation of matter. 

\subsubsection{Cylindrical nilHecke algebra}
With all the preliminaries finally out of the way, we compute the operator product algebra of the supersymmetric local operators inserted along the line defect. The only non-trivial operators to consider are the generalized monopole operators. We use the quantum-mechanical notation for the amplitudes introduced in section \ref{qmechbranes}. 

Consider the correlation function of a generalized monopole operator $\mathscr{O}_{s_{i, i + 1}}(t_1)$ associated to a simple reflection and an arbitrary polynomial $f(\varphi_1(t = 1, 0), \dots, \varphi_k(t = 1, 0))$ inserted at $t = 1$. From the discussion in section \ref{heckepara}, \ref{univpara}, the path integral computing this reduces to an equivariant integral over $\overline{\mathscr{M}}(s_{i, i + 1}; p) \simeq \mathbb{P}^1$, with integrand determined by \eqref{chernrootpara}: 
\begin{equation}
\begin{split}
    \bra{f} \mathscr{O}_{s_{i, i + 1}}(t_1) \ket{\varphi} & = \int_{\overline{\mathscr{M}}(s_{i, i + 1}; p_1)} f\Big(\mathscr{L}_1 \eval_{t = 1}, \dots, \mathscr{L}_k \eval_{t = 1} \Big) \\ & =\frac{f(\varphi_1, \dots, \varphi_i, \varphi_{i + 1}, \dots, \varphi_k) - f(\varphi_1, \dots, \varphi_{i + 1}, \varphi_i, \dots, \varphi_k)}{\varphi_i - \varphi_{i + 1}}. 
\end{split}
\end{equation}
We omitted the $c_1$'s in quoting \eqref{chernrootpara}, and the final equality is simply equivariant localization together with knowledge of the local weights of the universal line bundles at the fixed points from sections \ref{heckepara}, \ref{univpara}. Note that this enables an identification of the generalized monopole operator as an operator on functions (as usual, we consider functions as a right module for these operators): 
\begin{equation} \label{demazure}
\mathscr{O}_{s_{i, i + 1}}(t_1) \longrightarrow (1 - s_{i, i + 1}) \frac{1}{\varphi_i - \varphi_{i + 1}}. 
\end{equation}
It follows readily from this formula that $\mathscr{O}_{s_{i, i + 1}}^2 = 0$. On the other hand, we may check this geometrically using \eqref{charnpara} (to lighten the notation, we write only the dependence of $f$ on $\varphi_i, \varphi_{i + 1}$ when spelling out the localization formula):
\begin{equation}
\begin{split}
\bra{f} \mathscr{O}_{s_{i, i + 1}}(t_2) \mathscr{O}_{s_{i, i + 1}}(t_1) \ket{\varphi} & = \int_{\overline{\mathscr{M}}(s_{i, i + 1}, s_{i, i + 1}; p_1, p_2)} f\Big(\mathscr{L}_1 \eval_{t = 1}, \dots, \mathscr{L}_k \eval_{t = 1} \Big) \\
& = \frac{f(\varphi_i, \varphi_{i + 1})}{(\varphi_i - \varphi_{i + 1})^2} + \frac{f(\varphi_{i + 1}, \varphi_i)}{(\varphi_{i + 1} - \varphi_i)^2} \\
& + \frac{f(\varphi_{i + 1}, \varphi_i)}{(\varphi_{i + 1} - \varphi_i)(\varphi_i - \varphi_{i + 1})} + \frac{f(\varphi_i, \varphi_{i + 1})}{(\varphi_i - \varphi_{i + 1})(\varphi_{i + 1} - \varphi_i)} \\
& = 0. 
\end{split}
\end{equation}
By essentially the same argument, one may verify that the composition of two operators $\mathscr{O}_{s_{i, i +1}}(t_2) \mathscr{O}_{s_{j, j + 1}}(t_1)$ associated to two different simple reflections comes out the same if computed algebraically using \eqref{demazure} or geoemtrically with \eqref{charnpara}. 

In the analogous way, one finds for the affine reflection 
\begin{equation}
\bra{f} \mathscr{O}_{\widehat{s}_{1, k}}(t_1) \ket{\varphi} = \frac{f(\varphi_1, \dots, \varphi_k) - f(\varphi_k + \varepsilon, \dots, \varphi_1 - \varepsilon)}{\varphi_k - \varphi_1 + \varepsilon}. 
\end{equation}
In particular, if we set $\varepsilon = 0$ then this formula looks just like the finite reflections. The algebra generated by these generalized monopole operators $\mathscr{O}_{s_{i, i + 1}}$ and the variables $\varphi_i$ is a classical object in geometric representation theory called the cylindrical version of the nilHecke algebra. It has a diagrammatic presentation using strands that we will not make extensive use of in this paper, but will explain a natural origin of the combinatorics in the following section. 

\subsection{Coupling to matter: the KLRW algebras}
With the basics of the setup with line operators understood, we now do things a bit more systematically and include matter in the picture. What we will find for the algebra of local operators along the defect is a so-called cylindrical KLRW algebra, again acting in a natural polynomial module if we use the ideas of section \ref{qmechbranes}. 

Line operators in theories with matter can be a bit awkward to work with directly, due to the need to explicitly prescribe singular behavior in the matter fields near the line operator. A much more economical way to handle them is to realize them via orbifold constructions, as explained in \cite{Nekrasov_2018}. This trick is widely used e.g. in instanton counting to efficiently compute surface defect partition functions in theories with matter. 

The basic mathematical principle underlying this is well-known and somewhat classical: sheaves with parabolic structure (the holomorphic analogs of bundles with singular connections \eqref{lineopsing}) are in one-to-one correspondence with sheaves defined on orbifolds by a cyclic group $\mathbb{Z}_p$ acting with a codimension two fixed locus. The $\mathbb{Z}_p$-equivariant sheaves on the prequotient can be interpreted on the quotient as parabolic sheaves.

The orbifolds together with the definition of the Coulomb branch given in \cite{bfn} provides a fairly direct way to make contact with the theory of Bezrukavnikov and Kaledin \cite{bk2007}, \cite{kaledin} on quantizations in characteristic $p$ and tilting generators, elucidated in the context of Coulomb branches by Webster \cite{webster2019}. We will not discuss the relation to characteristic $p$ quantization in detail in this work. 

We will first explain how to reinterpret the computations done by pedestrian reasoning in section \ref{puregaugepara} by combining the orbifold construction with the discussion in section \ref{instmon}. In terms of orbifolds, the inclusion of matter becomes obvious. To gain some additional intuition for the results we explain them in terms of the ``KLRW diagrams''. We will then write down the KLRW algebras---this is, to our first knowledge, the first direct derivation of them using gauge theory methods following \cite{gw}. By compactifying to the effective sigma model as in section \ref{branes}, we arrive at the relation between KLRW algebras and Coulomb branches \cite{webster2019}. 

\subsubsection{Reinterpretation via orbifolds} \label{orbifolds}
We would like to understand how to repeat the calculations performed in section \ref{puregaugepara} in the language of orbifolds. The basic idea is the same as in section \ref{instmon}: the local geometry of the moduli spaces near the fixed points can be modeled by a certain computation with instantons on $\mathbb{C}^2$, and by taking suitable invariants in the character incorporating both the $\mathbb{Z}_p$ and $U(1)$ actions we arrive at the correct weights in the tangent spaces for the moduli spaces describing the Bogomolny equations with a line operator. 

We would like to locally model the relevant moduli spaces of bundles on $[0, 1] \times \mathbb{C}$ with a line singularity \eqref{lineopsing}, as arising from $\mathbb{Z}_p$-equivariant bundles on $[0, 1] \times \mathbb{C}$ with $\mathbb{Z}_p$ acting in the obvious way in the second factor. Actually, we want a bit more than this: we would like to incorporate the local operators at the junction of line defects of section \ref{puregaugepara} into this picture. 

This is naturally done using the correspondence of section \ref{instmon} and lifting back up to $\mathbb{C}^2$. In this local model, the preimage of the $t$-axis is the nodal curve $z_1z_2 = 0$ in $\mathbb{C}^2$. The component with $z_2 \neq 0$ corresponds to $t < 0$, and the component with $z_1 \neq 0$ corresponds to $t > 0$. A line defect along the $t$-axis lifts in this picture to a configuration of intersecting surface defects on the $z_1$ and $z_2$ planes. The possibility of inserting a local operator at the junction of two line defects corresponds to the freedom to insert a local operator at the origin $z_1 = z_2 = 0$ of $\mathbb{C}^2$. Instantons with such surface defects can be engineered by an orbifold construction \cite{Nekrasov_2018}; in the present situation, since there are two intersecting surface defects, one wishes to consider $\mathbb{Z}_p \times \mathbb{Z}_p$ acting on $\mathbb{C}^2$, with generators $(\omega_1, \omega_2) \in \mathbb{Z}_p \times \mathbb{Z}_p$ acting by $(\omega_1, \omega_2) \cdot (z_1, z_2) = (\omega_1z_1, \omega_2 z_2)$. 

Thus we are claiming that $U(1) \times \mathbb{Z}_p \times \mathbb{Z}_p$-equivariant instantons on $\mathbb{C}^2$ can be used as a local model for solutions to the Bogomolny equations with a junction of line defects along the $t$-axis. For our computations, it suffices to show this on the level of the characters of the tangent spaces at the fixed points, which we can do by direct calculation. 

Consider the character arising from the contribution of the origin of $\mathbb{C}^2$ to the index for instantons: 
\begin{equation}
-\sum_{i, j = 1}^k \frac{\widetilde{a}_i/\widetilde{a}_j}{(1 - \widetilde{q}_1)(1 - \widetilde{q}_2)}.
\end{equation}
We have written all variables as tilded to signify that they denote the equivariant parameters for the theory on the prequotient by $\mathbb{Z}_p \times \mathbb{Z}_p$. We will first take $\mathbb{Z}_p \times \mathbb{Z}_p$-invariants in this formula, and then project to $U(1)$-invariants using contour integrals as in section \ref{loc}. The subsequent discussion will largely follow \cite{Nekrasov_2018}, adapted to the present context.

To get interesting surface operators, the gauge bundle must carry a nontrivial $\mathbb{Z}_p \times \mathbb{Z}_p$-equivariant structure. This is encoded (recall our gauge group is always $U(k)$) in a choice of two maps of finite sets $c_{1, 2}: \{ 1, \dots, k \} \to \{1, \dots, p\}$ where we think of the target as the set of irreducible representations of $\mathbb{Z}_p$. Then, in terms of $\mathbb{Z}_p \times \mathbb{Z}_p$-invariant variables $a_i$, we write $\widetilde{a}_i = a_i \widetilde{q}_1^{c_1(i)} \widetilde{q}_2^{c_2(i)}$, and $q_1 = \widetilde{q}_1^p$, $q_2 = \widetilde{q}_2^p$. 

Taking the $\mathbb{Z}_p \times \mathbb{Z}_p$-invariant part of the character is now an elementary exercise: 
\begin{equation}
\begin{split}
\Bigg( \sum_{i, j = 1}^k \frac{\widetilde{a}_i/\widetilde{a}_j}{(1 - \widetilde{q}_1)(1 - \widetilde{q}_2)} \Bigg)^{\mathbb{Z}_p \times \mathbb{Z}_p} & = \frac{1}{(1 - q_1)(1 - q_2)} \Bigg( \sum_{i, j = 1}^k \frac{a_i}{a_j} \sum_{\ell_1 = 0}^{p - 1} \sum_{\ell_2 = 0}^{p - 1} \widetilde{q}_1^{c_1(i) - c_1(j) + \ell_1}\widetilde{q}_2^{c_2(i) - c_2(j) + \ell_2} \Bigg)^{\mathbb{Z}_p \times \mathbb{Z}_p} \\
& = \frac{1}{(1 - q_1)(1 - q_2)} \Big( \sum_{\substack{c_1(j) \geq c_1(i) \\ c_2(j) \geq c_2(i)}} \frac{a_i}{a_j} + \sum_{\substack{c_1(j) < c_1(i) \\ c_2(j) \geq c_2(i)}} \frac{a_i}{a_j} q_1 \\
& + \sum_{\substack{c_1(j) \geq c_1(i) \\ c_2(j) < c_2(i)}} \frac{a_i}{a_j} q_2 + \sum_{\substack{c_1(j) < c_1(i) \\ c_2(j) < c_2(i)}} \frac{a_i}{a_j} q_1 q_2 \Big). 
\end{split}
\end{equation}
Now we project to $U(1)$-invariants exactly as in section \ref{tangent}. We write $q_1 = q^{1/2} x$, $q_2 = q^{1/2} x^{-1}$, and $a_i = e^{i\varphi_i} q^{\mu_i/2} x^{\mu_i}$. After taking the residue at infinity in $x$, one finds 
\begin{equation} \label{orbifoldcharvect}
\begin{split}
T_{[(w, \mu)]} \overline{\mathscr{M}}((w, \mu); p) = & \sum_{\substack{c_1(j) \geq c_1(i) \\ c_2(j) \geq c_2(i)}} e^{i(\varphi_i - \varphi_j)}(1 + q + \dots + q^{\mu_i - \mu_j - 1}) \theta_{\mu_i > \mu_j} \\
& + \sum_{\substack{c_1(j) < c_1(i) \\ c_2(j) \geq c_2(i)}} e^{i(\varphi_i - \varphi_j)}(1 + q + \dots + q^{\mu_i - \mu_j}) \theta_{\mu_i \geq \mu_j} \\
& + \sum_{\substack{c_1(j) \geq c_1(i) \\ c_2(j) < c_2(i)}} e^{i(\varphi_i - \varphi_j)}(q + \dots + q^{\mu_i - \mu_j - 1}) \theta_{\mu_i > \mu_j + 1} \\
& + \sum_{\substack{c_1(j) < c_1(i) \\ c_2(j) < c_2(i)}} e^{i(\varphi_i - \varphi_j)}(q + \dots + q^{\mu_i - \mu_j})\theta_{\mu_i > \mu_j}. 
\end{split}
\end{equation}
The symbol $\theta_X = 1$ if the condition $X$ is satisfied, and vanishes otherwise. 

\subsubsection{}
In the first line, we have identified this as the character of the tangent space to a moduli space of the type considered in section \ref{heckepara}, associated to a particular element of the affine Weyl group. This element is obtained in the following geometric way. Thinking in terms of the circle of irreducible representations of $\mathbb{Z}_p$, so long as we choose $p$ large enough and $c_{1, 2}$ injective, one can represent the corresponding configuration as a pair of $p$-gons (viewed, for $p$ large, as an approximation to circles), with $1, \dots, k$ sitting at the vertices according to the maps $c_{1, 2}$. The change of ordering of $\{1, \dots, k\}$ on the $\mathbb{Z}_p$ circle (viewed as with a marked point at the trivial irreducible representation) in the assignments $c_1$, $c_2$ determines an element $w$ of the Weyl group, which together with the cocharacter $\mu$ determines an element $(w, \mu)$ of the affine Weyl group. 

Thinking carefully through the geometry of the orbifold construction, one will realize that this describes precisely the fixed point contribution of a local operator at the junction of the defect \eqref{lineopsing} with itself, determined by the corresponding element of the affine Weyl group. In appendix \ref{affineflags} we verify that this formula matches exactly the character of the tangent space to the Iwahori orbit at the corresponding affine Weyl group element; in light of our discussion in section \ref{heckepara}, this gives a quick algebra-geometric proof of the validity of our main assertions. In appendix \ref{affineflags} we also verify that this formula reproduces the weights used in the localization formulas for $\mathscr{O}_{s_{i, i+1}}$ and $\mathscr{O}_{\widehat{s}_{1, k}}$ above.  

The point of view on the affine Weyl group as arising from the orbifold construction makes direct geometric contact with another way of realizing it, in terms of diagrams of strands crossing on a cylinder. One views the $c_2$ and $c_1$ $\mathbb{Z}_p$-circles (to avoid possible confusion, one should view these circles as really the Pontryagin duals to $\mathbb{Z}_p$, but we will not belabor the terminology) as the two boundaries of the cylinder, and draws $k$ strands on the cylinder. The values $c_2(i)$ and $c_1(i)$ determine the initial and final position the $i$-th strand, and the value $\mu_i$ determines its winding number. Associating each generator of the nilHecke algebra to such a diagram on the cylinder (the variables $\varphi_i$ are represented by dots on the strands), one finds the typical diagrammatic presentation of the algebra. For us, the diagrams emerge naturally from the geometry of the junction of surface defects on $\mathbb{C}^2$. 

\subsubsection{Line operators with matter}
Now there is a clear virtue of engineering the line defect via an orbifold: it is determined entirely combinatorially from the analogous computation with no defect at all. The only disadvantage in the present setting is that, since the global geometry is $[0, 1] \times \mathbb{C}$ rather than $\mathbb{R}^3$, the orbifold model is valid only locally in the moduli space. Fortunately, the only moduli spaces we need to consider directly in this problem are either $\mathbb{P}^1$ or $\{ pt \}$, and in either case a local computation at a fixed point is enough to determine the answer entirely.

In particular, one may study local operators at the junction of two line defects for theories with matter (again, for simplicity we consider SQCD, $G = U(k)$ with $n$ fundamental hypers). Line operators with matter are specified by a singularity \eqref{lineopsing} in the gauge field, together with a singularity in the matter fields $Q, \widetilde{Q}$ as they approach $z = 0$. Rather than specifying the singular behavior of the matter fields explicitly, one can engineer the desired behavior with a $\mathbb{Z}_p \times \mathbb{Z}_p$ orbifold in the local model on $\mathbb{C}^2$, just as in the previous section. The only new data entering the orbifold construction are additional choices of functions $c^{m}_{1, 2}: \{ 1, \dots, n \} \to \{ 1, \dots, p \}$ specifying how $\mathbb{Z}_p \times \mathbb{Z}_p$ embeds into the flavor symmetry group, and in principle an additional cocharacter embedding the $U(1)$ circle fiber of $\mathbb{C}^2 \to \mathbb{R}^3$ into the flavor symmetry. The combinatorial data in the local model is enough to describe entirely the discrete data labeling the defects and the operators at their junctions, as all these features are visible in a small neighborhood of the junction.

\subsubsection{}
We will in fact only consider the case where $c_1^{m} = c_2^{m} := c^m$, this map is injective, and the flavor symmetry cocharacter is trivial. Then the inequivalent defects which are described this way are labeled combinatorially by chambers $C$ encoding the positions of the $k$ marked points on the $\mathbb{Z}_p$ circle associated to the gauge group, relative to the $n$ fixed marked points associated to the matter. In terms of the strand diagrams on the cylinder appearing in the final paragraph of the previous section, the image of $c_1^m = c_2^m$ is drawn as $n$ vertical lines, in some different color than the strands associated to the $k$ ``movable'' points. The chambers just correspond to the possible ways to place the $k$ points into the ``bins'' in the $\mathbb{Z}_p$ circle specified by the $n$ marked points associated to the matter fields. This chamber structure labeling the defects is familiar in other applications of orbifold defects, e.g. in the construction of stable envelopes \cite{tamagni}.  

We wish to consider the supersymmetric local operators which can appear at the junction of two such defects labeled by chambers $C$ and $C'$. As the diagrammatics on the cylinder makes clear, to generate the whole algebra of such operators, it suffices to consider (in addition to those already considered in section \ref{puregaugepara}) an operator associated to a single point $ i \in \{1, \dots, k\}$ crossing a single line associated to $\alpha \in \{1, \dots, n \}$, in either the positive or negative directions. Call the associated operators at the junctions of the orbifold defects $\mathscr{O}^{\pm}_{(i, \alpha)}$. For the $+$ operator, in terms of the combinatorial data, we have $c_2(i) < c^m(\alpha)$ and $c_1(i) > c^m(\alpha)$, with $c_2(j) = c_1(j)$ for $j \neq i$---to get the $-$ operator, just reverse the inequalities. 

One would like to compute amplitudes such as (using again our quantum-mechanical notation) 
\begin{equation}
    \bra{f} \mathscr{O}^{\pm}_{(i, \alpha)}(t_1) \ket{\varphi}
\end{equation}
by localization. Because none of the points $\{1, \dots, k \}$ cross each other in the $\mathbb{Z}_p$-circle, translating back to our earlier explicit geometric description, the parameters $\alpha_i$ of \eqref{lineopsing} do not vary along the line defect---it follows that the moduli space of solutions to the Bogomolny equations in this situation is just a point. The matter just contributes an obstruction bundle over this point, and the only nontrivial task in computing the above correlators is to determine the $U(1)_\varepsilon \times T$-weights in that obstruction bundle. 

This is readily done by essentially repeating the calculation of the previous section, starting with the formula 
\begin{equation}
\sum_{i = 1}^k \sum_{\alpha = 1}^n \frac{\widetilde{a}_i/\widetilde{m}_\alpha}{(1 - \widetilde{q}_1)(1 - \widetilde{q}_2)}
\end{equation}
for the contribution of the origin of $\mathbb{C}^2$ to the equivariant index of the Dirac complex. We take $U(1) \times \mathbb{Z}_p \times \mathbb{Z}_p$ invariants following the same steps as before; this leads to the formula for the character of $\mathscr{F}$ (which is a vector bundle over a point)
\begin{equation} \label{mattercharorbifold}
    \mathscr{F} = \sum_{\substack{c^m(\alpha) < c_1(j) \\ c^m(\alpha) \geq c_2(j)}} e^{i(\varphi_j - m_\alpha)}. 
\end{equation}
The answer is much simpler than \eqref{orbifoldcharvect} because all of the cocharacters associated to the $U(1)$ we quotient by are trivial. Then it is clear that, for the $+$ operator 
\begin{equation}
    \mathscr{F}^+_{(i, \alpha)} = e^{i(\varphi_i - m_\alpha)}
\end{equation}
and for the $-$ operator 
\begin{equation}
    \mathscr{F}^-_{(i, \alpha)} = 0. 
\end{equation}
Then it readily follows that 
\begin{equation}
\begin{split}
    \bra{f}\mathscr{O}^+_{(i, \alpha)} \ket{\varphi} & = \text{Euler}_{U(1)_\varepsilon \times T}(\mathscr{F}) f(\varphi_1, \dots, \varphi_k) = (\varphi_i - m_\alpha) f(\varphi_1, \dots, \varphi_k) \\
    \bra{f}\mathscr{O}^-_{(i, \alpha)} \ket{\varphi} & = \text{Euler}_{U(1)_\varepsilon \times T}(\mathscr{F}) f(\varphi_1, \dots, \varphi_k) = f(\varphi_1, \dots, \varphi_k).
\end{split}
\end{equation}
It is understood in these formulas that $f$ is placed at the endpoint of the line operator corresponding to the $C'$ chamber, and that the $t = 0$ intersects a line operator in the $C$ chamber.

Another consequence of \eqref{mattercharorbifold} is that the obstruction bundle is zero on the moduli spaces studied in section \ref{puregaugepara}, since in that case only the $\alpha_i$ parameters vary but the chamber $C$ of the defect does not---in diagrammatic terms, none of the points $\{1, \dots, k\}$ cross the points $\{1, \dots, n \}$. Thus there are no nonzero terms in \eqref{mattercharorbifold}. 

\subsubsection{KLRW algebra}
Combining the remarks of the previous two sections, we arrive at the following picture. In the 3d $\mathscr{N} = 4$ theory with a $U(k)$ vector multiplet and $n$ fundamental hypers, we have constructed a natural collection of line operators $L_C$ labeled by the chambers $C$ described in the previous section, which may be engineered by orbifolds. We consider their formal direct sum $T := \bigoplus_C L_C$ and aim to describe the algebra $\text{End}(T)$ of all local operators at all possible junctions. It is generated by the operators $\mathscr{O}_{s_{i, i + 1}}^{(C)} := \mathscr{O}_{(i, i + 1)} \in \text{End}(L_C)$ (and the one associated to the affine reflection) studied in section \ref{puregaugepara}, together with the operators $\mathscr{O}^{\pm}_{(i, \alpha)}$ studied in the previous section. For operators of the ``pure gauge theory type'', we get one such operator for each choice of $C$, since these operators do not interact with matter in any interesting way. The operators $\mathscr{O}^\pm_{(i, \alpha)}$ are in $\text{Hom}(L_C, L_{C'})$ or $\text{Hom}(L_{C'}, L_C)$, depending on the sign in $\mathscr{O}^{\pm}_{(i, \alpha)}$, where $C'$ is obtained from $C$ by moving the $i$-th strand in the positive direction across the line $\alpha$. 

The explicit computation of the correlation functions of the form $\bra{f} \mathscr{O}(t_1) \ket{\varphi}$ is rather trivial, as the path integral collapses to an equivariant integral over $\mathbb{P}^1$ or a single point, and the analysis of the previous section was enough to fix the behavior of the obstruction bundle. 

The only final new feature to keep in mind is that the algebra $\text{End}(\bigoplus_C L_C)$ does not quite act on functions of the variables $(\varepsilon, \varphi_i)$, but rather on collections of functions $f_C$ indexed by the possible chambers $C$. We will change our quantum mechanical notation to $\bra{f; C}$, $\ket{\varphi; C}$ to remember the line operator $L_C$ ending on the respective boundary. 

Then we may summarize our results via (we set $\varepsilon = m_\alpha = 0$ for simplicity, to match with more standard presentations in the literature, in particular the one relevant for \cite{minaamodel}):
\begin{equation}
\begin{split}
    \bra{f; C} \mathscr{O}_{(i, i + 1)}(t_1) \ket{\varphi; C'} & = \delta_{C, C'} \frac{f_C(\varphi_1, \dots, \varphi_i, \varphi_{i + 1}, \dots \varphi_k) - f_C(\varphi_1, \dots, \varphi_{i + 1}, \varphi_i, \dots, \varphi_k)}{\varphi_i - \varphi_{i + 1}} \\
    \bra{f; C} \mathscr{O}_{(1, k)}(t_1) \ket{\varphi; C'} & = \delta_{C, C'} \frac{f_C(\varphi_1, \dots, \varphi_k) - f_C(\varphi_k, \dots, \varphi_1)}{\varphi_k - \varphi_1} \\
    \bra{f; C} \mathscr{O}^{+}_{(i, \alpha)}(t_1) \ket{\varphi; C'} & = \delta_{C = C'+(i, \alpha)} \varphi_i f_C(\varphi_1, \dots, \varphi_k) \\
    \bra{f; C} \mathscr{O}^{-}_{(i, \alpha)}(t_1) \ket{\varphi; C'} & = \delta_{C = C'-(i, \alpha)}f_C(\varphi_1, \dots, \varphi_k). 
\end{split}
\end{equation}
We use the notation $C' = C\pm(i, \alpha)$ to mean that $C'$ is obtained from $C$ by moving the $i$-th strand across the $\alpha$-th red line in the positive or negative sense. 

This explicitly identifies the actions of all operators $\mathscr{O}$ on functions, via the analog of our quantum-mechanical discussion in section \ref{qmechbranes}. The algebra generated by $\mathscr{O}$'s and $\varphi_i$'s is called the cylindrical KLRW algebra associated to the framed $A_1$ quiver, with circle node of rank $k$ and framing node of rank $n$ (this just corresponds to the fact that we can view SQCD as $A_1$ quiver gauge theory). It can be presented diagrammatically, using the combinatorics of cylinders that we made reference to in order to organize our orbifold constructions. We will not use this diagrammatics in the present paper. 

Readers who have made it this far will note that all of the above formulas and their derivations have a straightforward generalization to arbitrary quiver gauge theories; we will not write the formulas explicitly here to keep the notation light. A consequence of following the quantum mechanical approach of section \ref{qmechbranes} is that we have constructed not just the KLRW algebra for the $A_1$ quiver, but also the natural polynomial module for it. 

Also following the logic of section \ref{qmechbranes}, one can contemplate the cigar reduction of this setup. When we set $\varepsilon = 0$, each $L_C$ determines a coherent sheaf $\mathcal{T}_C$ on the Coulomb branch, and the algebra of $\mathscr{O}$ and $\varphi$ operators may be identified as the global endomorphism algebra of $\mathcal{T} := \bigoplus_C \mathcal{T}_C$. This was shown rigorously in \cite{webster2019} by comparing to the techniques in \cite{bk2007}, \cite{kaledin}. These vector bundles are mirror in the two-dimensional sense to the ``$T$-branes'' or straight line Lagrangians appearing in \cite{minaamodel}, a fact established in \cite{minapengyixuan}. While the gauge theory setup of this paper has little to offer insofar as explaining the two-dimensional mirror symmetry phenomenon, we hope we have given a reasonably complete account of the results of \cite{webster2019} and the formalism of \cite{bfn} from a more differential-geometric point of view.

\newpage 
\begin{appendices}
\section{More on boundary conditions} \label{bcdetails}
In this appendix we fill in a few details on the boundary conditions discussed in section \ref{bc} which were left implicit there. 

In the main body, we gave the boundary conditions for the bosonic fields, but we must specify what happens to the fermionic fields as well. The boundary conditions for the fermions must be compatible with the topological supersymmetry $\delta$, ensure the vanishing of surface terms arising in the variation of the action functional, and set half the fermions to zero at each boundary. This turns out to determine the boundary conditions uniquely. 

Let us start with the $t = 0$ boundary. We have the Dirichlet conditons $A \eval_{t = 0} = 0$ and $\varphi \eval_{t = 0} = \text{const}$, thus compatibility with $\delta A = \psi$ and $\delta \overline{\varphi} = \eta$ forces $\psi \eval_{t = 0} = 0$ and $\eta \eval_{t = 0} = 0$. 

The fermionic kinetic terms in the action are 
\begin{equation}
\frac{1}{e^2} \int_W \tr \Big( i \chi \wedge D_A \psi + i \chi \wedge \star D_A \psi_3 + \psi \wedge \star D_A \eta \Big) 
\end{equation}
leading upon variation to the surface terms 
\begin{equation}
\frac{1}{e^2} \int_{\partial W} \tr \Big(i \chi \wedge \delta \psi +  i (\star \chi) \wedge \delta \psi_3 + (\star \psi) \wedge \delta \eta \Big). 
\end{equation}
At the $t = 0$ end, the $\psi$ and $\eta$ boundary conditions ensure the vanishing of the first and the last term; since the bosonic field $\sigma$ obeys a Neumann condition at $t = 0$ the compatibility with the topological supersymmetry variation $\delta \sigma = \psi_3$ forces $\star \chi \eval_{t = 0} = 0$. 

At $t = 1$, the real scalar $\sigma$ is fixed while $A$ and $\varphi$ remain free. Thus compatibility with $\delta \sigma = \psi_3$ requires $\psi_3 \eval_{t = 1} = 0$. Vanishing of the surface terms requires $\chi \eval_{t = 1} = 0$ and $\star \psi \eval_{t = 1} = 0$. 

Thus the complete list of boundary conditions is, at $t = 0$
\begin{alignat}{2}
& A(t = 0, z, \bar{z}) = 0   &&   \psi \eval_{t = 0} = 0  \nonumber \\
& D_t \sigma(t = 0, z, \bar{z}) = 0 \quad && \star \chi \eval_{t = 0} = 0 \\
& \varphi(t = 0, z, \bar{z}) = \text{diag}(\varphi_1, \dots, \varphi_k) \quad && \eta \eval_{t = 0} = 0 \nonumber 
\end{alignat}
and at $t = 1$
\begin{alignat}{2}
& i_{\partial_t} F_A(t = 1, z, \bar{z}) = 0  \quad &&   \chi \eval_{t = 1} = 0  \nonumber \\
& \sigma(t = 1, z, \bar{z}) = 0 \quad && \psi_3 \eval_{t = 1} = 0 \\
& D_t \varphi(t = 1, z, \bar{z}) = 0 \quad && \star \psi \eval_{t = 1} = 0. \nonumber 
\end{alignat}

It is straightforward to check that these coincide with the full half-BPS Dirichlet and Neumann boundary conditions in the physical $\mathscr{N} = 4$ theory \cite{Bullimore_2016}, written in the twisted notation. 

\newpage
\section{Details on universal bundle} \label{universalappendix}
This appendix is devoted to recalling standard constructions related to the universal bundle $\mathscr{E}$ over $\mathscr{M} \times W$, where $W = \mathbb{R}^2 \times [0, 1]$ denotes spacetime and $\mathscr{M}$ denotes the moduli space of solutions to the Bogomolny equations (with singularities) on $W$. The main goal is to explain why the observables given by the gauge-invariant polynomials in $\varphi$ may be identified with the characteristic classes of the universal bundle on the localization locus. Since much of this material is quite standard in topological field theory \cite{atiyahjeffrey} we will be rather brief and sketchy, emphasizing only those features new in our particular situation. 

Suppose one is given a family $(A(x; m), \sigma(x; m))$ of solutions to $F_A + \star D_A \sigma = 0$, where $x$ is shorthand for coordinates on $W$ and $m$ is shorthand for coordinates on the moduli space $\mathscr{M}$ of solutions. We begin by recalling how to describe the tangent space to $\mathscr{M}$ at the point $m$. Let $(\delta A, \delta \sigma)$ be a tangent vector to the space of connections and adjoint scalars. There is a natural metric on that space given by 
\begin{equation}
\norm{(\delta A, \delta \sigma)}^2 = \int_W \tr( \delta A \wedge \star \delta A + \delta \sigma \wedge \star \delta \sigma). 
\end{equation}
A pair $(\delta A, \delta \sigma)$ represents a tangent vector to the moduli space provided it solves the linearized Bogomolny equations $D_A \delta A + \star (D_A \delta \sigma + \comm{\delta A}{\sigma}) = 0$, and is orthogonal to the tangent space to the orbit of the group of gauge transformations. An infinitesimal gauge transformation with generator $\varepsilon$ acts by $\delta A = D_A \varepsilon$, $\delta \sigma = \comm{\sigma}{\varepsilon}$; thus the gauge fixing condition is in components $D^\mu \delta A_\mu + \comm{\sigma}{\delta \sigma} = 0$, and $\delta A_t (t = 1) = 0$. The last condition arises due to a surface term in integration by parts. There is no surface term from $t = 0$ because gauge transformations are taken to be trivial at that boundary. 

Let $m^I$, $I = 1, \dots, \dim \mathscr{M}$ denote local coordinates on $\mathscr{M}$. We may decompose the derivatives of $A(x; m)$ and $\sigma(x; m)$ along the family as 
\begin{equation}
\begin{split}
\pdv{A}{m^I} & = a_I + D_A \varepsilon_I \\
\pdv{\sigma}{m^I} & = b_I + \comm{\sigma}{\varepsilon_I}
\end{split}
\end{equation}
where $a_I, b_I$ satisfy the gauge fixing conditions and $\varepsilon_I$ is the generator of a compensating gauge transformation. The compensating gauge transformation combines with the gauge field to form a connection 
\begin{equation}
\mathscr{A} = A_\mu(x; m) dx^\mu + \varepsilon_I(x; m) dm^I
\end{equation}
on the universal bundle $\mathscr{E}$ over $W \times \mathscr{M}$. Its curvature decomposes as
\begin{equation}
\mathscr{F} = \Phi + \Psi + F_A
\end{equation}
where $\Phi \in \Omega^2(\mathscr{M}) \otimes \Omega^0(W)$ and $\Psi \in \Omega^1(\mathscr{M}) \otimes \Omega^1(W)$. Let $\nabla$ denote the covariant derivative corresponding to $\mathscr{A}$, so in particular $\nabla_\mu = D_\mu$. Then clearly 
\begin{equation}
\Psi = \comm{\nabla_I}{D_\mu} dm^I \wedge dx^\mu = \Big( \pdv{A_\mu}{m^I} - D_\mu \varepsilon_I \Big) dm^I \wedge dx^\mu = a_{I \mu} dm^I \wedge dx^\mu. 
\end{equation}
We would like to get a useful characterization of $\Phi$. It turns out such a characterization is via a differential equation that it solves in the $W$ directions. Note that 
\begin{equation}
D_\mu \comm{\nabla_I}{\nabla_J} = \comm{\comm{D_\mu}{\nabla_I}}{\nabla_J} + \comm{\nabla_I}{\comm{D_\mu}{\nabla_J}} = \comm{\nabla_J}{a_{I \mu}} + \comm{a_{J\mu}}{\nabla_I}
\end{equation}
Differentiating again, we find 
\begin{equation}
\begin{split}
\Delta_A \comm{\nabla_I}{\nabla_J} = D^\mu D_\mu \comm{\nabla_I}{\nabla_J} & = - \comm{a^\mu_J}{a_{I\mu}} + \comm{\nabla_J}{D^\mu a_{I \mu}} - \comm{a_{J \mu}}{a^\mu_I} + \comm{D^\mu a_{J \mu}}{\nabla_I} \\
& = 2\comm{a^\mu_I}{a_{J \mu}} + \comm{\nabla_J}{\comm{b_I}{\sigma}} + \comm{\comm{b_J}{\sigma}}{\nabla_I}.
\end{split}
\end{equation}
In passing to the final equality we used the gauge fixing condition $D^\mu a_{I \mu} + \comm{\sigma}{b_I} = 0$. 

Now we recall that $b_I = \nabla_I \sigma$ and observe
\begin{equation}
\begin{split}
\comm{\nabla_J}{\comm{b_I}{\sigma}} + \comm{\comm{b_J}{\sigma}}{\nabla_I} & = \nabla_J(\comm{b_I}{\sigma}) - \nabla_I(\comm{b_J}{\sigma}) \\
& = \comm{\nabla_J \nabla_I \sigma}{\sigma} + \comm{\nabla_I \sigma}{\nabla_J \sigma} - \comm{\nabla_I \nabla_J \sigma}{\sigma} - \comm{\nabla_J \sigma}{\nabla_I \sigma} \\
& = -\comm{ \comm{\comm{\nabla_I}{\nabla_J}}{\sigma}}{\sigma} + 2\comm{b_I}{b_J}.
\end{split}
\end{equation}
If we define $\Psi_3 := b_I dm^I \in \Omega^1(\mathscr{M}) \otimes \Omega^0(W)$, these calculations show that (here we take $\star$ in the $W$ directions only)
\begin{equation}
\Delta_A \Phi + \comm{\comm{\Phi}{\sigma}}{\sigma} = \comm{\Psi}{\star \Psi} + \comm{\Psi_3}{\Psi_3}. 
\end{equation}
In addition, $\Phi$ satisfies the boundary conditions $\Phi \eval_{t = 0} = 0$ (because $\varepsilon_I$ vanishes identically at $t = 0$) and $D_t \Phi \eval_{t = 1} = 0$ (because $a_{tI} \eval_{t = 1} = 0$). It is easy to show that the linear operator acting on $\Phi$ has no zero mode with these boundary conditions, thus this equation determines $\Phi$. 

Now the essential point is that after discarding an irrelevant $\delta$-exact term, the part of the action which determines the $\varphi$ equation of motion is 
\begin{equation}
\frac{1}{e^2} \int_W \tr \Big( D_A \varphi \wedge \star D_A \overline{\varphi} + \comm{\sigma}{\varphi} \star \comm{\sigma}{\overline{\varphi}} + \psi \wedge \star \comm{\psi}{\overline{\varphi}} + \psi_3 \wedge \star \comm{\psi_3}{\overline{\varphi}} \Big)
\end{equation}
and thus the equation of motion is 
\begin{equation}
\Delta_A \varphi + \comm{\comm{\varphi}{\sigma}}{\sigma} = \comm{\psi}{\star \psi} + \comm{\psi_3}{\psi_3}
\end{equation}
which we recognize as identical to the equation characterizing the universal curvature $\Phi$. In addition the equations of motion for $\psi$ and $\psi_3$ imply that they satisfy the linearized Bogomolny equations and gauge fixing conditions, in other words behave like $a_I dm^I$ and $b_I dm^I$ above. Note in particular that the gauge fixing condition sheds a new light on the boundary condition $\star \psi \eval_{t = 1} = 0$ discussed in the previous appendix. Thus the zero modes of $(A, \sigma, \psi, \psi_3)$ fit together to describe differential forms on the moduli space. . 

While this analysis is all classical, the fact that variation of $e^2$ is a $\delta$-exact deformation ensures that the semiclassical limit $e^2 \to 0$ is exact, and the one-loop determinants cancel by supersymmetry. Therefore, to identify $\varphi$ on the localization locus, one may simply replace it by the solution to its classical equation of motion. This means that the observables given by gauge-invariant polynomials in $\varphi$ descend to explicit closed differential forms on the moduli space.  

We showed that $\varphi$ and $\Phi$ solve essentially the same equation, though there is a subtlety in comparing them due to boundary conditions. As explained above, $\Phi \eval_{t = 0} = 0$,  but we impose $\varphi \eval_{t = 0} = \text{diag}(\varphi_1, \dots, \varphi_k)$. 

The topological consequence of this is that the gauge invariant polynomials in $\Phi$ descend in cohomology to the characteristic classes of the universal bundle $\mathscr{E}$, while the gauge invariant polynomials in $\varphi$ become (on the localization locus) the $G$-equivariant characteristic classes with equivariant parameters $\varphi_i$. In the presence of $\Omega$-deformation, these are further deformed to the $G \times U(1)_\varepsilon$-equivariant characteristic classes of $\mathscr{E}$. 

\newpage
\section{On affine flag manifold} \label{affineflags}
In this appendix we review in some detail the Schubert cells in the affine Grassmannian and affine flag manifold. This will in particular give a more rigorous justification for the informal treatment in section \ref{heckepara}. For notational simplicity, in this appendix all groups will be complex reductive, so in particular $G$ will coincide with what is called $G_{\mathbb{C}}$ in the rest of the paper, $T$ means the complex maximal torus, et cetera. 

\subsection{Affine Grassmannian and localization weights}
As a warm-up, we begin with the discussion of the affine Grassmannian 
\begin{equation}
    \text{Gr}_G = G(\mathscr{K})/G(\mathscr{O})
\end{equation}
where $\mathscr{K} = \mathbb{C}((z))$, $\mathscr{O} = \mathbb{C}[[z]]$, as in section \ref{grg}. The torus $\mathbb{C}^\times_\hbar \times T$ acts on $\text{Gr}_G$, where $T$ acts in the natural way and $\mathbb{C}^\times_\hbar$ acts by scaling the loop variable $z$. We will be interested in fixed points, attracting manifolds, tangent spaces, etc, associated to this torus action. This material is standard in the geometric representation theory literature, but it will be interesting to compare with the differential-geometric analysis of section \ref{tangent}. 

First consider the torus $\mathbb{C}^\times_\hbar$. For simplicity we write formulas for $G = GL_k$, but there are obvious generalizations to arbitrary reductive groups. A point of the fixed locus of $\mathbb{C}^\times_\hbar$ in $\text{Gr}_G$ corresponds to a solution of the equation 
\begin{equation} \label{loopfixed}
    g(\hbar z) = g(z) \hbar^\mu
\end{equation}
where $\hbar^\mu := \text{diag}(\hbar^{\mu_1}, \dots, \hbar^{\mu_k})$ for integers $\mu_1 \geq \dots \geq \mu_k$ (which should be viewed as a dominant coweight of $G$). The choices of $\mu$ label the connected components of the fixed locus. The general solution to this equation is $g(z) = g z^{\mu}$ for a constant matrix $g$. Viewed as a set of points in $G(\mathscr{K})/G(\mathscr{O})$, it is clear that these comprise a homogeneous space for $G$. To figure out which one, we need to find the stabilizer $\text{Stab}(\mu) \subset G$ of $z^\mu$. This is easy: $p \in \text{Stab}(\mu)$ provided 
\begin{equation}
p z^\mu = z^\mu h(z)
\end{equation}
for $h(z) \in G(\mathscr{O})$. Since we are in $GL_k$, upon taking matrix coefficients this is the same as $h_{ij}(z) = z^{-(\mu_i - \mu_j)}p_{ij}$. This is in $G(\mathscr{O})$ only if $p_{ij} = 0$ for those $i, j$ such that $\mu_i > \mu_j$. This defines a parabolic subgroup, $P_\mu = \text{Stab}(\mu)$, which for generic $\mu$ is a Borel (that is, if all $\mu_i$ are distinct, this is just the condition that $p_{ij}$ is a triangular matrix). We proved that the fixed locus is a disjoint union of generalized flag varieties,
\begin{equation}
    (\text{Gr}_G)^{\mathbb{C}^\times_\hbar} = \bigsqcup_{\mu \in \Lambda^+_{\text{cochar}}} G/P_{\mu}. 
\end{equation}
Taking attracting manifolds, we get a cell decomposition of $\text{Gr}_G$ by standard Morse-theoretic/Bialynicki-Birula arguments. It is clear that the attracting manifolds with respect to the chamber $|\hbar| \ll 1$ in $\mathbb{C}^\times_\hbar$ are precisely the same as left $G(\mathscr{O})$ orbits of the points $z^\mu$, so the corresponding stratification of the affine Grassmannian reads 
\begin{equation}
\text{Gr}_G = \bigsqcup_{\mu \in \Lambda^+_{\text{cochar}}} G(\mathscr{O}) z^\mu G(\mathscr{O})/G(\mathscr{O}) = \bigsqcup_{\mu \in \Lambda^+_{\text{cochar}}} \text{Gr}^\mu_G. 
\end{equation}
From this point of view we also learn that $\text{Gr}^\mu_G$ is an affine bundle over $G/P_\mu$, and that the $\mathbb{C}^\times_\hbar \times T$ fixed points in $\text{Gr}^\mu_G$ are in natural correspondence with the $T$ fixed points in $G/P_\mu$, which are in turn in natural correspondence with points in the Weyl orbit $\lambda \in W \cdot \mu$. The explicit point in $\text{Gr}_G$ corresponding to $\lambda$ is (the right $G(\mathscr{O})$-coset of) $z^\lambda$. 

Using the description of $\text{Gr}^\mu_G$ as an attracting manifold gives in addition a way to compute the character of its tangent space at a point $z^\lambda$. As $\text{Gr}_G$ is a homogeneous space for the group $G(\mathscr{K})$, its tangent space at the point $z^\lambda$ may be presented as a quotient of the Lie algebra by the Lie algebra of the stabilizing subgroup of $z^\lambda$:
\begin{equation}
    T_{z^\lambda}\text{Gr}_G \simeq \mathfrak{g}((z))/\text{Lie}(\text{Stab}(z^\lambda)).
\end{equation}
To avoid confusion with earlier notations, here $\text{Stab}(z^\lambda) \subset G(\mathscr{K})$. Moreover this identification is $\mathbb{C}^\times_\hbar \times T$-equivariant if we let $\mathbb{C}^\times_\hbar$ act in the obvious way and $T$ act via the adjoint action. 

By essentially the same considerations as above, one identifies $\text{Stab}(z^\lambda) \simeq z^{\lambda} G(\mathscr{O}) z^{-\lambda}$. Then, writing $\sum_\alpha$ for the sum over roots of $\mathfrak{g}$, and writing $(\hbar, t)$ for equivariant variables of $\mathbb{C}^\times_\hbar \times T$, 
\begin{equation}
\begin{split}
T \text{Gr}_G \eval_{z^\lambda} & = \sum_\alpha  \sum_{m_\alpha \in \mathbb{Z}} \hbar^{m_\alpha} t^\alpha - \sum_{\alpha} \sum_{m_\alpha \geq 0} \hbar^{m_\alpha + \alpha \cdot \lambda} t^\alpha \\
& = \sum_\alpha t^\alpha \sum_{m_\alpha < \alpha \cdot \lambda} \hbar^{m_\alpha}. 
\end{split}
\end{equation}
We are also writing this equation in $K$-theory, identifying the tangent space at fixed point with its $\mathbb{C}^\times_\hbar \times T$-character. 

The tangent space to the $\mathbb{C}^\times_\hbar$-attracting manifold at the point $z^\lambda$ in the chamber $|\hbar| \ll 1$ just consists of the terms in this expression containing nonnegative powers of $\hbar$, so we immediately conclude 
\begin{equation}
T \text{Gr}^\mu_G \eval_{z^\lambda} = \sum_{\substack{\alpha \, \, \, s.t. \\ \alpha \cdot \lambda > 0}} t^\alpha (1 + \hbar + \dots + \hbar^{\alpha \cdot \lambda - 1})
\end{equation}
which is in exact agreement with \eqref{char1} for $G = GL_k$, up to obvious changes in notation. 

\subsection{Affine flag manifold}
When written in this language, it is immediate to generalize this discussion to the affine flag variety, which is the relevant space for describing the localization loci in section \ref{klrw} upon translation to algebraic geometry. Fixing a choice of Borel subgroup $B \subset G$, define the Iwahori subgroup $\mathscr{I} \subset G(\mathscr{O})$ as 
\begin{equation}
    \mathscr{I} := \{ g \in G(\mathscr{O}) | g(0) \in B \}. 
\end{equation}
The affine flag variety is defined as the quotient 
\begin{equation}
    \mathscr{F}\ell := G(\mathscr{K})/\mathscr{I}. 
\end{equation}
Just as $\text{Gr}_G$ may be viewed as the moduli space of Hecke modifications of a fixed trivial bundle on $\mathbb{D} = \text{Spec} \mathbb{C}[[z]]$ at $z = 0$ of unspecified type, the affine flag manifold may be viewed as the moduli space of Hecke modifications of a \textit{parabolic} bundle, of unspecified type. 

For $G = GL_k$, this means the following. When we quotient by $\mathscr{I}$ instead of $G(\mathscr{O})$, we allow only gauge transformations which preserve a choice of fixed flag (since we haved fixed $B$, and do not consider it up to conjugacy) in the fiber of a rank $k$ vector bundle over $z = 0$. This means we can meaningfully distinguish choices of flag in the fiber over $z = 0$, which describe different choices of reduction of the structure group of the bundle to $B$ at $z = 0$. Thus, transition functions $g(z)$ considered only up to equivalence in $\mathscr{I}$ describe isomorphism classes of bundles together with a parabolic structure at $z = 0$. It is clear by comparison with section \ref{klrw} and \cite{gw} that the affine flag variety plays precisely the same role for describing disorder operators in the presence of a line defect as the affine Grassmannian does for the usual monopole operators. 

In general, a point of $\mathscr{F}\ell$ corresponds to a $G$-bundle on the formal disk $\mathbb{D}$, isomorphic to the trivial bundle away from $z = 0$, and with a reduction of the structure group to $B$ at $z = 0$, considered up to isomorphism. There is an additional $G/B$ worth of degrees of freedom in the modification at $z = 0$, corresponding to the possibility of moving the flag in the fiber around. More precisely, there is a map $\mathscr{F}\ell \to \text{Gr}_G$ with fiber $G/B$, as is obvious also from the quotient description.

We would like to get a stratification of the affine flag manifold analogous to the one we found for the affine Grassmannian. For this it is useful to use the whole torus $\mathbb{C}^\times_\hbar \times T$, and not just the loop rotation part.

The torus $\mathbb{C}^\times_\hbar \times T$ acts naturally on $\mathscr{F}\ell$. We would like to determine the $\mathbb{C}^\times_\hbar \times T$-fixed locus. Letting $(\hbar, t) \in \mathbb{C}^\times_\hbar \times T$, points in $\mathscr{F}\ell^{\mathbb{C}^\times_\hbar \times T}$ correspond to solutions of the equation
\begin{equation}
tg(\hbar z) = g(z) \hbar^\mu t' 
\end{equation}
for some $\mu = \text{diag}(\mu_1, \dots, \mu_k)$, $t' \in T$. We no longer impose any inequalities on $\mu_i$, since we are free to conjugate our compensating transformation only in $\mathscr{I}$ and not in all of $G(\mathscr{O})$. The general solution to this is $g(z) = w z^\mu$ where $w \in W$ conjugates $w^{-1} t w = t'$. The data $(w, \mu)$
naturally determine an element of the affine Weyl group $W_{\text{aff}} = W \ltimes \Lambda_{\text{cochar}}$, and we find 
\begin{equation}
    \mathscr{F}\ell^{\mathbb{C}^\times_\hbar \times T} = \bigsqcup_{(w, \mu) \in W_{\text{aff}}} wz^\mu \mathscr{I}/\mathscr{I}. 
\end{equation}
To get a stratification of $\mathscr{F}\ell$ itself, we just need to take attracting manifolds. If we pick the chamber $|\hbar| \ll |t_1| < \dots < |t_k| < 1$, the attracting manifolds of the fixed points $wz^\mu$ are exactly the same as orbits under the Iwahori, so we conclude 
\begin{equation}
    \mathscr{F}\ell = \bigsqcup_{(w, \mu) \in W_{\text{aff}}} \mathscr{I} wz^\mu \mathscr{I}/\mathscr{I}. 
\end{equation}
The cells in this stratification, each isomorphic to an affine space, are precisely the moduli spaces of Hecke modifications of a parabolic bundle of type $(w, \mu)$ described informally in section \ref{heckepara}. Their closures provide natural, though in general singular, compactifications of these moduli spaces.

We can use the same trick to compute the character of the tangent spaces to the Iwahori orbits at the fixed points $wz^\mu$. Namely, we may identify the stabilizer $\text{Stab}(wz^\mu) \simeq wz^\mu \mathscr{I} (wz^\mu)^{-1} \simeq z^{w(\mu)} w\mathscr{I}w^{-1} z^{-w(\mu)}$. Then, by the same logic as in the previous section, we have the formula 
\begin{equation}
T\mathscr{F}\ell \eval_{wz^\mu} = \sum_\alpha t^\alpha \sum_{m_\alpha \in \mathbb{Z}} \hbar^{m_\alpha} - \sum_{\alpha > 0} \hbar^{\alpha \cdot \mu} t^{w \cdot \alpha} - \sum_\alpha \sum_{m_\alpha \geq 0} \hbar^{\alpha \cdot \mu + m_\alpha + 1} t^{w \cdot \alpha}.
\end{equation}
Here $\alpha > 0$ means $\alpha \in \text{Lie}(B)$, and we used $\text{Lie}(\mathscr{I}) = \text{Lie}(B) \oplus z \mathfrak{g}[[z]]$. This further simplifies to 
\begin{equation} \label{aflchar}
\begin{split} 
    T\mathscr{F}\ell \eval_{w z^\mu} & = \sum_\alpha t^{w \cdot \alpha} \sum_{m_\alpha \leq \alpha \cdot \mu} \hbar^{m_\alpha} - \sum_{\alpha > 0} \hbar^{\alpha \cdot \mu} t^{w \cdot \alpha} \\
    & = \sum_{\alpha > 0} t^{w \cdot \alpha} \sum_{m_\alpha < \alpha \cdot \mu} \hbar^{m_\alpha} + \sum_{\alpha < 0} t^{w \cdot \alpha} \sum_{m_\alpha \leq \alpha \cdot \mu} \hbar^{m_\alpha} \\
    & = \sum_{\alpha > 0} \Big( t^{w \cdot \alpha} \sum_{m_\alpha < \alpha \cdot \mu} \hbar^{m_\alpha} + t^{-w \cdot \alpha} \sum_{m_\alpha \leq -\alpha \cdot \mu} \hbar^{m_\alpha} \Big).
\end{split}
\end{equation}
To pick out the terms corresponding to the attractive directions requires slightly more care, but amounts to the following recipe for the chamber we have chosen. We discard any term with a negative power of $\hbar$, keep terms with positive powers of $\hbar$, and for the constant terms in $\hbar$ we keep only the terms such that the exponent of $t$ is a positive root. This leads to the formula 

\begin{equation} 
\begin{split}
T(\mathscr{I} wz^\mu \mathscr{I}/\mathscr{I}) \eval_{wz^\mu} & = \sum_{\alpha > 0, w \cdot \alpha > 0} t^{w \cdot \alpha}(1 + \hbar + \dots + \hbar^{\alpha \cdot \mu - 1}) \theta_{\alpha \cdot \mu > 0} + \sum_{\alpha > 0, w \cdot \alpha < 0} t^{w \cdot \alpha} (\hbar + \dots + \hbar^{\alpha \cdot \mu - 1}) \theta_{\alpha \cdot \mu > 1} \\
& + \sum_{\alpha < 0, w \cdot \alpha < 0} t^{w \cdot \alpha} (\hbar + \dots + \hbar^{\alpha \cdot \mu}) \theta_{\alpha \cdot \mu > 0} + \sum_{\alpha < 0, w \cdot \alpha > 0} t^{w \cdot \alpha}(1 + \hbar + \dots + \hbar^{\alpha \cdot \mu}) \theta_{\alpha \cdot \mu \geq 0}
\end{split}
\end{equation}
where the symbol $\theta_X = 1$ if the condition $X$ is satisfied and vanishes otherwise. This clearly agrees with \eqref{orbifoldcharvect} up to obvious changes in notation, when $G = GL_k$.

In particular, specializing to $1 \in \mathbb{C}^\times_\hbar \times T$, it is straightforward to verify that this leads to the following formula for the dimension of the Schubert cell:
\begin{equation}
    \ell(w, \mu) := \dim(\mathscr{I} wz^\mu \mathscr{I}/\mathscr{I}) = \sum_{\alpha > 0, w \cdot \alpha > 0} |\alpha \cdot \mu| + \sum_{\alpha > 0, w \cdot \alpha < 0} |\alpha \cdot \mu - 1|
\end{equation}
which is also the length function on the affine Weyl group. Generators of the affine Weyl group are the elements of length $1$ (and the identity which has length zero). For $G = GL_k$, the elements with $\mu = 0$ and length $1$ are just the elements of $W$ which change the sign of exactly one positive root, which are the simple reflections $s_{i, i + 1}$ for $i = 1, \dots, k - 1$. There is one more generator of $W_{\text{aff}}$, given by the reflection in an affine root $s_{1k} z^{(1, 0, \dots, 0, -1)}$ ($s_{1k}$ denotes the permutation in $S_k$ exchanging $1$ and $k$ while preserving all the other letters). It is easy to check from the above formula that this element has length $1$. Actually, for the affine Weyl group of $G = GL_k$, there is also a $\pi_1(GL_k) \simeq \mathbb{Z}$ worth of length zero elements corresponding to powers of $z^{(1, \dots, 1)}$, but these will not play an important role for us. 

For the affine Weyl group elements of length $1$, the Schubert cells $\mathscr{I}wz^\mu\mathscr{I}/\mathscr{I} \simeq \mathbb{C}$, and their closures $\overline{\mathscr{I}wz^\mu \mathscr{I}/\mathscr{I}} \simeq \mathbb{P}^1$. In section \ref{heckepara} we argued that the path integral of 3d $\mathscr{N} = 4$ gauge theory in the presence of a codimension two defect, and with our usual choice of boundary conditions, reduces to an integral over these orbit closures taken in $\mathbb{C}^\times_\hbar \times T$-equivariant cohomology. To determine the localization weights at the fixed points we may therefore use the character formula obtained above. 

For the simple reflections, the part of \eqref{aflchar} tangent to the Iwahori orbit reads 
\begin{equation}
    T(\mathscr{I}s_{i, i + 1} \mathscr{I}/\mathscr{I}) \eval_{s_{i, i + 1}} = \sum_{\alpha > 0, w \cdot \alpha < 0} t^{- w \cdot \alpha} = \frac{t_i}{t_{i + 1}}. 
\end{equation}
Likewise, for the affine reflection, we obtain (note the only term with a nonzero contribution is that with $\alpha = (1, 0, \dots, 0, -1)$):
\begin{equation}
    T(\mathscr{I} s_{1k}z^{(1, 0, \dots, 0, -1)}\mathscr{I}/\mathscr{I}) \eval_{s_{1k} z^{(1, 0, \dots, 0, -1)}} = \hbar \frac{t_k}{t_1}. 
\end{equation}
This completes the more rigorous computation of the weights in the tangent spaces stated in \ref{heckepara}. 
\end{appendices}

\newpage

\printbibliography
    
\end{document}